\documentclass[twocolumn,prd,superscriptaddress,longbibliography]{revtex4-2}
\pdfoutput=1
\usepackage{bm}
\usepackage{graphicx}
\usepackage{booktabs}
\usepackage{slashed}
\usepackage{hyperref}
\usepackage{amssymb,amsmath,amsfonts}
\usepackage{color}
\usepackage[utf8]{inputenc}
\usepackage[caption=false]{subfig}

\newcommand{\vx}{\vec{x}}

\def \beq  {\begin{equation}}
\def \eeq  {\end{equation}}
\def \beqar {\begin{eqnarray}}
\def \eeqar {\end{eqnarray}}
\def\sqr#1#2{{\vcenter{\vbox{\hrule height.#2pt
\hbox{\vrule width.#2pt height#1pt \kern#1pt
\vrule width.#2pt}\hrule height.#2pt}}}}

\def\vx {{\vec x}}

\def\vf {{\varphi}}

\def\Tr {{\rm Tr}}

\def\vx {{\vec x}}

\def\del {\partial}

\def\bz {{\bar{z}}}

\def\A {{\cal A}}

\def\G {{\cal G}}

\def\vf {{\varphi}}

\def\half{\textstyle{1\over 2}}

\def\tdc{T_{\rm deconf}}

\begin{document}

\title{Fractional topological charge in $SU(N)$ gauge theories without dynamical quarks}

\author{V. P. Nair}
\affiliation{Department of Physics, City College of New York, CUNY, New York, NY 10031}
\author{Robert D. Pisarski}
\affiliation{Department of Physics, Brookhaven National Laboratory, Upton, NY 11973}

\begin{abstract}
  In $SU(N)$ gauge theories without dynamical quarks, we discuss how configurations with fractional
  $\mathbb{Z}_N$ magnetic charge also have fractional
  topological charge, $\sim 1/N$, and dominate topologically nontrivial fluctuations in the confining vacuum.
  They are not solutions of the classical equations of motion,
  but arise as quantum solutions of the effective Lagrangian, whose size  is essentially fixed,
  on the order of the confinement scale. We give both a general mathematical analysis and illustrative solutions.
  We discuss strong evidence for this from numerical simulations on the lattice, and suggest definitive tests.
  We also speculate how these objects change with the introduction of dynamical quarks, and their effects
  especially at low temperature and nonzero density.
\end{abstract}

\maketitle

\section{Introduction}
\label{sec:intro}

It is well known that topologically non-trivial configurations play an essential role in
Quantum ChromoDynamics (QCD).
For $J^P = 0^-$ mesons, such configurations must be present in order to split the iso-singlet $\eta'$ meson
from the octet of pions, kaons, and the $\eta$ meson \cite{Witten:1979vv,Veneziano:1979ec,tHooft:1986ooh}.
They also affect the spectrum of mesons with higher spin \cite{Giacosa:2017pos}, 
and contribute to the proton and photon structure functions in polarised deep-inelastic scattering
\cite{Veneziano:1989ei,Shore:1990zu,Shore:1991dv,Shore:1997tq,Narison:1998aq,Bass:2004xa,Shore:2007yn,Tarasov:2020cwl,Tarasov:2021yll}.

In weak coupling for a $SU(N)$ gauge theory, the dominant configurations are instantons,
which are self-dual solutions to the classical equations of motion.
By asymptotic freedom instanton effects can be reliably computed
at high temperature, $T$, or quark chemical potential, $\mu_{\rm qk}$
\cite{Gross:1980br,KorthalsAltes:2014dkx,KorthalsAltes:2015zpx,Pisarski:2019upw,Rennecke:2020zgb,Boccaletti:2020mxu,Nogradi2023}
\footnote{\label{density_comment} The computation at $T\neq 0$ is complete to one loop order, while that at 
  $\mu_{\rm qk} \neq 0$ is lacking, although doable \cite{Nogradi2023}.}
The action for a single instanton with unit topological charge is $= 8 \pi^2/g^2$.  
Since by asymptotic freedom the running coupling constant, $g^2(T) \sim 8 \pi^2/(c \log(T))$,
\footnote{For an $SU(N)$ gauge theory coupled to $N_f$ flavors of massless quarks,
  $c = (11 N - 4 N_f C_f)/3$, $C_f = (N^2-1)/(2N)$.  The factor of $T^4$ in Eq. (\ref{highT_top_sucep})
  arises from the integral over the instanton scale size, $\rho$.}
the topological susceptibility falls off sharply at high temperature,
\begin{equation}
  \chi_{\rm top}(T)\sim T^4 \, \exp(- 8 \pi^2/g^2(T)) \sim 1/T^{c-4} \;\; , \;\; T \rightarrow \infty \; .
  \label{highT_top_sucep}
\end{equation}
Remarkably, numerical simulations in lattice
QCD find that this power law holds down to temperatures as low as $\approx 300$~MeV
\cite{Borsanyi:2015cka,GrillidiCortona:2015jxo,Bonati:2015vqz,Borsanyi:2016ksw,Frison:2016vuc,Petreczky:2016vrs,Taniguchi:2016tjc,Lombardo:2020bvn,Jahn:2021qrp,Borsanyi:2021gqg,Chen:2022fid}
\footnote{
The overall magnitude of the topological susceptibility is about an order of magnitude greater than the
instanton result at one loop order, but surely it is necessary to include
the corrections to the instanton density at two loop order for $T \neq 0$.  
Multi-instanton configurations can also contribute at low $T$ \cite{Rennecke:2020zgb}.}.
This is valid when $\mu_{\rm qk} = 0$; at the end of the paper we discuss what might occur
for cold, dense quarks, where $T \ll \mu_{\rm qk} \neq 0$.

For $T < 300$~MeV, in QCD the topological susceptibility is
not that of a dilute instanton gas. To understand the what generates the topological
susceptibility at low temperature and in vacuum, it is useful to consider a
$SU(N)$ gauge theory without dynamical quarks.
In this case, a global $\mathbb{Z}_N$ symmetry is
spontaneously broken above the temperature for deconfinement, $T_{\rm deconf}$ \cite{Gaiotto:2014kfa}.
Numerical simulations of the pure gauge theory find that the dependence on $N$ is rather weak.
Forming a dimensionless ratio between the topological susceptibility and the square of the string tension,
at zero temperature $\chi_{\rm top}(0)/\sigma^2$, only varies by $\approx 10\%$ between $N=3$
\cite{Alles:1996nm,Durr:2006ky,Luscher:2010ik,Xiong:2015dya,Jahn:2018dke,Giusti:2018cmp}
and higher $N$
\cite{Lucini:2001ej,DelDebbio:2006yuf,Vicari:2008jw,GarciaPerez:2009mg,Fodor:2009nh,Fodor:2009ar,Panagopoulos:2011rb,Lucini:2012gg,Bonati:2013tt,Bonati:2015sqt,Ce:2016awn,Bonati:2016tvi,Kitano:2017jng,Itou:2018wkm,Bonanno:2018xtd,Bonati:2018rfg,Bonati:2019kmf,Bonanno:2020hht,Kitano:2021jho,Bennett:2022gdz}.  
These results suggest that as $N \rightarrow \infty$, the topological susceptibility does not vary with temperature
in the confined phase.
Numerical simulations find that the
deconfining phase transition is of first order for three or more colors, with
$\tdc \approx 270 $~MeV for $N = 3$
\cite{Boyd:1996bx,Borsanyi:2012ve,Shirogane:2016zbf,Caselle:2018kap,Borsanyi:2022xml}
For $N \geq 3$, $\chi_{\rm top}(T)$ jumps at $\tdc$, and then
falls off rapidly with increasing $T$, dominated by instantons above some temperature close to $\tdc$.

It is difficult to see how the topological susceptibility could be due to instantons in the confined phase
at large $N$ 
\footnote{
  There could be a dense liquid or crystal of instantons \cite{Carter:2001ih,Liu:2018znq}.
  However, as noted by Witten \cite{Witten:1978bc}, for 
  the topological susceptibility not to be exponentially suppressed in $N$, it is necessary
  to show that the action of such a liquid or crystal vanishes at $\sim N$, and so instead is $\sim 1$.
  Consider how this could occur semi-classically. The effective action for a classical instanton of size $\rho$
  is $\sim N$, and is a power series in $\lambda = g^2 N$:
  $${\cal S}_{\rm eff}(\rho \Lambda) = N \; \frac{8 \pi^2}{\lambda} \left( 1 + \lambda \log(\rho \Lambda) + \ldots \right),$$
  up to terms $\sim N^0$, where $\Lambda$ is the renormalization mass scale.
  At large $N$, one minimizes the effective action with respect to $\rho$, with the dominant configuration
  an instanton of a single scale size, $\rho_\infty$.  This is not sufficient, though:
  it is necessary that the effective action at this scale size
  {\it vanishes}, with ${\cal S}_{\rm eff}(\rho_\infty \Lambda) = 0$, up to corrections $\sim 1$.
  It is not obvious how this might come about.
  }.
Holding $g^2 N \equiv \lambda$ fixed as $N \rightarrow \infty$, the action of a single instanton
in the partition function is
exponentially suppressed, $\sim \exp(- (8 \pi^2/\lambda) N)$.
This quandary was recognized originally by 
Witten \cite{Witten:1979vv} and Veneziano \cite{Veneziano:1979ec}, who argued nevertheless that in vacuum
the topological susceptibility is not exponentially suppressed at large $N$ \cite{Witten:1998uka}.

The most natural possibility is that there are objects with fractional topological charge $\sim 1/N$, whose
contribution directly survives at infinite $N$, $\sim \exp(-8 \pi^2/\lambda)$.
On a torus, 't Hooft constructed explicit solutions with fractional topological charge $\sim 1/N$,
with twists in $Z(N)$ electric and magnetic charge
\cite{tHooft:1980kjq,tHooft:1981nnx,vanBaal:1982ag,Sedlacek:1982cd,Nash:1982kp,Killingback:1984en}.
Since they depend upon $Z(N)$ twisted boundary conditions, however, they are limited to finite volume.

From numerical simulations on the lattice, Gonzalez-Arroyo and Martinez \cite{Gonzalez-Arroyo:1995ynx},
and then with Montero \cite{Gonzalez-Arroyo:1995isl},
used cooling techniques to isolate configurations with nontrivial $Z(N)$ electric and magnetic charge.
They argue that in infinite volume, that objects with fractional topological charge
dominate, and produce a topological susceptibility
and string tension of the correct magnitude.  See, also, Refs.
\cite{Gonzalez-Arroyo:1996eos,GarciaPerez:2000aiw,Gonzalez-Arroyo:2019wpu,DasilvaGolan:2022jlm,Gonzalez-Arroyo:2023kqv}.

At nonzero temperature in the deconfined phase, 
analytically Kraan, van Baal, Lee, and Lu (KvBLL)
\cite{Lee:1997vp,Lee:1998vu,Lee:1998bb,Kraan:1998pm,Kraan:1998sn,GarciaPerez:1999hs,Diakonov:2002fq,Eto:2004rz,Eto:2006pg,Eto:2006mz,Bruckmann:2009nw,Diakonov:2009jq,Poppitz:2008hr,Anber:2021upc}
showed that at nonzero temperature
instantons can be viewed as made of $N$ constituents, each with topological charge $1/N$.
The constituents of KvBLL instantons have
nontrival holonomy at spatial infinity, though, and thus cannot be pulled arbitrarily far apart.

A useful limit is to study gauge theories on a femto-slab, where one spatial dimension, $L$,
is very small, with $L \Lambda_{\rm QCD} \ll 1$, 
where $\Lambda_{\rm QCD}$ is the renormalization mass scale of QCD.
Over large distances the theory reduces to one in $2+1$ dimensions
\cite{Polyakov:1976fu,Dunne:2000vp,Kogan:2002au,Kovchegov:2002vi}.
On a femto-slab, semi-classical techniques demonstrate
that monopole-instantons with topological charge $\sim 1/N$ are ubiquitous
\cite{Unsal:2006pj,Unsal:2007vu,Unsal:2007jx,Unsal:2008ch,Shifman:2008ja,Simic:2010sv,Anber:2011gn,Poppitz:2012sw,Anber:2013doa,Aitken:2017ayq,Unsal:2020yeh,sym14010180,Tanizaki:2022ngt}.
Presumably this survives when the size of the slab increases to distances where $L \Lambda_{\rm QCD} \sim 1$.

In this paper we consider configurations with fractional topological charge $\sim 1/N$ in both
the $\mathbb{CP}^{N-1}$ model in $1+1$ dimensions
\cite{Gross:1977wu,Witten:1978bc,DAdda:1978vbw,DiVecchia:1979pzw,Berg:1979uq,Fateev:1979dc,Rothe:1980rp,Zhitnitsky:1988zd,Eto:2004rz,Eto:2006pg,Eto:2006mz,Ahmad:2005dr,Lian:2006ky,Bruckmann:2007zh,Brendel:2009mp,Unsal:2020yeh},
and for $SU(N)$ gauge theories, without dynamical quarks, in
$3+1$ dimensions.

The $\mathbb{CP}^{N-1}$ model is useful.  Instantons with
integral topological charge are solutions of the classical equations of motion
\cite{Atiyah:1978ri}.  For both models,
the classical action is invariant under a scale symmetry, which implies that instantons have a scale size, $\rho$,
which ranges from zero to infinity.  In Sec. \ref{sec:cpn},
we construct solutions which stationary points of a quantum
effective action, whose size is manifestly non-perturbative, on the order of the confinement
scale.  They only exist as multi-valued solutions.

In Sec. \ref{sec:math} we give a general mathematical analysis, extending that of Refs.
\cite{tHooft:1980kjq,tHooft:1981nnx,vanBaal:1982ag,Sedlacek:1982cd,Nash:1982kp,Killingback:1984en}.
The crucial element is that the configuration space of a pure $SU(N)$ gauge theory
has a global $Z(N)$ symmetry, which is absent when dynamical quarks are present.
We demonstrate generally how objects with integral topological charge are composed of those
with fractional $Z(N)$ magnetic charge.
As in the $\mathbb{CP}^{N-1}$ model, the relevant solutions are necessarily multivalued;
similarly, we also suggest that in gauge theories that their size is fixed, on the order of the confinement scale.

In the deconfined phase we outline the construction of a solution
with topological charge $1/N$ in the deconfined phase in Sec. \ref{sec:simple}.
It is distinct from KvBLL instantons,
which have non-trivial holonomy at spatial infinity, and integral magnetic charge.  Instead,
our solution, a type of $Z(N)$ dyon,
has trivial holonomy at spatial infinity, but being multivalued, has fractional $Z(N)$ magnetic charge.

In Sec. \ref{sec:lattice}  we discuss results from numerical simulations on the lattice.
Notably, with significant effort the $N$-dependence of higher moments of fluctuations in the topological charge can
be measured
\cite{DelDebbio:2006yuf,Vicari:2008jw,Panagopoulos:2011rb,Bonati:2015sqt,Bonati:2016tvi,Bonati:2018rfg,Bonati:2019kmf,Bonanno:2020hht}.
For example, the kurtosis coefficient $b_2$
is the ratio of fluctuations between the fourth and second moments in the topological charge.
We discuss how results by 
Bonanno, Bonati, and D'Elia \cite{Bonanno:2020hht}, on the $N$-dependence of $b_2$,
strongly suggest that there is a dense liquid of fractionally charged objects.
While intriguing, it is an indirect measurement.  We then review how objects with fractional topological charge
can be measured directly on the lattice, following Edwards, Heller, and Narayanan \cite{Edwards:1998dj}.

We conclude with a discussion of how quarks and $\mathbb{Z}_N$ dyons might interact in QCD.
In this we depend crucially upon recent results from the lattice
by Biddle, Kamleh, and Leinweber \cite{Biddle:2022acd,Biddle:2022zgw,Leinweber:2022dpp,Biddle:2023lod},
and speculate how $\mathbb{Z}_N$ dyons might generate topologically nontrivial fluctuations
in cold, dense quark matter.

Some of our conclusions are familiar.
To measure a system whose total topological charge is fractional,
$\mathbb{Z}_N$  twisted boundary conditions must be used
\cite{tHooft:1980kjq,tHooft:1981nnx,vanBaal:1982ag,Sedlacek:1982cd,Nash:1982kp,Killingback:1984en,Gonzalez-Arroyo:1995ynx,Gonzalez-Arroyo:1995isl,Gonzalez-Arroyo:1996eos,GarciaPerez:2000aiw,Gonzalez-Arroyo:2019wpu,DasilvaGolan:2022jlm,Gonzalez-Arroyo:2023kqv}.
Even so, following Gonzalez-Arroyo and Martinez
\cite{Gonzalez-Arroyo:1995ynx,Gonzalez-Arroyo:1995isl,Gonzalez-Arroyo:1996eos,GarciaPerez:2000aiw,Gonzalez-Arroyo:2019wpu,DasilvaGolan:2022jlm,Gonzalez-Arroyo:2023kqv}, and as on a femto-slab
\cite{Polyakov:1976fu,Dunne:2000vp,Kogan:2002au,Kovchegov:2002vi,Unsal:2006pj,Unsal:2007vu,Unsal:2007jx,Unsal:2008ch,Shifman:2008ja,Simic:2010sv,Anber:2011gn,Poppitz:2012sw,Anber:2013doa,Aitken:2017ayq,Unsal:2020yeh,sym14010180,Tanizaki:2022ngt},
the vacuum is a condensate of objects with fractional topological charge.

In other ways we differ from previous analysis.
The multi-valued nature of our configurations is rather unlike known solutions, although we suggest tests to
distinguish them from, {\it e.g.}, KvBLL instantons.  
While we argue that our configurations dominate the topological susceptibility,
our methods are not adequate to demonstrate that they produce confinement, as in Refs.  
\cite{Gonzalez-Arroyo:1995ynx,Gonzalez-Arroyo:1995isl,Gonzalez-Arroyo:1996eos,GarciaPerez:2000aiw,Gonzalez-Arroyo:2019wpu,DasilvaGolan:2022jlm,Gonzalez-Arroyo:2023kqv}.   Lastly, for our configurations, which are on the order
of the confinement scale, the really crucial question is {\it how} dense the condensate is, and whether they
can distinguished from other fluctuations.   In this
analyzing how the confingurations evaporate as the
temperature is raised, especially near the deconfining transition, will be essential.

\section{$\mathbb{CP}^{N-1}$ model}
\label{sec:cpn}

Consider a nonlinear sigma model in two spacetime dimensions, with the target space the complex projective space
$\mathbb{CP}^{N-1}$. The target space is formed by
$N$ complex variables $z^i$, the so-called homogeneous
coordinates, identifying $z^i \sim w \, z^i$, $w \in \mathbb{C} - \{ 0\}$.
The magnitude of $w$ can be fixed by setting $\sum_{i=1}^N \bz^i z^i = \bz\cdot z = 1$,
and the phase of $w$ removed by gauging the overall $U(1)$ symmetry.
The Lagrangian density is
\begin{equation}
  {\cal L} = \frac{1}{g^2} \sum_{i = 1}^N \; |D_\mu z^i(x)|^2 \;\; ; \;\; D_\mu = \partial_\mu - i A_\mu(x) \; .
  \label{lag_cpn}
\end{equation}
where the $z^i$'s satisfy $\bz\cdot z = 1$ at each point in two spacetime dimensions, $x_\mu$.
The gauge field $A_\mu$ ensures that the $z$'s are also invariant under local $U(1)$ transformations,
$z^i(x) \rightarrow {\rm e}^{i \alpha(x)} z^i(x)$
\footnote{Notice that the constraint $\bz\cdot z = 1$ defines a sphere $S^{2 N -1}$,
while the gauge symmetry removes a phase, i.e., $S^1$, which leaves $S^{2 N -1}/ S^1$.
This is another way to define $\mathbb{CP}^{N-1} $.}.
The absence of a kinetic term for the gauge field $A_\mu(x)$ 
ensures that no new degrees of freedom are introduced via the gauging.
Classically $A_\mu$ can be eliminated by its equation of motion,
\beq
A_\mu = - {i \over 2} \left( {{\bz^i \del_\mu z^i  - \del_\mu \bz^i \, z^i }\over \bz \cdot z }\right) \; ,
\label{lag-cpn2}
\eeq
so that ${\cal L}$ can be re-expressed entirely in terms of the $z^i$ and $\bz^i$.
The only coupling constant  in Eq. (\ref{lag_cpn}) is $g^2$, which is dimensionless.

The Lagrangian in Eq. (\ref{lag_cpn}) is obviously invariant under global $SU(N)$ transformations,
$z^i \rightarrow U^{i}_{~j} \, z^j$, where $U \in SU(N)$.
Elements of the center of $SU(N)$, which is ${\mathbb{Z}_N}$, are special.  These are
the $U_k = {\rm e}^{2 \pi i k/N} {\bf 1}$, where $k = 0, 1\ldots (N-1)$,
underwhich the $z^i$'s transform as $z^i \rightarrow e^{2 \pi i k/N} z^i$.
Since the $z^i$ are homogeneous coordinates, though, this ${\mathbb{Z}_N}$ rotation
can be eliminated by a global $U(1)$ rotation.
This reduces the full global symmetry to $SU(N)/{\mathbb{Z}_N}$ \cite{Witten:1978bc,DAdda:1978vbw,Unsal:2020yeh}.

The topological winding number is
\begin{equation}
  Q = \frac{1}{2 \pi} \int d^2 x \; \epsilon^{\mu \nu} \partial_\mu A_\nu \; .
\end{equation}
For fields where $z^i$ approaches a constant at infinity, $Q$ is an integer.
All classical configurations with $ Q \neq 0$ are known \cite{DAdda:1978vbw}. 
Keeping $g^2 N$ fixed as $N \rightarrow \infty$, the value of the classical action is uniformly $\sim N$.  The fluctuations
about arbitrary instanton configurations have also been computed.  While they simplify for
$N=2$, where it reduces to an $O(3)$ model \cite{Gross:1977wu,Berg:1979uq,Fateev:1979dc},
for $N > 2$ the integration over the collective coordinates of the instantons is not tractable.  Even so,
at large $N$ they appear to be exponentially suppressed.

As discussed in the seminal papers \cite{Witten:1978bc,DAdda:1978vbw},
the large $N$ analysis in the quantum theory
can be carried out by
introducing a Lagrange multiplier field $\lambda(x)$ to impose the constraint 
$\bz\cdot z - 1 = 0$ and then integrating out
the $z^i$
fields. This leads to an effective action for $A_\mu$ and $\lambda$,
\begin{equation}
  {\cal S}_{\rm eff} = N \, {\rm tr} \, \log \left( -D_\mu^2 + i \lambda \right) - i \int d^2 x \; \frac{\lambda(x)}{g^2}
  \; .
\label{eff_act_cpn}
\end{equation}
The corresponding equations of motion are
\begin{equation}
  N \; {\rm tr} \; D_\mu^{\rm cl} \frac{1}{-(D^{\rm cl}_\mu)^2 + m^2(x)} = 0 \; ,
  \label{eqn-A}
\end{equation}
and
\begin{equation}
  N \; {\rm tr} \; \frac{1}{-(D^{\rm cl}_\mu)^2 + m^2(x)} - \frac{i}{g^2} = 0 \; ,
  \label{eqn-lamb}
\end{equation}
for arbitrary solutions $A_\mu(x) = A_\mu^{\rm cl}(x)$ and $i \lambda(x) = m^2(x)$.  In vacuum
$A_\mu^{\rm cl}=0$ and $m^2(x)$ is constant, with the dynamically generated mass
$m$ related to the coupling constant $g^2$ through dimensional transmutation \cite{Witten:1978bc,DAdda:1978vbw}.

The quantum dynamics of the model, defined by Eq.
(\ref{eff_act_cpn}), is rather different from that expected from the classical analysis 
of Eq. (\ref{lag_cpn}). Classically, the constraint
$\bz \cdot z = 1$ necessarily breaks the $SU(N)$ global symmetry, and the $z$ fields are massless.
In contrast, the quantum vacuum is invariant under the $SU(N)/{\mathbb{Z}_N}$ symmetry, in accord 
the Mermin-Wagner-Coleman theorem. The expectation values of the
$z^i$ vanish at infinity, and are massive fields.

As shown by Witten, Eq. (16) of Ref. \cite{Witten:1978bc}, an effective theory for
the quantum theory can be written in a derivative expansion,
\beq
{\cal S}_{\rm eff} = \int \vert (\del_\mu -i A_\mu ) Z_{\rm qu}^i \vert^2 - m^2  \bar{Z}_{\rm qu}^i Z_{\rm qu}^i
- {N \over 48 \pi m^2} F_{\mu\nu}F^{\mu\nu} + \cdots
\label{eff_act-2}
\eeq
Here the $Z_{\rm qu}^i$ are effective fields to describe the low energy behavior of Eq. (\ref{eff_act_cpn}),
and are {\it not} the original $z^i$ fields of Eq. (\ref{lag_cpn}): the $Z_{\rm qu}^i$
are massive fields, which vanish at infinity.  As mentioned above,
the mass squared $m^2$ is dynamically generated, fixed by the solution of
Eqs. (\ref{eqn-A}) and ({\ref{eqn-lamb}). 

There are many other terms which contribute to the derivative expansion in Eq. (\ref{eff_act-2}).
These include terms with higher (covariant) derivatives of $Z_{\rm qu}^i$; higher derivatives of $A_\mu$, which
by gauge invariance must enter as powers of $F_{\mu \nu}^2$; and lastly, derivatives of
the constraint field $\lambda(x)$.  None of these higher order terms qualitatively change our
discussion of the stationary points of Eq. (\ref{eff_act-2}).
We stress that in this effective theory, the global symmetry of
$SU(N)/{\mathbb{Z}_N}$ is unbroken, so by confinement
the only allowed states are $\mathbb{Z}_N$-invariant \cite{Witten:1978bc}.

We now turn to consider
topologically nontrivial field configurations of the quantum effective action in Eq. (\ref{eff_act-2}).
A general analysis of how to proceed in general field theories is outlined in the Appendix.
Adopting polar coordinates $(r,\varphi)$ in two dimensions, at large $r$ the solution for the gauge field must satisfy
\begin{equation}
  A_\mu dx^\mu \sim Q  d \varphi \; ; \; {\rm i.e.,} ~A_\varphi \sim {Q \over r} \; \; {\rm as~}r \rightarrow \infty \; ,
  \label{asymp_behavior_int}
\end{equation}
so that $\int F \sim Q \neq 0$.
This should be accompanied by a suitable
ansatz for $m^2(x)$ is a function of $r$, but
we do not elaborate on this since it is not important for the main thread of our arguments.
Determining the solution of the nonlocal equations of motion in
Eqs. (\ref{eqn-A}) and (\ref{eqn-lamb}) is not elementary. 
But there is one aspect of any such solution which is worth of note, and which
in fact is a recurrent point throughout our analysis.
While the classical action is invariant under scale transformations, at large $N$ the quantum effective
action is not. Thus while Eq. (\ref{asymp_behavior_int}) fixes the behavior at infinity, the nature of
the full solution varies over a distance $\sim 1/m$.

We next turn to the possibility of configurations with fractional topological charge.
On a femto-slab classical instantons were constructed by Unsal \cite{Unsal:2020yeh}; their
size is necessarily on the order of the width of the slab.
In contrast, we consider quantum instantons
in vacuum.  As a first step, consider the spherically symmetric configuration
\beq
F_{12} = \begin{cases}
{2 / (N a^2)} \hskip .1in&r < a\\
0& r>a
\end{cases}
\label{vortex1}
\eeq
This corresponds to $Q = \int (F/2\pi) = 1/N$, and the gauge potential
\beq
A_\mu dx^\mu = - {1\over N \pi a^2} \int d^2x'\, {\epsilon_{\mu\nu} (x- x')^\nu \over \vert x- x'\vert^2} \rho(x')\, dx^\mu
\label{vortex2}
\eeq
where $\rho (x')$ is equal to $1$ in a small disk of radius $a$, and zero elsewhere.
This configuration is a slightly thickened vortex, with 
Eq. (\ref{vortex2}) consistent with the asymptotic behavior
of Eq. (\ref{asymp_behavior_int}), except that now the topological charge is fractional, with $Q = 1/N$.
The contribution of this configuration to the action (\ref{eff_act-2}) is
\beq
{N \over 48 \pi m^2} \int F^2 = {1\over 6 N m^2 a^2}
\label{vortex3}
\eeq
The action for the higher terms will be similarly suppressed, since they must involve powers of $F_{\mu \nu}$.

The ansatz of Eq. (\ref{vortex2}) may be written for $\rho$ with support around the origin, as
\beq
A_\mu dx^\mu = f(r) d\varphi = {1\over N} \begin{cases}
(r^2 /a^2)\, d\varphi \hskip .1in& r\leq a\\
d\varphi& r>a\\
\end{cases}
\label{vortex3a}
\eeq

Turning to the $Z_{\rm qu}^i$-dependent part of the action, the only point of subtlety
is about the phase of $Z_{\rm qu}^i$. With the background of Eqs. (\ref{vortex1}) and (\ref{vortex2}),
the parallel transport of $Z_{\rm qu}^i$ in a full circle around the origin
(or the location of the vortex) gives $Z_{\rm qu}^i \rightarrow e^{2 \pi i/ N} Z_{\rm qu}^i$.
The phase may also be viewed as the Aharonov-Bohm phase acquired by
$Z_{\rm qu}^i$ in a circuit around the vortex.
While the $Z_{\rm qu}^i$ are not single-valued, this phase can be removed
by an $SU(N)$ transformation in its center $\mathbb{Z}_N$.

We can now supplement the ansatz Eqs. (\ref{vortex2}) or (\ref{vortex3a})
with a suitable ansatz for $Z^i_{\rm qu}$, such as
\beq
Z_{\rm qu}^1 = e^{i \varphi /N} h(r) , \hskip .1in Z^i_{\rm qu} = 0, \hskip .1in
i = 2, 3, \cdots, N,
\label{vortex4}
\eeq
or any $SU(N)/{\mathbb{Z}_N}$ transformation of this.
We have incorporated the aperiodicity in $\varphi$ mentioned above,
namely, $Z_{\rm qu}^i(r,2 \pi) = {\rm e}^{2 \pi i/N} Z_{\rm qu}^i(r,0)$. This multi-valuedness
is where we differ from previous analysis by Berg and L\"uscher \cite{Berg:1979uq} and
Fateev, Frolov, and Schwarz \cite{Fateev:1979dc}.

Taking the matter part of the action as in Eq. (\ref{eff_act-2}), we find
\beq
S_{\rm eff} = 2 \pi \int dr \; r \left[ \left( {\del h \over \del r}\right)^2
+ {h^2 \over r^2} \left( f - {1\over N} \right)^2 + m^2 h^2
\right] +\cdots
\label{vortex5}
\eeq
The behavior of $h(r)$ for small and large values of $r$ can be inferred from the equation of motion for $h$, namely,
\beq
- {1\over r} {\del \over \del r} \left( r {\del h \over \del r} \right)
+ \left( f - {1\over N} \right)^2 {h \over r^2} + m^2 h +\cdots = 0
\label{vortex6}
\eeq
By examining the small $r$ and large $r$ limits of this equation, we 
can see that
\beq
h(r) \sim \begin{cases}
r^{1\over N} \hskip .2in& r \rightarrow 0\\
e^{- m r} & r \rightarrow \infty\\
\end{cases}
\label{vortex7}
\eeq
Notice that $h$ vanishes exponentially as $r \rightarrow \infty$.
This is a significant point. While the gauge part of the configuration
(\ref{vortex3a}) is like an Abrikosov-Nielsen-Olesen vortex, the asymptotic
behavior of $Z^i_{\rm qu}$ is very different. We may also note that the vanishing of $Z^i_{\rm qu}$ 
at spatial infinity is consistent with the fact that any configuration of finite action should reproduce
vacuum behavior at spatial infinity.

Introducing a scale factor $r_0$, a simple ansatz consistent with 
Eq. (\ref{vortex7}) is
\beq
h(r) = C \, { u^{1\over N} \over 1+ u^{1\over N}} e^{-\mu u},
\hskip .1in u = {r\over r_0}, ~\mu = m r_0
\label{vortex8}
\eeq
It is easy to verify that $S_{\rm eff}$ is finite with this ansatz and that
the term in Eq. (\ref{vortex5}) involving both $f$ and $h$ depends on 
$a$. Along with the gauge field contribution
in Eq. (\ref{vortex3}), we get a nonlinear expression involving
$a$ and $C$. Treating these as variational parameters, 
we can obtain values which minimize the action, at least within the
class of ans\"atze in Eqs. (\ref{vortex3a}), (\ref{vortex4}), and ({\ref{vortex8}).

A few comments are in order at this point.  Notice that, even if $C =0$,
we do have a vortex-like configuration in Eq. (\ref{vortex3a}).
Although extremization with respect to $a$ with just this term leads to
$a \rightarrow \infty$, there are terms with higher powers of $F$ in the action,
indicated by ellipsis in Eq. (\ref{eff_act-2}).
Including them and extremizing will lead to a finite value for $a$,
which can only be set by the single dimensionful parameter in the model, the mass $m$.
The inclusion of higher $Z_{\rm qu}^i$-dependent terms produces
terms which are of order $C^4$ and higher. Thus we expect that
extremization including such terms gives finite values to
both $a$ and $C$.  As noted, this is equivalent to solving the nonlocal equations of motion
in Eqs. (\ref{eff_act_cpn}) and (\ref{eqn-A}).

To frame this more generally, at large $N$, we can again consider expanding 
Eq. (\ref{eff_act_cpn}) in powers of $A_\varphi$ which is of order
$1/N$, based on our ansatz.  As for the solution with integral
topological charge, and as in the example above,
as a solution of the quantum action the size is $\sim 1/m$.  The term linear in
$A_\varphi$ vanishes by the equation of motion for the gauge potential.  Taking $m^2(x) = m^2$, then,
the term in the action $\sim 1$ automatically vanishes. 
The expansion of the effective action to quadratic
order in $A_\varphi$, i.e., as in Eq. (\ref{eff_act-2}),
shows that the nonzero contribution of the $A$-part of the action will be of order $1/N$.

This demonstrates that there are configurations with fractional topological charge, $\sim 1/N$.
We have not computed the exact configuration, for reasons we now discuss.  At nonzero $\theta$,
the energy of the vacuum is an even function in $\theta$
\cite{Witten:1979vv,Veneziano:1979ec,DelDebbio:2006yuf,Vicari:2008jw,Bonati:2016tvi},
\begin{equation}
  E(\theta) - E(0) = \frac{\chi}{2} \, \theta^2 \left( 1 + b_2 \, \theta^2 + \ldots \right) \; ;
  \label{energy_theta}
\end{equation}
$\chi$ is the topological susceptibility, $\chi = \langle Q^2 \rangle/V$, where $V$ is the volume of space-time.
The second coefficient, $b_2$, is the kurtosis of the topological charge,
\begin{equation}
  b_2 = - \, \frac{1}{12} \; \frac{\langle Q^4\rangle - 3 \langle Q^2 \rangle^2}{\langle Q^2 \rangle} \; .
  \label{definition_b2}
\end{equation}
For both $\chi$ and $b_2$, all expectation values are computed at $\theta = 0$.  

The topological susceptibility is a dimensional quantity, and so by dimensional transmutation $\chi \sim m^2$.
The $N$-dependence of the coefficients can be understood by assuming that fluctuations
in the topological charge are fractional, $\Delta Q \sim 1/N$.  Since there are
$N$ ways of inserting a charge $1/N$ in the theory, $\chi \sim N(1/N)^2 \sim 1/N$.  Similarly,
$b_2 \sim (1/N)^4/(1/N)^2 \sim 1/N^2$, {\it etc.}  We did not compute the exact configurations with fractional
topological charge because that can be computed from the free energy in a constant background field for $F_{\mu \nu}$
\cite{Vicari:2008jw,Bonati:2016tvi}.
Thus the $\theta$-dependence is certainly described by a dense liquid of fractionally charged instantons.  

\section{Towards fractional instantons in 4d}
\label{sec:math}
\mathchardef\mhyphen="2D

We now turn to nonabelian gauge theories in four dimensions.
One of the key steps in understanding configurations of fractional topological charge is the identification of what is meant by the gauge group. Although this question
has been analyzed before, it is useful to collect some of the basic ideas here.
We will first consider the boundary values for gauge transformations based on the Gauss law (or the nature of the test functions to be used in implementing the Gauss law) and how these are related to
charge quantization conditions. 
This will clarify the nature of the configuration space
and will naturally lead to the possibility of fractional topological charges.
\setcounter{secnumdepth}{2}
\subsection{Gauss law in the $E$-representation}

We consider the gauge theory in the $A_0 =0$ gauge. 
We must then impose the Gauss law on the wave functions.
Quantization conditions on the electric charge will be important for us, so it is 
more appropriate to consider wave functions in the representation 
which are eigenstates of the electric field operators $E^a$. 
In other words, the wave functions are functionals of the electric field.
The Gauss law operator is given by
\beq
G^a(x)= \nabla_i E_i^a + f^{abc} A^b_i E^c_i \label{1}
\eeq
where $f^{abc}$ are the structure constants of the Lie algebra of $G$.
The gauge potential and the electric field obey the usual commutation rule
$[A^a_i(x), E^b_j(y)]= i \delta^{ab}\delta_{ij}\delta (x-y)$, so that, in
the $E$-representation, $A^a_i = i (\delta / \delta E^a_i)$.
The physical wave functions $\Psi$ are selected by the condition 
that the Gauss law operator must annihilate them.
This condition can be written as
\begin{eqnarray}
  &\int_M& \theta^a(x) G^a(x) ~\Psi = \nonumber \\
&=&  \int_M \theta^a(x)\left[ \nabla_iE^a_i \nonumber 
    -i f^{abc} E^b_i {\delta \over \delta E^c_i}\right] \Psi =0 \\
  \label{2}
\end{eqnarray}
(The integral is over the spatial manifold $M$.)
This law should be required only for test functions $\theta^a(x)$
obeying certain conditions; the nature of these conditions
will be clear  from the following discussion.
Treating $\theta^a(x)$ as an infinitesimal group parameter, (\ref{2}) may be
written as
\beqar
\delta \Psi &\equiv& \Psi( U^{-1}EU) -\Psi(E)\nonumber\\
&=&-\left[ i \int_M \theta^a(x) \nabla_iE^a_i \right]~\Psi
\label{3}
\eeqar
where $E_i= T^a E^a_i,~ U=\exp(iT^a \theta^a )\approx 1+iT^a \theta^a$, 
$T^a$ being hermitian matrices which form a basis for the Lie algebra of
$G$, with $[T^a,T^b]= if^{abc}T^c$. For the fundamental representation, we write
$T^a = t^a$ and normalize 
them by $\Tr (t^at^b) =\half \delta^{ab}$.
The quantity $U^{-1}E_iU$ is the gauge transform of $E_i$ and hence $\delta \Psi$ 
measures the change of $\Psi$ under a gauge transformation with parameter 
$\theta^a(x)$. Obviously, if $\Psi$ is a solution to (\ref{3}), then so is
$\Psi\, f(E)$ where $f(E)$ is a gauge-invariant function of $E_i$.
The general solution to (\ref{3}) may therefore be written as $\Psi= \rho\,
\Phi (E)$, where $\Phi (E)$ is an arbitrary gauge-invariant function and
$\rho$ is a particular solution to
\beq
\delta \rho +\left[ i \int_x \theta^a(x) \nabla_iE^a_i \right]~\rho =0 \label{4}
\eeq

A finite transformation, and the corresponding variation of $\rho$, can be
obtained by composition of infinitesimal transformations. Assume that, for
an electric field $E_i$, we have started from the identity and built up a 
finite transformation $U$. At this point, the electric field is given by 
${\cal E}_i=U^{-1}E_iU$. A further infinitesimal transformation would be 
given by $iT^a \theta^a = U^{-1}\delta U$. Thus (\ref{4}), written for 
an arbitrary point on the space of $U$'s, becomes
\beqar
\delta \rho + 2 \int_M \Tr (\nabla_i{\cal E}_i~ U^{-1}\delta U) \rho &=&0\nonumber\\
\delta (\log \rho ) = -2  \int_M \Tr (\nabla_i{\cal E}_i ~U^{-1}\delta U) 
&\equiv& \Omega 
\label{5}
\eeqar
One can integrate this equation along a curve in the space of $U$'s from 
the identity to $U$ to obtain the change of $\rho$ under a finite 
transformation. With $\delta$ interpreted as a derivative 
on the space of $U$'s, $U^{-1}\delta U$ is a covariant vector
(or one-form) and the result of the integration is generally path-dependent.
For the result to be independent of the path of integration, the curl of $ \Tr (\nabla\cdot {\cal E} \,U^{-1}\delta U)$, viewed as a covariant vector or
as a one-form on the space of the $U$'s, must vanish. Thus the
the integrability condition for (\ref{5}), or 
the path-independence for the change in $\rho$, becomes
\beq
\delta \Omega = \delta \left[ -2 \int \Tr (\nabla_i{\cal E}_i U^{-1}\delta U)
\right] =0
\label{5a}
\eeq
Here we take $\delta$ to signify the exterior derivative, so that
$\delta$ acting on a one-form (or covariant vector) gives the curl.
We now write $\Omega =\Omega_1 +\Omega_2$ with
\beqar
\Omega_1&=& 2\int_M\Tr ({\cal E}_i \nabla_i(U^{-1}\delta U))\nonumber\\
&=& 2 \int_M\Tr \left[ E_i \left( \nabla_i (\delta U U^{-1} ) - [\nabla_i U U^{-1}, 
\delta U U^{-1} ] \right) \right]\nonumber\\
\nonumber\\
\Omega_2&=& -2 \oint_{\partial M}\Tr ({\cal E}_i U^{-1}\delta U )dS^i \nonumber \\
&=& -2 \oint_{\partial M}\Tr (E_i \delta U  ~U^{-1})dS^i
\label{6}
\eeqar
It is easily checked, using $\delta (\delta U U^{-1} ) = (\delta U U^{-1})^2$, {\it without the need of any integration-by-parts on $M$}, that
$\delta \Omega_1 =0$. 
For the second term, we find
\beq
\delta \Omega_2= -2 \oint_{\partial M} \Tr ( E_i \delta U U^{-1} 
\delta U U^{-1} ) dS^i \label{7}
\eeq
This is in general not zero. Indeed if $U$ is constant on $\partial M$,
$\delta \Omega_2 =-2 \Tr[Q(\delta UU^{-1})^2],~Q=\oint E_idS^i$.
In this case, $\delta \Omega_2$ has the form of the coadjoint orbit
two-form on $G/H$, where $H\subset
G$ is the subgroup which commutes with the charge $Q$. This form, well-known as the basis for the Borel-Weil-Bott theory on group representations,
is a nondegenerate
two-form on $G/H$. In order to have $\delta \Omega =0$, we must therefore
implement the Gauss law only for those $U$'s which obey the restriction
\beq
\oint_{\partial M} \Tr [ E_i \delta UU^{-1}] dS^i =0 \label{8a}
\eeq
This is basically the cocycle condition
which allows us to build up finite transformations using sequences of
infinitesimal transformations.
If $E_i$ on $\partial M$ can be arbitrary, this
condition (\ref{8a})  would require fixing $U$ to some
value, say, $U_\infty$
on $\partial M$. (If $U_\infty$ is held fixed, $\delta U_\infty =0$, so that the 
requirement (\ref{8a}) is trivially satisfied.)
This clarifies the nature of the test functions $\theta^a$
in (\ref{4}) in imposing the Gauss law: {\it The test functions must be so chosen that
they lead to $U_\infty$ on $\del M$}.

The key question for us is then:
What are the allowed values of $U_\infty$? This will be determined by
the charge quantization conditions. But before we take up this issue,
a
comment on the asymptotic behavior of $U$ is in order.
Although we argued using constant $U$ on $\del M$, generically, we cannot impose the Gauss law for $U$'s which are not constant on $\del M$ as well,
since $\delta \Omega_2 $ will not vanish for such cases.
In fact, $U$'s which are not constant
on $\del M$ correspond to degrees of freedom which
are physical and generate the
``edge modes" of a gauge theory. If we consider the boundary to be at
spatial infinity, such edge modes are irrelevant. This will be the case 
for our analysis in this paper.

Returning to constant values of $U$ on $\del M$, and
the identification of the possible values of $U_\infty$, we start with the
question:
How does $\Psi$ change under
 transformations which go to a constant $U \neq U_\infty$.
It is easily seen that the action of a general infinitesimal
transformation
\beq
\delta A^a_i = -\del_i \theta^a - f^{abc} A_i^b \theta^c, \hskip .3in
\delta E^a_i = - f^{abc} E^b_i \theta^c
\label{G1}
\eeq
is given by
\beq
\delta \Psi = \left[ i \int_M D_i\theta^a(x) ~E^a_i \right]~\Psi 
= \exp\left[ iQ^a \theta^a (r= \infty )\right] ~\Psi\label{G2}
 \eeq
where $Q^a$ is the electric charge $Q^a = \oint E^a_i dS_i$. Thus transformations which go to a constant $\neq U_\infty$ act as a Noether symmetry, under which the charged states undergo a phase transformation. If the only charges in the theory correspond to the adjoint representation of $G$ and its products, i.e., if the states are invariant under ${\mathbb Z}_N \in SU(N)$,
then the wave functions are {\it invariant} for those $U$'s which go to an element of the center
${\mathbb Z}_N$ at spatial infinity.
We have seen that 
we can implement the Gauss law only for transformations which go to a fixed element
$U_\infty$ at spatial infinity. Now we see that the allowed choices for
$U_\infty$ correspond to an element of the center ${\mathbb Z}_N$.

To recapitulate briefly,  we have seen that the true gauge transformations of the theory, in the sense of corresponding to a redundancy of description,
are of the form $U(\vx )$ with:\\
a) $U \rightarrow$ a constant $U_\infty$ at spatial infinity\\
b) $U_\infty \in \mathbb{Z}_N$ for a theory with charges which are
$\mathbb{Z}_N$-invariant.
\subsection{Charge quantization and $U_\infty$: An alternate argument}

There is another way to arrive at the conclusion of the previous subsection,
namely, by a direct analysis of the charge quantization conditions.
Notice that, for $U$'s obeying (\ref{8a}), we can write $\Omega$ as
\beqar
\Omega &=& 2\int_M \Tr [ E_i U \nabla_i (U^{-1}\delta U) U^{-1}]\nonumber\\
&=& 2\int_M \Tr [ E_i \delta (\nabla_i U U^{-1})] \nonumber \\
&=& \delta \left(
2\int_M \Tr [ E_i \nabla_iUU^{-1}]\right)
\label{9}
\eeqar
Using this and integrating (\ref{5}) from the identity to $U$, we obtain
\beq
\rho( U^{-1}EU) = \rho (E) \exp\left(2 \int_M \Tr (E_i\nabla_iUU^{-1})\right)
\label{10}
\eeq
This equation will be important for us; it will have a key role in subsequent analysis. So another comment and another derivation will be appropriate
before proceeding. One concern about (\ref{10})  might be that we have used integration from the identity to $U$. 
In three spatial dimensions, 
since $\Pi_3 (G)={\mathbb Z}$, there are $U$'s which are not connected to the identity. Even though the derivation given above does not quite make it clear,
the result (\ref{10}) 
holds even for $U$'s which are 
not in the connected component. This can be seen by the following 
alternate derivation borrowed from \cite{Jackiw:1996ec}.
\beqar
\Psi (E) &=&\int d\mu (A) \exp\left[{ 2\int_M \Tr (E_iA_i)}\right]~ \Psi (A)\nonumber\\
&=& \int d\mu (A) \exp \left[{2\int_M \Tr (E_iA_i)}\right]  \nonumber \\
&& \;\;\;\;\;\;\;\; \times \; \Psi (U^{-1}AU+U^{-1}\nabla U)\nonumber\\
&=&\exp(-2\int_M \Tr (E_i\nabla_iUU^{-1})) \nonumber \\
&& \times \int d\mu (A) 
\exp\left[{2\int_M \Tr (U^{-1}E_iUA_i)}\right]~ \Psi (A)\nonumber\\
&=& \exp(-2\int_M \Tr (E_i\nabla_i UU^{-1}))~ \Psi (U^{-1}EU)
\label{12}
\eeqar
where we have first used the gauge invariance of the wave functions in the
$A$-representation (i.e. $ \Psi (A) = \Psi (U^{-1}AU+U^{-1}\nabla U)$) and then changed the variable of integration from
$A$ to $U^{-1}AU +U^{-1}\nabla U$. With $\Psi = \rho\, \Phi (E)$, (\ref{12})
gives (\ref{10}). (This derivation is simpler but the earlier analysis does 
reveal some interesting aspects of imposing the Gauss law.)

Equation (\ref{10}) contains certain charge quantization requirements which 
can be used to see why the boundary values of $U$ can be an element
of the center, rather than strictly being the identity.
We can show that $\rho (E)$ of (\ref{10})
will vanish unless certain conditions are satisfied by $E_i$. 
For this, it is adequate to examine some special configurations.
The basic strategy is to choose an electric field configuration and
a $U$ which commutes with the chosen configuration for $E_i$.
Equation (\ref{10}) then gives an identity of the form 
$\rho =\rho e^{i\lambda}$ where the
phase $\lambda$ is given by the integral $2\int_M \Tr (E_i\nabla_iUU^{-1})$.
This would imply that $\rho$ must vanish unless the phase is an integral 
multiple of $2\pi$; this is the constraint for the chosen type of field 
configuration.
For simplicity, 
we shall use $G=SU(2)$ for the example below; generalization to other 
groups is straightforward.

For our example, we choose polar coordinates $(r,\theta , \vf )$ and take 
\beq 
\begin{split}
E_{\theta} &=E_{\vf}=0, \hskip .2in E_r= {\sigma_3 \over 2}~{q\over 4\pi r^2} \\
U &= \exp (i\sigma_3 f(r))
\end{split}
\label{cq1}
\eeq
This field corresponds to a point charge at $r=0$.
To avoid the singularity, we shall remove the point $r=0$ from $M$. Thus
the boundary $\partial M$ consists of a small sphere around $r=0$ and the sphere
at spatial infinity.
Even though $U$ is not constant in space, we have chosen it to commute with
the given $E_i$. Evaluating the phase factor in (\ref{10}), we obtain
\beq
\rho = \rho \exp \left( 2i~ (\Delta f) ~q \right)
\label{cq2}
\eeq
where $\Delta f = f(\infty) - f(0)$.
As for the values of $f(0), ~f(\infty )$, they should be integral
multiples of $\pi$ to be consistent with the trivial action of $U$ on states at the 
boundaries. 
If we require $U$ to go to the identity (and not just an element of the center)
at the boundary, $\Delta f = 2 \pi n$,
$n \in {\mathbb Z}$.
Equation (\ref{cq2}) then tells us that we can have nonzero $\rho$ for
$q = \half n$. The Gauss law for, say fermion sources,
may be written as 
\beq
\nabla \cdot E^a  + f^{abc} A^b\cdot E^c = {\bar \psi} T^a \psi
\label{cq3}
\eeq
For the fundamental representation, this gives, for a point source with
$T^3$-charge,
$E^3 = {1\over 2} ({1/4\pi r^2})$. This is consistent with the quantization of
$q$. On the other hand, if we allow $U$ to go to $-1$, then we only need
$\Delta f = \pi n$. Correspondingly, (\ref{cq2}) tells us that $q$ should be quantized as $q =n$. Equation (\ref{cq3}) also tells us that this is consistent with
sources transforming under ${\mathbb Z}_2$-invariant representations.

The result of the arguments presented here is that
wave functions are {\it invariant} under gauge transformations
which go to an element of the center in theories where the
charges are in $\mathbb{Z}_N$-invariant representations.
Such transformations therefore characterize the redundancy
of the variables $(A_i, E_i)$ in the theory.

The configuration
we have used for obtaining charge quantization
has a divergent kinetic energy
$T={\half } \int E^2$. 
It is possible to find nonsingular configurations which lead to the same result; it is just that the argument will be a little more elaborate.
\subsection{Nature of the configuration space}

The $E$-representation of the wave functions was useful in elucidating the
nature of the allowed boundary values for $U$. However, for the
analysis and formulation of ans\"atze for the configurations with fractional topological charge, the $A$-representation is more appropriate, so 
this is the representation we will use for the rest of this paper.

We can now formalize the situation with the gauge transformations
as follows.
Staying within the $A_0 =0$ gauge, let
\beqar
{\mathcal A} &\equiv&  \{ {\rm Set ~of ~all ~gauge ~potentials}~ A_i \}\nonumber\\
&\equiv&\{ {\rm Set ~of ~all ~Lie\mhyphen algebra\mhyphen valued ~vector ~fields}\nonumber\\
&&\hskip .1in {\rm on ~space}~ {\mathbb R}^3\}\nonumber
\eeqar
Further, let 
\beqar
{\mathcal G} &\equiv&\{ {\rm Set ~of ~all} ~g(\vx ): {\mathbb R}^3 \rightarrow SU(N),
~{\rm such ~that}\nonumber\\
&&\hskip .2in g(\vx ) \longrightarrow ~{\rm constant} \in SU(N) ~{\rm as} ~\vert \vx \vert \longrightarrow \infty\}
\nonumber\\
{\mathcal G}_\omega &\equiv& \{ {\rm Set ~of ~all} ~g(\vx ): {\mathbb R}^3 \rightarrow SU(N),
~{\rm such ~that}\nonumber\\
&&\hskip .2in g(\vx ) \longrightarrow \omega \in {\mathbb Z}_N ~{\rm as} ~\vert \vx \vert \longrightarrow \infty\}
\nonumber
\eeqar
Evidently, ${\mathcal G}/{\mathcal G}_1 = SU(N)$, the set of rigid transformations or the set of constant boundary values for elements $g$ in ${\mathcal G}$.  Our discussion of the Gauss law shows that the gauge group, namely, the 
set of transformations which leave the wave functions invariant, is given by ${\mathcal G}_1$
 in a theory without ${\mathbb Z}_N$-invariance. However, in a theory with ${\mathbb Z}_N$-invariance, 
 ${\mathcal G}_\omega$ leaves $\Psi$ invariant for any $\omega$, so that the gauge group is
 ${\mathcal G}_* = \cup_{\omega \in {\mathbb Z}_N} {\mathcal G}_\omega$. Since the difference
 between ${\mathcal G}_1$ and ${\mathcal G}_\omega$ is in the boundary value, we may
 also consider any element of ${\mathcal G}_\omega$ to be of the form
 $g (\vx )~ \omega$, where $g (\vx )$ goes to the identity at spatial infinity.
 
The physical configuration space, for theories with charges in the fundamental representation, i.e., without $\mathbb{Z}_N$-invariance,
is given by ${\mathcal A}/ {\mathcal G}_1$. It is easy to see that this space is multiply connected.
Consider a sequence of configurations $A_i (\vx, \tau )$ with $0\leq \tau \leq 1$ given by
\beqar
A_i (\vx, \tau ) &=&A_i (\vx ) (1-\tau ) + \tau\, A^g_i (\vx )  \; , \nonumber\\
A_i^g (\vx )&= &  g^{-1} A_i (\vx ) g + g^{-1} \del_i g  \; ,\label{G3}
\eeqar
where $g (\vx ) \in {\mathcal G}_1$. Thus $g (\vx ) \rightarrow 1$ at spatial infinity. 
The starting point and ending point of this sequence of gauge fields are gauge-equivalent, so that (\ref{G3}) gives a closed curve in ${\mathcal A}/{\mathcal G}_1$. If this curve is contractible, then we will be able to transform the entire sequence into gauge-equivalent configurations, writing
\beq
A_i (\vx, \tau ) =  g^{-1}(\vx ,\tau)  A_i (\vx ) g(\vx, \tau)  + g^{-1}(\vx, \tau)  \del_i g(\vx, \tau)
\label{G4}
\eeq
The transformations $g (\vx , \tau )$ give a homotopic deformation of the identity
(at $\tau =0$) to $g(\vx ) $ at $\tau =1$. The homotopy classes of transformations
$g \in {\mathcal G}_1$ are characterized by the winding number
\beq
Q [g] = {1\over 24\pi^2} \int \Tr (g^{-1} dg )^3\label{G5}
\eeq
Thus if $g$ is chosen to have nonzero winding number, then we do not have the possibility
(\ref{G4}), leading to the conclusion that there are noncontractible paths in ${\mathcal A}/{\mathcal G}_1$. In other words, if  $g (\vx )$ has nonzero winding number, the configuration (\ref{G3})
traces out a noncontractible path in ${\mathcal A}/{\mathcal G}_1$ as $\tau$ changes from $0$ to $1$.
The usual instanton is an example of such a path,
which, although it is not captured by the simple parametrization
given in (\ref{G3}), is deformable to
(\ref{G3}).
In general, the noncontractible paths
are topologically nontrivial configurations with nonzero instanton number,
but not necessarily self-dual (or antiself-dual).
In fact, evaluating the instanton number on the configurations (\ref{G3}), we find
\beqar
\nu [A] &\equiv& -{1\over 8\pi^2}\int_{M \times [0,1]} \Tr ( F ~ F)\nonumber\\
&=& {1\over 24\pi^2} \int_{M, \tau =1}\Tr (g^{-1} dg )^3
\label{G6}
\eeqar
where we used the fact that $g$ goes to the identity at spatial infinity.
\subsection{Fractional values of $\nu$}

It is now easy to see how one may get fractional values of $\nu$. We consider a path in the space  of gauge potentials ${\cal A}$ of the form (\ref{G4}), say with $g = U(\vx, \tau )$, where
$U(\vx, 1)$ is such that it goes to $\omega = \exp (2\pi i /N)$ as $\vert \vx \vert \rightarrow
\infty$. In other words, $U(\vx, 1) \in {\cal G}_\omega$. Therefore the path
$A^U = U^{-1} A U + U^{-1} \nabla U$ is closed in the ${\mathbb Z}_N$-invariant theory. The instanton number of this configuration can be evaluated explicitly, but before doing that, a comment is in order.
The configuration $U^{-1} A U + U^{-1} \nabla U$ looks similar to
(\ref{G4}), but there is an important difference. In (\ref{G4}), 
$g(\vx, \tau )$ gives a homotopy between the identity and $g(\vx )$,
so that $A_i (\vx , \tau)$ is gauge-equivalent to $A_i (\vx )$ for any
value of $\tau$. Further, the value of $g(\vx, \tau )$ as
$\vert \vx \vert \rightarrow \infty$ is
identity. To get a noncontractible path, one needs to consider
$A_i$ which depend on $\tau$ as in (\ref{G3}) (or in the usual self-dual
instanton configurations). In the present case, the boundary value
of $U$ changes from the identity to $\omega$, 
so that at $\tau \neq 0, 1$, $U$ is not an element of
$\G_1$ or $\G_\omega$. This is why the configurations
 $ U^{-1} A U + U^{-1} \nabla U$ can still give a nonzero $\nu$.

Turning to details, it is useful to have an explicit construction of such a $U(\vx, \tau )$. 
Let $t^a$, $a =  1, 2, \cdots, (N^2-1)$,  denote a basis of hermitian $N\times N$ matrices for the Lie algebra
of $SU(N)$, normalized so that $\Tr (t^a t^b) = \half \delta^{ab}$.
We can take $t^{N^2-1}$ to be diagonal and given by
\beq
(t^{N^2-1})_{ij} = \sqrt{N\over 2(N-1)} \left\{ \begin{matrix}
{1\over N} \delta_{ij} &~& i,j = 1, 2, \cdots, (N-1)\\
{1\over N}- 1&~&  i=j=N\\
\end{matrix}
\right.
\label{G7}
\eeq
This is the $SU(N)$ version of the usual hypercharge matrix.
It is easy to see that 
\beq
g = \exp( i 2\pi \tau \sqrt{2(N-1)/N} ~ t_{N^2-1})
\label{G7a}
\eeq
is a path from $g=1$ to $g =\omega$ in $SU(N)$ as $\tau$ varies from zero to $1$.
Thus it is a closed path in the pure $SU(N)$ gauge theory.
Keeping in mind that instantons are essentially in an $SU(2)$ subgroup of $SU(N)$, we define
the $N\times N$ matrix 
\beq
Y_{ij} = \left\{ \begin{matrix}
{1\over N} \delta_{ij}& i,j =  1, 2, \cdots, (N-2)\\
{1\over 2} (\sigma\cdot {\hat x} )_{ij} + \left( {1\over N} - {1\over 2} \right)_{ij} & i,j = N-1, N\\
\end{matrix}
\right.
\label{G8}
\eeq
We can then define 
\beq
U(\vx, \tau ) = \exp (i Y \Theta (r,\tau ))\label{G9}
\eeq
with $\Theta (r, 0) =0$, $\Theta (0,\tau ) =0$ and
$\Theta (\infty, \tau ) = 2\pi \tau$. 
(One example of such a function is
$\Theta (r, \tau) = 2\pi \tau r/(r+r_0)$. There are obviously infinitely 
many $\Theta$'s consistent with the required boundary behavior.) This gives a spherically symmetric ansatz
for an element of ${\cal G}_\omega$.
It is easy to verify that $U(\infty, \tau)$ traces out a path from the identity to 
$\omega$ in $SU(N)$. Also,
since $U(\infty, \tau) \rightarrow 1, \omega$ at
$\tau =0, 1$, it qualifies as a gauge transformation at the two ends in the
$SU(N)/{\mathbb Z}_N$ theory.

Returning to the configurations
$A_i^U = U^{-1} A_i U + U^{-1} \del_i U$
in the space of potentials, we see that this corresponds to a closed
path in $\A/ \G_\omega$.
Since $U$ depends on $\vx, \tau$, but $A_i$ depends only on $\vx$,  
\beqar
{\cal F} &=& dA^U + A^U A^U = U^{-1} F U + d\tau {\del \over \del \tau} A^U\nonumber\\
&=& U^{1} (F - Da ) U\label{G10}
\eeqar
 where $F$ involves only the spatial components of the field strength tensor
 and $a = d\tau~ {\dot U} U^{-1}$. (For this calculation, $\tau$ can also be viewed
 as the time coordinate, so that $Da$ is essentially the electric field.)
From (\ref{G10}), 
 \beqar
 \nu  &=& - {1\over 8\pi^2} \int \Tr ( (F- Da ) (F- Da) = {1\over 4\pi^2} \int \Tr (Da F )\nonumber\\
 &=& {1\over 4\pi^2} \oint \Tr (a F)\label{G11}
 \eeqar
 The indicated boundary integration is over spatial infinity and over all $\tau$.
 This shows that we will need a nonzero magnetic flux to obtain a nonzero value
 for $\nu$. Therefore, we consider monopole-like configurations with the asymptotic
 behaviour 
 \beqar
 F &=& {1\over 2} F_{ij} dx^i \wedge dx^j\nonumber\\
 &\rightarrow& -{ i\over 2} (\sigma \cdot {\hat x}  )~ {M\over 2} ~\epsilon_{ijk} {{\hat x}^k \over r^2}
 dx^i\wedge dx^j
 \label{G12}
 \eeqar
 where $\sigma_i$ are in the $2\times 2$ block of $i, j = (N-1), N$ viewed as
 an $N\times N$ matrix. $M$ is (electric charge $e$ times) the monopole charge.
 We then find
 \beq
 \nu = M
 \label{G13}
 \eeq
 $M$ must be quantized according to the Dirac quantization condition.
 This condition, for a general gauge group is the Goddard-Nuyts-Olive
 (GNO) quantization condition \cite{Goddard:1976qe} and amounts to the following.
 If the electric charges correspond to representations of $G$, then the magnetic charges
 $M$ take values in the dual group ${\tilde G}$. For our case, we note that
 the GNO dual of $SU(N)$ is $SU(N)/ {\mathbb Z}_N$. Thus if the electric charges
 are ${\mathbb Z}_N$-invariant, taking values corresponding to $SU(N)/ {\mathbb Z}_N$
 representations, then the fundamental charges of $SU(N)$ are allowed values for
 $M$. They are thus quantized in units of  $1/N$. 
 
 Thus we see that we can indeed obtain fractional values of $\nu$.
 The problem however, is that for the nonsingular 't Hooft-Polyakov (`t H-P) monopoles, the quantization condition is not quite the Dirac (or GNO) condition. In fact, for the case of $SU(2)$, 
 $M$ is an integer for `t H-P monopoles, whereas the GNO condition would suggest
 that it is possible to get $M = {\half}$.
 It is, however, possible to construct nonsingular configurations of separated GNO monopoles which
 have a total flux consistent with the `t Hooft-Polyakov condition.
 These will look like some split versions of the `t H-P monopole.
 
 We can construct an ansatz for the split monopole for the case of $SU(2)$ as follows.
 Let $A_D$ be the Dirac form of the monopole given by
 \beq
 A_D = {({\hat x}_1 d{\hat x}_2 - {\hat x}_2 d{\hat x}_1 ) \over  (1+ {\hat x}_3)} \label{G14}
 \eeq
The `t Hooft-Polyakov form of the monopole is then given by
\beqar
A &=& (1- K(r) ) \left[  g^{-1} i \left({\sigma_3\over 2} \right) A_D~ g + g^{-1} d g \right]\nonumber\\
&=&  i \left({ \sigma^a \over 2}\right) ~(1- K) ~\epsilon_{abc}  {x^b\over r^2} dx^c
\label{G15}
\eeqar
where $g$ is the matrix
\beq
g = {1\over \sqrt{1+z\bz}} \left[ \begin{matrix} 
1 & z \\
-\bz &1\\
\end{matrix}
\right] \label{G16}
\eeq
 and $z = \tan (\theta /2)~ e^{-i\vf}$ and
 \beq
 {\hat x}_1 = {z +\bz \over 1+ z\bz}, \hskip .2in
 {\hat x}_2 =  {i (z-\bz ) \over 1+ z\bz}, \hskip .2in
 {\hat x}_3 = {1 - z\bz \over 1+ z\bz}
 \label{G17}
 \eeq
 We may also note that $g^{-1} \sigma_3~ g = \sigma \cdot {\hat x}$.
The function $K(r)$ vanishes exponentially outside of the core of the monopole,
and $1 - K(r) \sim r^2$ for small $r$.
 The advantage of writing it as in (\ref{G15}) is that, for large $r$, we can trivially calculate $F$ as
 \beq
 F  = i g^{-1} {\sigma_3\over 2}  g ~ dA_D = 
{i\over 2} \sigma\cdot {\hat x} ~ \sin\theta ~d\theta d\vf
 \label{G18}
 \eeq
Thus $F^a = - {\hat x}^a  \sin\theta ~d\theta d\vf$, with
$\int F^a {\hat x}^a = - 4\pi$. We can now modify this ansatz with some of the flux piped away from the monopole by a vortex. We consider an Abelian vortex given by
\beq
A_v = {1\over 2}~ f (\rho, x_3)~ {x_1 dx_2 - x_2 dx_1 \over \rho^2}
\label{G19}
\eeq
where $\rho^2 = x_1^2 + x_2^2$. This is a vortex along the $x_3$-axis. We also have
$f (\rho, x_3) \rightarrow 1$ as $\rho$ becomes large, essentially
outside the core of the vortex. The factor of $\half$ tells us that the flux carried by this vortex is
$2\pi /2$; it is a ${\mathbb Z}_N$-vortex, for $N=2$. We will consider a vortex of finite length
$L$ by taking as an ansatz
\beq
f( \rho, x_3) =  {1\over 2} \tanh \lambda \rho ~ [ \tanh {\tilde \lambda} x_3~-~
\tanh {\tilde \lambda} (x_3- L)] 
\label{G20}
\eeq
This function vanishes exponentially for $x_3 \ll 0$ and for $x_3 \gg L$.
The core of the vortex has an extent in $\rho$ of the order of $1/\lambda$.
Our modified ansatz is now given by
\beq
A = (1- K(r) ) \left[  g^{-1} i \sigma_3 (A_D -A_v)~ g + g^{-1} d g \right]
\label{G21}
\eeq

Consider a large sphere of radius $R$ much larger than the core of the monopole and the core of the vortex.  If $R\ll L$, then the sphere intersects the vortex. The flux may be computed by
taking $K \rightarrow 0$ , so that
\beq
F = { i \over 2} \sigma \cdot {\hat x}  (d A_D - 2 ~ dA_v )
\label{G22}
\eeq
The flux is then $-(4\pi - 2\pi) = -2\pi$. This is what we expect for a GNO monopole, and is equivalent to $M = \half$. 
If we consider a sphere of radius much larger than $L$, then the contribution
from $A_v$ is zero, since $f$ vanishes and we get $- 4\pi$ for the total flux.
In this sense, we can view the configuration (\ref{G21}) as
a split monopole.

The relevance of the split monopole can be understood from the following question: In a calculation or simulation of the vacuum-to-vacuum transition amplitude, can we see configurations with fractional values of $\nu$?  For this it is useful to write $\nu$ in terms of the Chern-Simons integral
\beq
S_{CS}(M) = -{1\over 8\pi^2} \int_M \Tr (A ~dA + {2\over 3} A^3 )
\label{G23}
\eeq
The topological charge $\nu$, which is the integral of
the exterior derivative of the Chern-Simons over spacetime,
can then be written as
\beqar
\nu &=&  S_{CS} (M, \tau =1) - S_{CS} (M, \tau =0)\nonumber\\
&&\hskip .1in + {1\over 8\pi^2} \oint_{\del M} \Tr (A_i E_j ) dx^0\wedge  dx^i \wedge dx^j.
\label{G24}
\eeqar
As the representative of the vacuum at $\tau =0$, we may take
$A_i =0$. The final configuration is also the vacuum, so it must be a gauge transform of $A_i =0$, say, $A_i = g^{-1} dg$.
Further, if we consider spatial boundary conditions (periodic, Dirichlet, etc.)
which lead to vanishing of the integral with the
electric flux on $\del M$,
we find
\begin{eqnarray}
  \nu &=&  S_{CS} (M, \tau =1) - S_{CS} (M, \tau =0) \nonumber \\
  &=& {1\over 24\pi^2} \int_M \Tr (g^{-1} dg)^3 = Q [g] \; . 
\label{G25}
\end{eqnarray}
Since $Q [g]$ is an integer, even for $g$'s such that
$g \rightarrow \omega$ on $\del M$,
we get integral values of $\nu$ in the vacuum-to-vacuum
amplitude. 
However, we can have configurations like
$A_i^U = U^{-1} A_i U + U^{-1} \del_i U$ where $A_i$ is a split monopole configuration as in (\ref{G21}). We get separated configurations, each of which in isolation may be considered as having a fractional value of $\nu$, but the total value of $\nu$ is integral.

\subsection{Simple solution}
\label{sec:simple}

We will now illustrate the analysis given above in a related but slightly different way and also comment on the situation with finite nonzero temperature $T$. It is convenient to frame this discussion in terms
of a nonzero $A_0$, by replacing
the field configuration $(A_0 =0, U^{-1} A_i U + U^{-1} \del_i U)$
by its gauge equivalent version
$(A_0 = {\dot U} U^{-1}, A_i )$. 
For $A_0$, at $r=\infty$ our choice is then
\begin{equation}
A_0 \; = \; \frac{2 \pi T}{N} \; \mathbf{k} \; .
\end{equation}
Here $\mathbf{k}$ is a diagonal $SU(N)$ matrix related to
$\mathbb{Z}_N$ transformations, so their elements are integers.  There are two choices,
\begin{eqnarray}
\mathbf{k}_1 \; &=& \;
\left(
\begin{array}{cc}
{\mathbf 1}_{N-1} & 0      \\
0             & -(N-1) \\
\end{array}
\right) \; ,
\\
\mathbf{k}_2 \; &=& \;
\left(
\begin{array}{ccc}
{\mathbf 1}_{N-2} & 0      & 0\\
0             & -(N-1) & 0 \\
0             & 0      & 1 \\
\end{array}
\right) \; .
\end{eqnarray}
These are obviously related to the matrix $(t^{N^2-1})_{ij} $ in Eq. (\ref{G7}).
For $\mathbf{k}$ equal to either $\mathbf{k}_i$, the Wilson line in the imaginary
time direction, $t$, is
\begin{equation}
\Omega \; = \; \exp \left( i \int^{1/T}_0 \; A_0 \; dt \right) 
\; = \; \exp\left( \frac{2 \pi i}{N} \; \mathbf{k} \right) \; ,
\end{equation}
has nontrivial holonomy, as these values represent $\mathbb{Z}_N$ degenerate vacum

For the spatial components, construct a split 't H-P monopole, as in the previous section.
Divide a sphere into an upper and a lower hemisphere,
with gauge potentials on each, $A^\pm$, and take
\begin{equation}
A^\pm_\phi \; = \; \frac{1}{2N r } \; \mathbf{m} \;
\frac{\left( \pm 1 - \cos\theta \right)}{\sin \theta} \; .
\label{zn_soln}
\end{equation}
To see this is a $\mathbb{Z}_N$ monopole, compute the the
Wilson line for a special closed path, $\vec{s}$.
Since the vector potential is specified by two patches, we compute
the Wilson line with $A^+$,
going around by $2 \pi$ in $\phi$; then, take the Wilson line
with $A^-$, running in the opposite direction:
\begin{eqnarray}
&&\exp \left( \; i \oint \vec{A}^+ \cdot d\vec{s} \; \right)
\; \left( \exp\left( \;
i \oint \vec{A}^- \cdot d\vec{s} \; \right) \right)^\dagger \nonumber \\
 &=& \exp\left( \frac{2 \pi i}{N} \; \mathbf{m} \right) \; .
\end{eqnarray}
This is manifestly gauge invariant, and $=1$ if the configuration
is trivial, $A^+ = A^-$.  For the $\mathbb{Z}_N$ monoopole, instead one
obtains a non-trivial element of $\mathbb{Z}_N$.  For this to be
true, $\mathbf{m}$ must be one of the two matrices, $\mathbf{c}_{1}$ or $\mathbf{c}_2$.

The above are the boundary conditions at spatial infinity, $r \rightarrow
\infty$.  At the origin, $r = 0$, we require all $A_\mu$'s to vanish,
at least like $\sim r^2$, so that $F_{\mu \nu} \sim r$ as $r \rightarrow 0$.

As argued in Sec. (\ref{sec:cpn}), in general we expect that this exists only as a quantum
instanton, on the order of the confinement scale.  At nonzero temperature, however,
$1/T$ provides an alternate length scale.  
While the solution is approximately self-dual over
distances $\sim 1/T$, because of the presence of the Debye screening mass, it is
not self-dual over larger distances.  This generates corrections $\sim \sqrt{g^2}$ to the action.

It is straightforward to compute the topological charge.  For large $r$,
\begin{equation}
A_0(r) \; = \;
\frac{2 \pi T}{N} \; \mathbf{k} \; - \; \frac{1}{2 N r} \; \mathbf{m} \; + \; \ldots
\end{equation}
For a static configuration,
\begin{equation}
Q \; = \;
\frac{1}{4 \pi^2} \int d^4 x \; \partial_i \; {\rm tr} \left( A_0 \; B_i
\right) = \frac{1}{N^2} \; \bf{m} \cdot \bf{k} \; .
\end{equation}
This was first derived by 't Hooft \cite{tHooft:1980kjq}.

There are only two cases to consider.  Either the $\mathbb{Z}_N$ charges
are the same, or they are different.  If they are the same,
$\bf{m} = \bf{k_1}$,
\begin{equation}
Q \; = \; \frac{N-1}{N} \; .
\end{equation}
If they charges are different, such as $\mathbf{m} = \mathbf{k}_1$ and $\mathbf{k}
= \mathbf{k}_2$, then
\begin{equation}
Q \; = \; - \; \frac{1}{N} \; .
\end{equation}

We conclude this section by discussing the relationship between the configuration above and that of
Kraan, van Baal, Lee, and Lu (KvBLL)
\cite{Lee:1997vp,Lee:1998vu,Lee:1998bb,Kraan:1998pm,Kraan:1998sn,GarciaPerez:1999hs,Diakonov:2002fq,Eto:2004rz,Eto:2006pg,Eto:2006mz,Bruckmann:2009nw,Diakonov:2009jq,Poppitz:2008hr,Anber:2021upc}
Like ours, their solution carries magnetic charge and has nontrivial holonomy.  Our ansatz, however, carries
$\mathbb{Z}_N$ magnetic charge, and so must be represented by a multivalued function, Eq. (\ref{zn_soln}), while
that of KvBLL has integral magnetic charge.
For our solution, the boundary condition for holonomy at spatial infinity ensures that it is a vacuum which is
degenerate with the vacuum.  Thus when one loop corrections are included, the action for our solution will remain
finite.  In contrast, for the solution of KvBLL, the holonomy is at a maximum of the holonomous potential.  When
one loop corrections are included, then, the action for a constituent with charge $1/N$
diverges as the spatial volume.  The action for an instanton with integral charge remains finite, which
is why on the quantum level, these constituents cannot be pulled apart.
In the next section, we discuss how to distinguish between our configurations and those of KvBLL.

\section{$\mathbb{Z}_N$ dyons on the lattice}
\label{sec:lattice}

In the previous section we argued that configurations with fractional magnetic charge can generate
fractional topological charge.  From this
the $N$-dependence of $E(\theta)$ in Eq. (\ref{energy_theta}) follows immediately,
and accords with general expectation \cite{Witten:1979vv,Veneziano:1979ec}.
Assuming that fluctuations in the topological charge are $\Delta Q \sim 1/N$, since there are $N^2$ ways of
inserting a fractional charge in an $SU(N)$ gauge theory, the topological susceptibility $\chi \sim N^2 (1/N)^2 \sim 1$.
As in the $\mathbb{CP}^{N-1} $ model, $b_2 = \widetilde{b}_2/N^2$, where $\widetilde{b}_2$ is a number of
order one.  

The precise value of $\widetilde{b}_2$ is rather interesting.
By numerical simulations on the lattice,
Bonanno, Bonati, and D'Elia \cite{Bonanno:2020hht} computed $b_2$ for $N=4$ and $6$, and compared to known
results for $N=3$.  They exclude a constant value of $b_2$, as expected for a dilute gas of instantons.
Instead, their results strongly favor $b_2 = \widetilde{b}_2/N^2$.  Comparing to Eq. (\ref{definition_b2}),
$\widetilde{b}_2 = -1/12 \approx -.08$ for a dilute gas of fractionally charged objects.  Instead,
Ref. \cite{Bonanno:2020hht} find a value which is more than twice as large, $\widetilde{b}_2 \approx -.19$.
This indicates that fractional instantons do not form a dilute gas, but a dense liquid.
That the vacuum of a $SU(N)$ gauge theory is complicated, with a dense liquid of $\mathbb{Z}_N$ dyons, is
to be expected.

While the value of $\widetilde{b}_2$ is most
suggestive, it does not comprise definitive evidence for fractional topological charge.
We then discuss a way of measuring fractional topological charge directly. 
In the continuum, an instanton in a $SU(N)$ gauge theory
with topological charge one, coupled to single massless Dirac quark in the fundamental
representation, has two zero modes, one for each chirality.  In the adjoint
representation, however, there are $2N$ zero modes.   Thus a single $\mathbb{Z}_N$ dyon has
two zero modes for a quark in the adjoint representation.

On the lattice, in the pure gauge theory one can use an external quark propagator
to look for isolated zero modes.  To ensure these are not lattice artifacts,
it is imperative to use a Dirac propagator with exact chiral symmetry, such
as the overlap operator
\cite{Ginsparg:1981bj,Kaplan:1992bt,Narayanan:1993zzh,Narayanan:1993sk,Narayanan:1993ss,Narayanan:1994gw,Luscher:1998pqa}.

Using an external quark propagator in the adjoint representation, then, one can look for isolated
$\mathbb{Z}_N$ dyons.  
This was first done by Edwards, Heller, and Narayanan \cite{Edwards:1998dj}, who 
found evidence for fractional topological charge.  Their lattices were coarse, however.  With present 
techniques and much finer lattices
it should be possible to establish the existence of fractional topological charge close to the continuum limit
\footnote{Fodor {\it et al.} performed simulations in a $SU(3)$ gauge theory
using an external quark propagator in the sextet representation
\cite{Fodor:2009nh}.  As they discuss, the sextet representation is sensitive to the presence of objects
with topological charge $1/5$, and for which they see no evidence.  However, this does not exclude
the appearance of objects with charge $1/3$.}.
With the overlap operator, one would look for configurations with (almost) zero modes;
from the eigenvector, one could estimate the position and size of the object.

We stress that unless there are boundary conditions which are twisted with respect to $\mathbb{Z}_N$
\cite{tHooft:1980kjq,tHooft:1981nnx}, then the net topological charge will always be integral.
Even so, it should be possible from the eigenvectors to see if N $\mathbb{Z}_N$ dyons 
are tightly bound into instantons, or if $\mathbb{Z}_N$ dyons and anti-dyons
form a dense liquid of objects with fractional topological charge.

In the confining phase of a gauge theory surely the worldlines of
the $\mathbb{Z}_N$ dyons are tangled, both with themselves and those of other dyons and anti-dyons.  
This is especially true if the the size of the dyons is on the order of the confinement scale.  This may help explain why
lattice studies by Horvath {\it et al.} do not find evidence for a simple instanton, concentrated about
a single point in spacetime, but for an extended structure
\cite{Horvath:2002yn,Ahmad:2005dr,Horvath:2005rv,Lian:2006ky,Thacker:2010zk,Thacker:2011sz,Alexandru:2021pap}.

The change in the behavior of $\mathbb{Z}_N$ dyons is especially interesting near the deconfining transition
temperature $T_d$.  $\mathbb{Z}_N$ dyons carry
$\mathbb{Z}_N$ magnetic charge.  This is allowed in the confined phase, where $\mathbb{Z}_N$ magnetic charge is unconfined.
In the deconfined phase, though, $\mathbb{Z}_N$ magnetic charge 
propagating in the temporal direction is confined.  Thus $\mathbb{Z}_N$ dyons are only relevant at best
in a narrow temperature region above $T_d$.  
As the temperature increases, so will the magnetic string tension, binding the
$\mathbb{Z}_N$ dyons with increasing strength into instantons with integral topological charge.
This window of temperature at $T \geq T_d$
where $\mathbb{Z}_N$ dyons are relevant could vanish as $N \rightarrow \infty$, which appears to be suggested by
numerical simulations on the lattice \cite{Bonati:2013tt}.

We note that at temperatures just above $T_d$, it should be possible
to distinguish between our configurations, and those of KvBLL,
by measuring the value of the Polyakov loop at the location of the near zero mode of the adjoint quark propagator.

Especially interesting to study would be the behavior of $\mathbb{Z}_2$ dyons in a $SU(2)$ gauge theory,
where the deconfining transition is of second order.

\section{$\mathbb{Z}_N$ dyons and quarks}
\label{sec:quarks}

We have concentrated exclusively on a gauge theory without dynamical quarks.  In this section we discuss
what might occur with their introduction.

For a $\mathbb{Z}_N$ magnetic monopole (or dyon), a Wilson loop in the fundamental
representation picks up a phase of $\exp(2 \pi i/N)$ as it encircles the worldline of the monopole.
The same is true for dynamical quarks in the fundamental representation, and so it is not obvious
how the monopole density changes as quarks are introduced.

There are recent results about the density of $\mathbb{Z}_N$ monopoles in the presence of dynamical quarks.
While the definition of $\mathbb{Z}_N$ monopoles, and so their density,
is gauge dependent, changes in the density should be meaningful and
gauge invariant.  Biddle, Kamleh, and Leinweber \cite{Biddle:2022acd,Biddle:2022zgw,Leinweber:2022dpp,Biddle:2023lod} have
studied the change in the density of $\mathbb{Z}_N$ monopoles as quarks are
introduced, and find that the monopole density {\it strongly} increases as the mass of the quarks {\it de}creases.

While the sign of the effect is unexpected, we can use this to
suggest how anomalous interactions might change as a function of the
temperature, $T$, and quark chemical potential, $\mu_{\rm qk}$.

At $\mu_{\rm qk} = 0$ and $T \neq 0$, the lattice finds that the
topological susceptiblity is consistent with a 
dilute gas of instantons, as in Eq. (\ref{highT_top_sucep}), for $T > 300$~MeV
\cite{Borsanyi:2015cka,GrillidiCortona:2015jxo,Bonati:2015vqz,Borsanyi:2016ksw,Frison:2016vuc,Petreczky:2016vrs,Taniguchi:2016tjc,Lombardo:2020bvn,Jahn:2021qrp,Borsanyi:2021gqg,Chen:2022fid}.
Notice that this temperature is close to the deconfining temperature for the pure glue theory,
of $T_d \approx 270$~MeV.
\footnote{We also comment that by looking
at the spectra of higher spin mesons, the region between $300 \leq T \leq 600$~MeV is still far from a perturbative
quark-gluon plasma \cite{Glozman:2022lda}.}

The lattice finds that the crossover temperature for chiral symmetry is at $T_\chi \approx 156 \pm 2$~MeV
\cite{HotQCD:2018pds,Borsanyi:2020fev,Guenther:2022wcr}.  Thus for $T_\chi \leq T \leq 300$~MeV, massless
quarks interact with what is surely a dense liquid of $\mathbb{Z}_N$ dyons.
At a temperature $T < T_\chi$, the topological susceptibility changes slowly with temperature, as massive
quarks, or equivalently hadrons, interact with this dense liquid of $\mathbb{Z}_N$ dyons.  Thus in QCD there
are demonstrably three regimes for the topological suceptibility.

Similarly, at low temperature and nonzero quark chemical potential, $\mu$, it is natural to suggest
that there are again three regimes for the topological susceptibility.  While this regime is not accessible
to classical computers because of the sign problem, eventually it will be measured using quantum computers.
Nevertheless, we can at least speculate.

Because the number of degrees of freedom for quarks and gluons at nonzero temperature is so much greater
than that at $T = 0$ and $\mu_{\rm qk} \neq 0$, 
estimates with a dilute instanton gas indicate that at zero temperature,
instantons do not dominate until {\it very} high
chemical potential, at least $\mu_{\rm qk} \sim 2$~GeV \cite{Pisarski:2019upw}.
We note that this bound uses the incomplete result for 
the instanton density at zero temperature and $\mu_{\rm qk} \neq 0$; a better bound would follow
from the full instanton density \cite{Nogradi2023}.

Consider then the opposite limit, moving up in the quark chemical potential.  The chemical potential has no
effect upon the free energy until it exceeds one-third the mass of the nucleon, minus the binding energy of nuclear
matter, at something like $\mu_{\rm qk} \sim 300$~MeV.  Assuming that this regime is like that for $\mu_{\rm qk} = 0$ and
$T< T_\chi$, the topological susceptibility presumably varies little as $\mu$ increases.  This should
hold until chiral symmetry is restored at $\mu_{\rm qk} = \mu_\chi$,
and the quarks are (essentially) massless.  {\it Then} the topological
susceptiblity should vary significantly, as the Fermi sea of massless quarks
interacts strongly with a dense liquid of $\mathbb{Z}_N$ dyons.
This includes both a chirally symmetric hadronic phase, a chirally symmetric quarkyonic phase
\cite{McLerran:2007qj,Lajer:2021kcz}, and perhaps even into the perturbative regime, for $\mu_{\rm qk} > 1$~GeV
\cite{Gorda:2018gpy,Gorda:2021znl,Gorda:2021kme}.
In the latter, for $1 \leq \mu_{\rm qk}  \leq 2$~GeV, color superconductivity is dominant near the Fermi surface,
but the effects of the axial anomaly can still affect the possible pairing mechanisms \cite{Pisarski:1999gq}.

In summary, while the interactions betweeen massless, dynamical quarks and a dense liquid of $\mathbb{Z}_N$
dyons dominate for $\mu_{\rm qk} = 0$ and the intermediate temperature region of $T_\chi \approx 156 \leq T \leq 300$~MeV,
the analogous regime for $T = 0$ could be {\it much} broader, from $\mu_\chi \leq \mu_{\rm qk} \leq 2$~GeV.

This suggestion is obviously conjecture, and so we do not bother with considering how the entire phase diagram
in $T$ and $\mu_{\rm qk}$ might fill out.  It does indicate that this phase diagram is exceedingly rich.

\acknowledgments
V.P.N. was supported in part by the U.S. National Science Foundation Grants No.
PHY-2112729 and No. PHY-1820271.
R.D.P. was supported by the U.S. Department of Energy under contract DE-SC0012704.
R.D.P. thanks A. Alexandru, O. Alvarez, A. Dumitru, U. Heller, I. Horvath, T. Izubuchi, J. Lenaghan,
N. Karthik, R. Narayanan, P. Petreczky, E. Poppitz, S. Sharma, R. Venugopalan, and M. Unsal for discussions.
Lastly, we thank C. Bonati, C. Bonanno, and M. D'Elia for discussions about their work.

\appendix*
\section{Critical points of the quantum action}
\label{sec:appendix}

In this Appendix we discuss how topologically nontrivial configurations arise
not as solutions to the classical equations of motion,
but as critical points of the effective quantum action.
Before discussing details of such configurations, it is useful to comment briefly on the role of such critical points.
Denoting the fields generically by the symbol $\vf$,
with Minkowski signature the effective action $\Gamma (\Phi )$ is given by the functional integral
\begin{equation}
e^{i \Gamma (\Phi )} = \int [d\vf] 
\exp\left( i S(\vf + \Phi ) - i \int \vf {\delta \Gamma \over \delta \Phi}
\right)
\label{eff_act-3}
\end{equation}
Consider then a solution $\Phi$ of ${\delta \Gamma \over \delta \Phi } = 0$
with boundary values $\phi \rightarrow \Phi_1$ as $t \rightarrow - \infty$,
$\Phi_2$ as $t \rightarrow +\infty$. From Eq. (\ref{eff_act-3}), $\Gamma (\Phi)$ evaluated on this solution is 
\begin{equation}
e^{i \Gamma (\Phi )} = \int [d\vf] 
\exp\left( i S(\vf + \Phi )  \right).
\label{eff_act-4}
\end{equation}
Independently, we can see that the  transition amplitude 
from a configuration $\Phi_1$ at $t \rightarrow -\infty$
to $\Phi_2$ at $t \rightarrow +\infty$ is 
\begin{eqnarray}
\langle \Phi_2 \vert  \Phi_1\rangle &=& \langle{\Phi_2}\vert e^{- i H (t -\tau)} \vert {\Phi_1} \rangle  \Big\vert_{t\rightarrow \infty, \tau \rightarrow -\infty}
\nonumber\\
&=&  \int [d\vf] \, e^{ i S (\vf)}\Big\vert_{\vf(t=-\infty) = \Phi_1, \, \vf (t=\infty) = \Phi_2}\nonumber\\
&=& \int [d\vf] \, e^{ i S (\vf + \Phi )}
\label{eff_act-5}
\end{eqnarray}
where we shift $\vf \rightarrow \vf + \Phi$ in the last line and integrate 
over $\vf$'s which vanish as $t \rightarrow \pm \infty$.
The boundary values $\Phi_1$, $\Phi_2$ for $\vf + \Phi$ are carried 
by $\Phi$. Comparing Eqs. (\ref{eff_act-4}) and (\ref{eff_act-5}), we see that
the solution of ${\delta \Gamma \over \delta \Phi } = 0$
gives the transition amplitude for $\Phi_1 \rightarrow \Phi_2$.
(This is essentially the result that the S-matrix is given by
$\Gamma$ evaluated on its critical points; here we are using
$\vf$-diagonal states, rather than specifying the incoming and outgoing states by spins and momenta of the particles.)
By the same reasoning, by obtaining solutions with specific boundary behavior at $t \rightarrow \pm \infty$,
or more generally, specific asymptotic behavior 
in spacetime, we get information about
transition amplitudes. The analysis in text on the critical points of
the effective action should be viewed with this interpretation in mind.
We will be using Euclidean signature in this section, as is 
appropriate for tunneling transitions.

A related point, perhaps worth emphasizing, is that the solution $\Phi$ should not be interpreted as
the vacuum expectation value of the quantum field.
(Notice that, to get the vacuum-to-vacuum amplitude, one has to
integrate Eq. (\ref{eff_act-5}) over $\Phi_1$, $\Phi_2$
after taking the product with the vacuum wave functions $\Psi_0 (\Phi_1)$
and $\Psi^*_0(\Phi_2)$. 
The configurations we discuss will have boundary behaviors which correspond to the vacuum
in the $SU(N)/{\mathbb{Z}_N}$ theory, but
are distinct before modding out by $\mathbb{Z}_N$.


\begin{thebibliography}{156}%
\makeatletter
\providecommand \@ifxundefined [1]{%
 \@ifx{#1\undefined}
}%
\providecommand \@ifnum [1]{%
 \ifnum #1\expandafter \@firstoftwo
 \else \expandafter \@secondoftwo
 \fi
}%
\providecommand \@ifx [1]{%
 \ifx #1\expandafter \@firstoftwo
 \else \expandafter \@secondoftwo
 \fi
}%
\providecommand \natexlab [1]{#1}%
\providecommand \enquote  [1]{``#1''}%
\providecommand \bibnamefont  [1]{#1}%
\providecommand \bibfnamefont [1]{#1}%
\providecommand \citenamefont [1]{#1}%
\providecommand \href@noop [0]{\@secondoftwo}%
\providecommand \href [0]{\begingroup \@sanitize@url \@href}%
\providecommand \@href[1]{\@@startlink{#1}\@@href}%
\providecommand \@@href[1]{\endgroup#1\@@endlink}%
\providecommand \@sanitize@url [0]{\catcode `\\12\catcode `\$12\catcode
  `\&12\catcode `\#12\catcode `\^12\catcode `\_12\catcode `\%12\relax}%
\providecommand \@@startlink[1]{}%
\providecommand \@@endlink[0]{}%
\providecommand \url  [0]{\begingroup\@sanitize@url \@url }%
\providecommand \@url [1]{\endgroup\@href {#1}{\urlprefix }}%
\providecommand \urlprefix  [0]{URL }%
\providecommand \Eprint [0]{\href }%
\providecommand \doibase [0]{https://doi.org/}%
\providecommand \selectlanguage [0]{\@gobble}%
\providecommand \bibinfo  [0]{\@secondoftwo}%
\providecommand \bibfield  [0]{\@secondoftwo}%
\providecommand \translation [1]{[#1]}%
\providecommand \BibitemOpen [0]{}%
\providecommand \bibitemStop [0]{}%
\providecommand \bibitemNoStop [0]{.\EOS\space}%
\providecommand \EOS [0]{\spacefactor3000\relax}%
\providecommand \BibitemShut  [1]{\csname bibitem#1\endcsname}%
\let\auto@bib@innerbib\@empty
\bibitem [{\citenamefont {Witten}(1979{\natexlab{a}})}]{Witten:1979vv}%
  \BibitemOpen
  \bibfield  {author} {\bibinfo {author} {\bibfnamefont {E.}~\bibnamefont
  {Witten}},\ }\bibfield  {title} {\bibinfo {title} {{Current Algebra Theorems
  for the U(1) Goldstone Boson}},\ }\href
  {https://doi.org/10.1016/0550-3213(79)90031-2} {\bibfield  {journal}
  {\bibinfo  {journal} {Nucl. Phys. B}\ }\textbf {\bibinfo {volume} {156}},\
  \bibinfo {pages} {269} (\bibinfo {year} {1979}{\natexlab{a}})}\BibitemShut
  {NoStop}%
\bibitem [{\citenamefont {Veneziano}(1979)}]{Veneziano:1979ec}%
  \BibitemOpen
  \bibfield  {author} {\bibinfo {author} {\bibfnamefont {G.}~\bibnamefont
  {Veneziano}},\ }\bibfield  {title} {\bibinfo {title} {{U(1) Without
  Instantons}},\ }\href {https://doi.org/10.1016/0550-3213(79)90332-8}
  {\bibfield  {journal} {\bibinfo  {journal} {Nucl. Phys. B}\ }\textbf
  {\bibinfo {volume} {159}},\ \bibinfo {pages} {213} (\bibinfo {year}
  {1979})}\BibitemShut {NoStop}%
\bibitem [{\citenamefont {'t~Hooft}(1986)}]{tHooft:1986ooh}%
  \BibitemOpen
  \bibfield  {author} {\bibinfo {author} {\bibfnamefont {G.}~\bibnamefont
  {'t~Hooft}},\ }\bibfield  {title} {\bibinfo {title} {{How Instantons Solve
  the U(1) Problem}},\ }\href {https://doi.org/10.1016/0370-1573(86)90117-1}
  {\bibfield  {journal} {\bibinfo  {journal} {Phys. Rept.}\ }\textbf {\bibinfo
  {volume} {142}},\ \bibinfo {pages} {357} (\bibinfo {year}
  {1986})}\BibitemShut {NoStop}%
\bibitem [{\citenamefont {Giacosa}\ \emph {et~al.}(2018)\citenamefont
  {Giacosa}, \citenamefont {Koenigstein},\ and\ \citenamefont
  {Pisarski}}]{Giacosa:2017pos}%
  \BibitemOpen
  \bibfield  {author} {\bibinfo {author} {\bibfnamefont {F.}~\bibnamefont
  {Giacosa}}, \bibinfo {author} {\bibfnamefont {A.}~\bibnamefont
  {Koenigstein}},\ and\ \bibinfo {author} {\bibfnamefont {R.~D.}\ \bibnamefont
  {Pisarski}},\ }\bibfield  {title} {\bibinfo {title} {{How the axial anomaly
  controls flavor mixing among mesons}},\ }\href
  {https://doi.org/10.1103/PhysRevD.97.091901} {\bibfield  {journal} {\bibinfo
  {journal} {Phys. Rev. D}\ }\textbf {\bibinfo {volume} {97}},\ \bibinfo
  {pages} {091901} (\bibinfo {year} {2018})},\ \Eprint
  {https://arxiv.org/abs/1709.07454} {arXiv:1709.07454 [hep-ph]} \BibitemShut
  {NoStop}%
\bibitem [{\citenamefont {Veneziano}(1989)}]{Veneziano:1989ei}%
  \BibitemOpen
  \bibfield  {author} {\bibinfo {author} {\bibfnamefont {G.}~\bibnamefont
  {Veneziano}},\ }\bibfield  {title} {\bibinfo {title} {{Is There a QCD Spin
  Crisis?}},\ }\href {https://doi.org/10.1142/S0217732389001830} {\bibfield
  {journal} {\bibinfo  {journal} {Mod. Phys. Lett. A}\ }\textbf {\bibinfo
  {volume} {4}},\ \bibinfo {pages} {1605} (\bibinfo {year} {1989})}\BibitemShut
  {NoStop}%
\bibitem [{\citenamefont {Shore}\ and\ \citenamefont
  {Veneziano}(1990)}]{Shore:1990zu}%
  \BibitemOpen
  \bibfield  {author} {\bibinfo {author} {\bibfnamefont {G.~M.}\ \bibnamefont
  {Shore}}\ and\ \bibinfo {author} {\bibfnamefont {G.}~\bibnamefont
  {Veneziano}},\ }\bibfield  {title} {\bibinfo {title} {{The U(1)
  {Goldberger-Treiman} Relation and the Two Components of the Proton 'Spin'}},\
  }\href {https://doi.org/10.1016/0370-2693(90)90272-8} {\bibfield  {journal}
  {\bibinfo  {journal} {Phys. Lett. B}\ }\textbf {\bibinfo {volume} {244}},\
  \bibinfo {pages} {75} (\bibinfo {year} {1990})}\BibitemShut {NoStop}%
\bibitem [{\citenamefont {Shore}\ and\ \citenamefont
  {Veneziano}(1992)}]{Shore:1991dv}%
  \BibitemOpen
  \bibfield  {author} {\bibinfo {author} {\bibfnamefont {G.~M.}\ \bibnamefont
  {Shore}}\ and\ \bibinfo {author} {\bibfnamefont {G.}~\bibnamefont
  {Veneziano}},\ }\bibfield  {title} {\bibinfo {title} {{The U(1)
  Goldberger-Treiman relation and the proton 'spin': A Renormalization group
  analysis}},\ }\href {https://doi.org/10.1016/0550-3213(92)90639-S} {\bibfield
   {journal} {\bibinfo  {journal} {Nucl. Phys. B}\ }\textbf {\bibinfo {volume}
  {381}},\ \bibinfo {pages} {23} (\bibinfo {year} {1992})}\BibitemShut
  {NoStop}%
\bibitem [{\citenamefont {Shore}\ and\ \citenamefont
  {Veneziano}(1998)}]{Shore:1997tq}%
  \BibitemOpen
  \bibfield  {author} {\bibinfo {author} {\bibfnamefont {G.~M.}\ \bibnamefont
  {Shore}}\ and\ \bibinfo {author} {\bibfnamefont {G.}~\bibnamefont
  {Veneziano}},\ }\bibfield  {title} {\bibinfo {title} {{Testing target
  independence of the `proton spin' effect in semiinclusive deep inelastic
  scattering}},\ }\href {https://doi.org/10.1016/S0550-3213(97)00751-7}
  {\bibfield  {journal} {\bibinfo  {journal} {Nucl. Phys. B}\ }\textbf
  {\bibinfo {volume} {516}},\ \bibinfo {pages} {333} (\bibinfo {year}
  {1998})},\ \Eprint {https://arxiv.org/abs/hep-ph/9709213}
  {arXiv:hep-ph/9709213} \BibitemShut {NoStop}%
\bibitem [{\citenamefont {Narison}\ \emph {et~al.}(1999)\citenamefont
  {Narison}, \citenamefont {Shore},\ and\ \citenamefont
  {Veneziano}}]{Narison:1998aq}%
  \BibitemOpen
  \bibfield  {author} {\bibinfo {author} {\bibfnamefont {S.}~\bibnamefont
  {Narison}}, \bibinfo {author} {\bibfnamefont {G.~M.}\ \bibnamefont {Shore}},\
  and\ \bibinfo {author} {\bibfnamefont {G.}~\bibnamefont {Veneziano}},\
  }\bibfield  {title} {\bibinfo {title} {{Topological charge screening and the
  'proton spin' beyond the chiral limit}},\ }\href
  {https://doi.org/10.1016/S0550-3213(99)00061-9} {\bibfield  {journal}
  {\bibinfo  {journal} {Nucl. Phys. B}\ }\textbf {\bibinfo {volume} {546}},\
  \bibinfo {pages} {235} (\bibinfo {year} {1999})},\ \Eprint
  {https://arxiv.org/abs/hep-ph/9812333} {arXiv:hep-ph/9812333} \BibitemShut
  {NoStop}%
\bibitem [{\citenamefont {Bass}(2005)}]{Bass:2004xa}%
  \BibitemOpen
  \bibfield  {author} {\bibinfo {author} {\bibfnamefont {S.~D.}\ \bibnamefont
  {Bass}},\ }\bibfield  {title} {\bibinfo {title} {{The Spin structure of the
  proton}},\ }\href {https://doi.org/10.1103/RevModPhys.77.1257} {\bibfield
  {journal} {\bibinfo  {journal} {Rev. Mod. Phys.}\ }\textbf {\bibinfo {volume}
  {77}},\ \bibinfo {pages} {1257} (\bibinfo {year} {2005})},\ \Eprint
  {https://arxiv.org/abs/hep-ph/0411005} {arXiv:hep-ph/0411005} \BibitemShut
  {NoStop}%
\bibitem [{\citenamefont {Shore}(2008)}]{Shore:2007yn}%
  \BibitemOpen
  \bibfield  {author} {\bibinfo {author} {\bibfnamefont {G.~M.}\ \bibnamefont
  {Shore}},\ }\bibfield  {title} {\bibinfo {title} {{The U(1)(A) Anomaly and
  QCD Phenomenology}},\ }\href@noop {} {\bibfield  {journal} {\bibinfo
  {journal} {Lect. Notes Phys.}\ }\textbf {\bibinfo {volume} {737}},\ \bibinfo
  {pages} {235} (\bibinfo {year} {2008})},\ \Eprint
  {https://arxiv.org/abs/hep-ph/0701171} {arXiv:hep-ph/0701171} \BibitemShut
  {NoStop}%
\bibitem [{\citenamefont {Tarasov}\ and\ \citenamefont
  {Venugopalan}(2020)}]{Tarasov:2020cwl}%
  \BibitemOpen
  \bibfield  {author} {\bibinfo {author} {\bibfnamefont {A.}~\bibnamefont
  {Tarasov}}\ and\ \bibinfo {author} {\bibfnamefont {R.}~\bibnamefont
  {Venugopalan}},\ }\bibfield  {title} {\bibinfo {title} {{Role of the chiral
  anomaly in polarized deeply inelastic scattering: Finding the triangle graph
  inside the box diagram in Bjorken and Regge asymptotics}},\ }\href
  {https://doi.org/10.1103/PhysRevD.102.114022} {\bibfield  {journal} {\bibinfo
   {journal} {Phys. Rev. D}\ }\textbf {\bibinfo {volume} {102}},\ \bibinfo
  {pages} {114022} (\bibinfo {year} {2020})},\ \Eprint
  {https://arxiv.org/abs/2008.08104} {arXiv:2008.08104 [hep-ph]} \BibitemShut
  {NoStop}%
\bibitem [{\citenamefont {Tarasov}\ and\ \citenamefont
  {Venugopalan}(2022)}]{Tarasov:2021yll}%
  \BibitemOpen
  \bibfield  {author} {\bibinfo {author} {\bibfnamefont {A.}~\bibnamefont
  {Tarasov}}\ and\ \bibinfo {author} {\bibfnamefont {R.}~\bibnamefont
  {Venugopalan}},\ }\bibfield  {title} {\bibinfo {title} {{Role of the chiral
  anomaly in polarized deeply inelastic scattering. II. Topological screening
  and transitions from emergent axionlike dynamics}},\ }\href
  {https://doi.org/10.1103/PhysRevD.105.014020} {\bibfield  {journal} {\bibinfo
   {journal} {Phys. Rev. D}\ }\textbf {\bibinfo {volume} {105}},\ \bibinfo
  {pages} {014020} (\bibinfo {year} {2022})},\ \Eprint
  {https://arxiv.org/abs/2109.10370} {arXiv:2109.10370 [hep-ph]} \BibitemShut
  {NoStop}%
\bibitem [{\citenamefont {Gross}\ \emph {et~al.}(1981)\citenamefont {Gross},
  \citenamefont {Pisarski},\ and\ \citenamefont {Yaffe}}]{Gross:1980br}%
  \BibitemOpen
  \bibfield  {author} {\bibinfo {author} {\bibfnamefont {D.~J.}\ \bibnamefont
  {Gross}}, \bibinfo {author} {\bibfnamefont {R.~D.}\ \bibnamefont
  {Pisarski}},\ and\ \bibinfo {author} {\bibfnamefont {L.~G.}\ \bibnamefont
  {Yaffe}},\ }\bibfield  {title} {\bibinfo {title} {{QCD and Instantons at
  Finite Temperature}},\ }\href {https://doi.org/10.1103/RevModPhys.53.43}
  {\bibfield  {journal} {\bibinfo  {journal} {Rev. Mod. Phys.}\ }\textbf
  {\bibinfo {volume} {53}},\ \bibinfo {pages} {43} (\bibinfo {year}
  {1981})}\BibitemShut {NoStop}%
\bibitem [{\citenamefont {Korthals~Altes}\ and\ \citenamefont
  {Sastre}(2014)}]{KorthalsAltes:2014dkx}%
  \BibitemOpen
  \bibfield  {author} {\bibinfo {author} {\bibfnamefont {C.~P.}\ \bibnamefont
  {Korthals~Altes}}\ and\ \bibinfo {author} {\bibfnamefont {A.}~\bibnamefont
  {Sastre}},\ }\bibfield  {title} {\bibinfo {title} {{Thermal instanton
  determinant in compact form}},\ }\href
  {https://doi.org/10.1103/PhysRevD.90.125002} {\bibfield  {journal} {\bibinfo
  {journal} {Phys. Rev. D}\ }\textbf {\bibinfo {volume} {90}},\ \bibinfo
  {pages} {125002} (\bibinfo {year} {2014})},\ \Eprint
  {https://arxiv.org/abs/1406.6911} {arXiv:1406.6911 [hep-th]} \BibitemShut
  {NoStop}%
\bibitem [{\citenamefont {Korthals~Altes}\ and\ \citenamefont
  {Sastre}(2015)}]{KorthalsAltes:2015zpx}%
  \BibitemOpen
  \bibfield  {author} {\bibinfo {author} {\bibfnamefont {C.~P.}\ \bibnamefont
  {Korthals~Altes}}\ and\ \bibinfo {author} {\bibfnamefont {A.}~\bibnamefont
  {Sastre}},\ }\bibfield  {title} {\bibinfo {title} {{Caloron correction to the
  effective potential in thermal gluodynamics}},\ }\href@noop {} {\  (\bibinfo
  {year} {2015})},\ \Eprint {https://arxiv.org/abs/1501.00474}
  {arXiv:1501.00474 [hep-th]} \BibitemShut {NoStop}%
\bibitem [{\citenamefont {Pisarski}\ and\ \citenamefont
  {Rennecke}(2020)}]{Pisarski:2019upw}%
  \BibitemOpen
  \bibfield  {author} {\bibinfo {author} {\bibfnamefont {R.~D.}\ \bibnamefont
  {Pisarski}}\ and\ \bibinfo {author} {\bibfnamefont {F.}~\bibnamefont
  {Rennecke}},\ }\bibfield  {title} {\bibinfo {title} {{Multi-instanton
  contributions to anomalous quark interactions}},\ }\href
  {https://doi.org/10.1103/PhysRevD.101.114019} {\bibfield  {journal} {\bibinfo
   {journal} {Phys. Rev. D}\ }\textbf {\bibinfo {volume} {101}},\ \bibinfo
  {pages} {114019} (\bibinfo {year} {2020})},\ \bibinfo {note} {and references
  therein},\ \Eprint {https://arxiv.org/abs/1910.14052} {arXiv:1910.14052
  [hep-ph]} \BibitemShut {NoStop}%
\bibitem [{\citenamefont {Rennecke}(2020)}]{Rennecke:2020zgb}%
  \BibitemOpen
  \bibfield  {author} {\bibinfo {author} {\bibfnamefont {F.}~\bibnamefont
  {Rennecke}},\ }\bibfield  {title} {\bibinfo {title} {{Higher topological
  charge and the QCD vacuum}},\ }\href
  {https://doi.org/10.1103/PhysRevResearch.2.033359} {\bibfield  {journal}
  {\bibinfo  {journal} {Phys. Rev. Res.}\ }\textbf {\bibinfo {volume} {2}},\
  \bibinfo {pages} {033359} (\bibinfo {year} {2020})},\ \Eprint
  {https://arxiv.org/abs/2003.13876} {arXiv:2003.13876 [hep-th]} \BibitemShut
  {NoStop}%
\bibitem [{\citenamefont {Boccaletti}\ and\ \citenamefont
  {Nogradi}(2020)}]{Boccaletti:2020mxu}%
  \BibitemOpen
  \bibfield  {author} {\bibinfo {author} {\bibfnamefont {A.}~\bibnamefont
  {Boccaletti}}\ and\ \bibinfo {author} {\bibfnamefont {D.}~\bibnamefont
  {Nogradi}},\ }\bibfield  {title} {\bibinfo {title} {{The semi-classical
  approximation at high temperature revisited}},\ }\href
  {https://doi.org/10.1007/JHEP03(2020)045} {\bibfield  {journal} {\bibinfo
  {journal} {JHEP}\ }\textbf {\bibinfo {volume} {03}},\ \bibinfo {pages}
  {045}},\ \Eprint {https://arxiv.org/abs/2001.03383} {arXiv:2001.03383
  [hep-ph]} \BibitemShut {NoStop}%
\bibitem [{\citenamefont {Nogradi}\ \emph {et~al.}(2023)\citenamefont
  {Nogradi}, \citenamefont {Papavasilliou},\ and\ \citenamefont
  {Pisarski}}]{Nogradi2023}%
  \BibitemOpen
  \bibfield  {author} {\bibinfo {author} {\bibfnamefont {D.}~\bibnamefont
  {Nogradi}}, \bibinfo {author} {\bibfnamefont {J.}~\bibnamefont
  {Papavasilliou}},\ and\ \bibinfo {author} {\bibfnamefont {R.~D.}\
  \bibnamefont {Pisarski}}} (\bibinfo {year} {2023}),\ \bibinfo {note} {work in
  progress}\BibitemShut {NoStop}%
\bibitem [{Note1()}]{Note1}%
  \BibitemOpen
  \bibinfo {note} {\label {density_comment} The computation at $T\protect \neq
  0$ is complete to one loop order, while that at $\mu _{\protect \rm qk}
  \protect \neq 0$ is lacking, although doable \cite
  {Nogradi2023}.}\BibitemShut {Stop}%
\bibitem [{Note2()}]{Note2}%
  \BibitemOpen
  \bibinfo {note} {For an $SU(N)$ gauge theory coupled to $N_f$ flavors of
  massless quarks, $c = (11 N - 4 N_f C_f)/3$, $C_f = (N^2-1)/(2N)$. The factor
  of $T^4$ in Eq. (\ref {highT_top_sucep}) arises from the integral over the
  instanton scale size, $\rho $.}\BibitemShut {Stop}%
\bibitem [{\citenamefont {Borsanyi}\ \emph
  {et~al.}(2016{\natexlab{a}})\citenamefont {Borsanyi}, \citenamefont
  {Dierigl}, \citenamefont {Fodor}, \citenamefont {Katz}, \citenamefont
  {Mages}, \citenamefont {Nogradi}, \citenamefont {Redondo}, \citenamefont
  {Ringwald},\ and\ \citenamefont {Szabo}}]{Borsanyi:2015cka}%
  \BibitemOpen
  \bibfield  {author} {\bibinfo {author} {\bibfnamefont {S.}~\bibnamefont
  {Borsanyi}}, \bibinfo {author} {\bibfnamefont {M.}~\bibnamefont {Dierigl}},
  \bibinfo {author} {\bibfnamefont {Z.}~\bibnamefont {Fodor}}, \bibinfo
  {author} {\bibfnamefont {S.~D.}\ \bibnamefont {Katz}}, \bibinfo {author}
  {\bibfnamefont {S.~W.}\ \bibnamefont {Mages}}, \bibinfo {author}
  {\bibfnamefont {D.}~\bibnamefont {Nogradi}}, \bibinfo {author} {\bibfnamefont
  {J.}~\bibnamefont {Redondo}}, \bibinfo {author} {\bibfnamefont
  {A.}~\bibnamefont {Ringwald}},\ and\ \bibinfo {author} {\bibfnamefont
  {K.~K.}\ \bibnamefont {Szabo}},\ }\bibfield  {title} {\bibinfo {title}
  {{Axion cosmology, lattice QCD and the dilute instanton gas}},\ }\href
  {https://doi.org/10.1016/j.physletb.2015.11.020} {\bibfield  {journal}
  {\bibinfo  {journal} {Phys. Lett. B}\ }\textbf {\bibinfo {volume} {752}},\
  \bibinfo {pages} {175} (\bibinfo {year} {2016}{\natexlab{a}})},\ \Eprint
  {https://arxiv.org/abs/1508.06917} {arXiv:1508.06917 [hep-lat]} \BibitemShut
  {NoStop}%
\bibitem [{\citenamefont {Grilli~di Cortona}\ \emph {et~al.}(2016)\citenamefont
  {Grilli~di Cortona}, \citenamefont {Hardy}, \citenamefont {Pardo~Vega},\ and\
  \citenamefont {Villadoro}}]{GrillidiCortona:2015jxo}%
  \BibitemOpen
  \bibfield  {author} {\bibinfo {author} {\bibfnamefont {G.}~\bibnamefont
  {Grilli~di Cortona}}, \bibinfo {author} {\bibfnamefont {E.}~\bibnamefont
  {Hardy}}, \bibinfo {author} {\bibfnamefont {J.}~\bibnamefont {Pardo~Vega}},\
  and\ \bibinfo {author} {\bibfnamefont {G.}~\bibnamefont {Villadoro}},\
  }\bibfield  {title} {\bibinfo {title} {{The QCD axion, precisely}},\ }\href
  {https://doi.org/10.1007/JHEP01(2016)034} {\bibfield  {journal} {\bibinfo
  {journal} {JHEP}\ }\textbf {\bibinfo {volume} {01}},\ \bibinfo {pages}
  {034}},\ \Eprint {https://arxiv.org/abs/1511.02867} {arXiv:1511.02867
  [hep-ph]} \BibitemShut {NoStop}%
\bibitem [{\citenamefont {Bonati}\ \emph
  {et~al.}(2016{\natexlab{a}})\citenamefont {Bonati}, \citenamefont {D'Elia},
  \citenamefont {Mariti}, \citenamefont {Martinelli}, \citenamefont {Mesiti},
  \citenamefont {Negro}, \citenamefont {Sanfilippo},\ and\ \citenamefont
  {Villadoro}}]{Bonati:2015vqz}%
  \BibitemOpen
  \bibfield  {author} {\bibinfo {author} {\bibfnamefont {C.}~\bibnamefont
  {Bonati}}, \bibinfo {author} {\bibfnamefont {M.}~\bibnamefont {D'Elia}},
  \bibinfo {author} {\bibfnamefont {M.}~\bibnamefont {Mariti}}, \bibinfo
  {author} {\bibfnamefont {G.}~\bibnamefont {Martinelli}}, \bibinfo {author}
  {\bibfnamefont {M.}~\bibnamefont {Mesiti}}, \bibinfo {author} {\bibfnamefont
  {F.}~\bibnamefont {Negro}}, \bibinfo {author} {\bibfnamefont
  {F.}~\bibnamefont {Sanfilippo}},\ and\ \bibinfo {author} {\bibfnamefont
  {G.}~\bibnamefont {Villadoro}},\ }\bibfield  {title} {\bibinfo {title}
  {{Axion phenomenology and $\theta$-dependence from $N_f = 2+1$ lattice
  QCD}},\ }\href {https://doi.org/10.1007/JHEP03(2016)155} {\bibfield
  {journal} {\bibinfo  {journal} {JHEP}\ }\textbf {\bibinfo {volume} {03}},\
  \bibinfo {pages} {155}},\ \Eprint {https://arxiv.org/abs/1512.06746}
  {arXiv:1512.06746 [hep-lat]} \BibitemShut {NoStop}%
\bibitem [{\citenamefont {Borsanyi}\ \emph
  {et~al.}(2016{\natexlab{b}})\citenamefont {Borsanyi} \emph
  {et~al.}}]{Borsanyi:2016ksw}%
  \BibitemOpen
  \bibfield  {author} {\bibinfo {author} {\bibfnamefont {S.}~\bibnamefont
  {Borsanyi}} \emph {et~al.},\ }\bibfield  {title} {\bibinfo {title}
  {{Calculation of the axion mass based on high-temperature lattice quantum
  chromodynamics}},\ }\href {https://doi.org/10.1038/nature20115} {\bibfield
  {journal} {\bibinfo  {journal} {Nature}\ }\textbf {\bibinfo {volume} {539}},\
  \bibinfo {pages} {69} (\bibinfo {year} {2016}{\natexlab{b}})},\ \Eprint
  {https://arxiv.org/abs/1606.07494} {arXiv:1606.07494 [hep-lat]} \BibitemShut
  {NoStop}%
\bibitem [{\citenamefont {Frison}\ \emph {et~al.}(2016)\citenamefont {Frison},
  \citenamefont {Kitano}, \citenamefont {Matsufuru}, \citenamefont {Mori},\
  and\ \citenamefont {Yamada}}]{Frison:2016vuc}%
  \BibitemOpen
  \bibfield  {author} {\bibinfo {author} {\bibfnamefont {J.}~\bibnamefont
  {Frison}}, \bibinfo {author} {\bibfnamefont {R.}~\bibnamefont {Kitano}},
  \bibinfo {author} {\bibfnamefont {H.}~\bibnamefont {Matsufuru}}, \bibinfo
  {author} {\bibfnamefont {S.}~\bibnamefont {Mori}},\ and\ \bibinfo {author}
  {\bibfnamefont {N.}~\bibnamefont {Yamada}},\ }\bibfield  {title} {\bibinfo
  {title} {{Topological susceptibility at high temperature on the lattice}},\
  }\href {https://doi.org/10.1007/JHEP09(2016)021} {\bibfield  {journal}
  {\bibinfo  {journal} {JHEP}\ }\textbf {\bibinfo {volume} {09}},\ \bibinfo
  {pages} {021}},\ \Eprint {https://arxiv.org/abs/1606.07175} {arXiv:1606.07175
  [hep-lat]} \BibitemShut {NoStop}%
\bibitem [{\citenamefont {Petreczky}\ \emph {et~al.}(2016)\citenamefont
  {Petreczky}, \citenamefont {Schadler},\ and\ \citenamefont
  {Sharma}}]{Petreczky:2016vrs}%
  \BibitemOpen
  \bibfield  {author} {\bibinfo {author} {\bibfnamefont {P.}~\bibnamefont
  {Petreczky}}, \bibinfo {author} {\bibfnamefont {H.-P.}\ \bibnamefont
  {Schadler}},\ and\ \bibinfo {author} {\bibfnamefont {S.}~\bibnamefont
  {Sharma}},\ }\bibfield  {title} {\bibinfo {title} {{The topological
  susceptibility in finite temperature QCD and axion cosmology}},\ }\href
  {https://doi.org/10.1016/j.physletb.2016.09.063} {\bibfield  {journal}
  {\bibinfo  {journal} {Phys. Lett. B}\ }\textbf {\bibinfo {volume} {762}},\
  \bibinfo {pages} {498} (\bibinfo {year} {2016})},\ \Eprint
  {https://arxiv.org/abs/1606.03145} {arXiv:1606.03145 [hep-lat]} \BibitemShut
  {NoStop}%
\bibitem [{\citenamefont {Taniguchi}\ \emph {et~al.}(2017)\citenamefont
  {Taniguchi}, \citenamefont {Kanaya}, \citenamefont {Suzuki},\ and\
  \citenamefont {Umeda}}]{Taniguchi:2016tjc}%
  \BibitemOpen
  \bibfield  {author} {\bibinfo {author} {\bibfnamefont {Y.}~\bibnamefont
  {Taniguchi}}, \bibinfo {author} {\bibfnamefont {K.}~\bibnamefont {Kanaya}},
  \bibinfo {author} {\bibfnamefont {H.}~\bibnamefont {Suzuki}},\ and\ \bibinfo
  {author} {\bibfnamefont {T.}~\bibnamefont {Umeda}},\ }\bibfield  {title}
  {\bibinfo {title} {{Topological susceptibility in finite temperature ( 2+1
  )-flavor QCD using gradient flow}},\ }\href
  {https://doi.org/10.1103/PhysRevD.95.054502} {\bibfield  {journal} {\bibinfo
  {journal} {Phys. Rev. D}\ }\textbf {\bibinfo {volume} {95}},\ \bibinfo
  {pages} {054502} (\bibinfo {year} {2017})},\ \Eprint
  {https://arxiv.org/abs/1611.02411} {arXiv:1611.02411 [hep-lat]} \BibitemShut
  {NoStop}%
\bibitem [{\citenamefont {Lombardo}\ and\ \citenamefont
  {Trunin}(2020)}]{Lombardo:2020bvn}%
  \BibitemOpen
  \bibfield  {author} {\bibinfo {author} {\bibfnamefont {M.~P.}\ \bibnamefont
  {Lombardo}}\ and\ \bibinfo {author} {\bibfnamefont {A.}~\bibnamefont
  {Trunin}},\ }\bibfield  {title} {\bibinfo {title} {{Topology and axions in
  QCD}},\ }\href {https://doi.org/10.1142/S0217751X20300100} {\bibfield
  {journal} {\bibinfo  {journal} {Int. J. Mod. Phys. A}\ }\textbf {\bibinfo
  {volume} {35}},\ \bibinfo {pages} {2030010} (\bibinfo {year} {2020})},\
  \Eprint {https://arxiv.org/abs/2005.06547} {arXiv:2005.06547 [hep-lat]}
  \BibitemShut {NoStop}%
\bibitem [{\citenamefont {Jahn}\ \emph {et~al.}(2021)\citenamefont {Jahn},
  \citenamefont {Junnarkar}, \citenamefont {Moore},\ and\ \citenamefont
  {Robaina}}]{Jahn:2021qrp}%
  \BibitemOpen
  \bibfield  {author} {\bibinfo {author} {\bibfnamefont {P.~T.}\ \bibnamefont
  {Jahn}}, \bibinfo {author} {\bibfnamefont {P.~M.}\ \bibnamefont {Junnarkar}},
  \bibinfo {author} {\bibfnamefont {G.~D.}\ \bibnamefont {Moore}},\ and\
  \bibinfo {author} {\bibfnamefont {D.}~\bibnamefont {Robaina}},\ }\bibfield
  {title} {\bibinfo {title} {{Multicanonical reweighting for the QCD
  topological susceptibility}},\ }\href
  {https://doi.org/10.1103/PhysRevD.104.014502} {\bibfield  {journal} {\bibinfo
   {journal} {Phys. Rev. D}\ }\textbf {\bibinfo {volume} {104}},\ \bibinfo
  {pages} {014502} (\bibinfo {year} {2021})},\ \Eprint
  {https://arxiv.org/abs/2103.01069} {arXiv:2103.01069 [hep-lat]} \BibitemShut
  {NoStop}%
\bibitem [{\citenamefont {Borsanyi}\ and\ \citenamefont
  {Sexty}(2021)}]{Borsanyi:2021gqg}%
  \BibitemOpen
  \bibfield  {author} {\bibinfo {author} {\bibfnamefont {S.}~\bibnamefont
  {Borsanyi}}\ and\ \bibinfo {author} {\bibfnamefont {D.}~\bibnamefont
  {Sexty}},\ }\bibfield  {title} {\bibinfo {title} {{Topological susceptibility
  of pure gauge theory using Density of States}},\ }\href
  {https://doi.org/10.1016/j.physletb.2021.136148} {\bibfield  {journal}
  {\bibinfo  {journal} {Phys. Lett. B}\ }\textbf {\bibinfo {volume} {815}},\
  \bibinfo {pages} {136148} (\bibinfo {year} {2021})},\ \Eprint
  {https://arxiv.org/abs/2101.03383} {arXiv:2101.03383 [hep-lat]} \BibitemShut
  {NoStop}%
\bibitem [{\citenamefont {Chen}\ \emph {et~al.}(2022)\citenamefont {Chen},
  \citenamefont {Chiu},\ and\ \citenamefont {Hsieh}}]{Chen:2022fid}%
  \BibitemOpen
  \bibfield  {author} {\bibinfo {author} {\bibfnamefont {Y.-C.}\ \bibnamefont
  {Chen}}, \bibinfo {author} {\bibfnamefont {T.-W.}\ \bibnamefont {Chiu}},\
  and\ \bibinfo {author} {\bibfnamefont {T.-H.}\ \bibnamefont {Hsieh}},\
  }\bibfield  {title} {\bibinfo {title} {{Topological susceptibility in finite
  temperature QCD with physical $(u/d, s, c)$ domain-wall quarks}},\
  }\href@noop {} {\  (\bibinfo {year} {2022})},\ \Eprint
  {https://arxiv.org/abs/2204.01556} {arXiv:2204.01556 [hep-lat]} \BibitemShut
  {NoStop}%
\bibitem [{Note3()}]{Note3}%
  \BibitemOpen
  \bibinfo {note} {The overall magnitude of the topological susceptibility is
  about an order of magnitude greater than the instanton result at one loop
  order, but surely it is necessary to include the corrections to the instanton
  density at two loop order for $T \protect \neq 0$. Multi-instanton
  configurations can also contribute at low $T$ \cite
  {Rennecke:2020zgb}.}\BibitemShut {Stop}%
\bibitem [{\citenamefont {Gaiotto}\ \emph {et~al.}(2015)\citenamefont
  {Gaiotto}, \citenamefont {Kapustin}, \citenamefont {Seiberg},\ and\
  \citenamefont {Willett}}]{Gaiotto:2014kfa}%
  \BibitemOpen
  \bibfield  {author} {\bibinfo {author} {\bibfnamefont {D.}~\bibnamefont
  {Gaiotto}}, \bibinfo {author} {\bibfnamefont {A.}~\bibnamefont {Kapustin}},
  \bibinfo {author} {\bibfnamefont {N.}~\bibnamefont {Seiberg}},\ and\ \bibinfo
  {author} {\bibfnamefont {B.}~\bibnamefont {Willett}},\ }\bibfield  {title}
  {\bibinfo {title} {{Generalized Global Symmetries}},\ }\href
  {https://doi.org/10.1007/JHEP02(2015)172} {\bibfield  {journal} {\bibinfo
  {journal} {JHEP}\ }\textbf {\bibinfo {volume} {02}},\ \bibinfo {pages}
  {172}},\ \Eprint {https://arxiv.org/abs/1412.5148} {arXiv:1412.5148 [hep-th]}
  \BibitemShut {NoStop}%
\bibitem [{\citenamefont {Alles}\ \emph {et~al.}(1997)\citenamefont {Alles},
  \citenamefont {D'Elia},\ and\ \citenamefont {Di~Giacomo}}]{Alles:1996nm}%
  \BibitemOpen
  \bibfield  {author} {\bibinfo {author} {\bibfnamefont {B.}~\bibnamefont
  {Alles}}, \bibinfo {author} {\bibfnamefont {M.}~\bibnamefont {D'Elia}},\ and\
  \bibinfo {author} {\bibfnamefont {A.}~\bibnamefont {Di~Giacomo}},\ }\bibfield
   {title} {\bibinfo {title} {{Topological susceptibility at zero and finite T
  in SU(3) Yang-Mills theory}},\ }\href
  {https://doi.org/10.1016/S0550-3213(97)00205-8} {\bibfield  {journal}
  {\bibinfo  {journal} {Nucl. Phys. B}\ }\textbf {\bibinfo {volume} {494}},\
  \bibinfo {pages} {281} (\bibinfo {year} {1997})},\ \bibinfo {note} {[Erratum:
  Nucl.Phys.B 679, 397--399 (2004)]},\ \Eprint
  {https://arxiv.org/abs/hep-lat/9605013} {arXiv:hep-lat/9605013} \BibitemShut
  {NoStop}%
\bibitem [{\citenamefont {Durr}\ \emph {et~al.}(2007)\citenamefont {Durr},
  \citenamefont {Fodor}, \citenamefont {Hoelbling},\ and\ \citenamefont
  {Kurth}}]{Durr:2006ky}%
  \BibitemOpen
  \bibfield  {author} {\bibinfo {author} {\bibfnamefont {S.}~\bibnamefont
  {Durr}}, \bibinfo {author} {\bibfnamefont {Z.}~\bibnamefont {Fodor}},
  \bibinfo {author} {\bibfnamefont {C.}~\bibnamefont {Hoelbling}},\ and\
  \bibinfo {author} {\bibfnamefont {T.}~\bibnamefont {Kurth}},\ }\bibfield
  {title} {\bibinfo {title} {{Precision study of the SU(3) topological
  susceptibility in the continuum}},\ }\href
  {https://doi.org/10.1088/1126-6708/2007/04/055} {\bibfield  {journal}
  {\bibinfo  {journal} {JHEP}\ }\textbf {\bibinfo {volume} {04}},\ \bibinfo
  {pages} {055}},\ \Eprint {https://arxiv.org/abs/hep-lat/0612021}
  {arXiv:hep-lat/0612021} \BibitemShut {NoStop}%
\bibitem [{\citenamefont {Luscher}\ and\ \citenamefont
  {Palombi}(2010)}]{Luscher:2010ik}%
  \BibitemOpen
  \bibfield  {author} {\bibinfo {author} {\bibfnamefont {M.}~\bibnamefont
  {Luscher}}\ and\ \bibinfo {author} {\bibfnamefont {F.}~\bibnamefont
  {Palombi}},\ }\bibfield  {title} {\bibinfo {title} {{Universality of the
  topological susceptibility in the SU(3) gauge theory}},\ }\href
  {https://doi.org/10.1007/JHEP09(2010)110} {\bibfield  {journal} {\bibinfo
  {journal} {JHEP}\ }\textbf {\bibinfo {volume} {09}},\ \bibinfo {pages}
  {110}},\ \Eprint {https://arxiv.org/abs/1008.0732} {arXiv:1008.0732
  [hep-lat]} \BibitemShut {NoStop}%
\bibitem [{\citenamefont {Xiong}\ \emph {et~al.}(2016)\citenamefont {Xiong},
  \citenamefont {Zhang}, \citenamefont {Chen}, \citenamefont {Liu},
  \citenamefont {Liu},\ and\ \citenamefont {Ma}}]{Xiong:2015dya}%
  \BibitemOpen
  \bibfield  {author} {\bibinfo {author} {\bibfnamefont {G.-Y.}\ \bibnamefont
  {Xiong}}, \bibinfo {author} {\bibfnamefont {J.-B.}\ \bibnamefont {Zhang}},
  \bibinfo {author} {\bibfnamefont {Y.}~\bibnamefont {Chen}}, \bibinfo {author}
  {\bibfnamefont {C.}~\bibnamefont {Liu}}, \bibinfo {author} {\bibfnamefont
  {Y.-B.}\ \bibnamefont {Liu}},\ and\ \bibinfo {author} {\bibfnamefont {J.-P.}\
  \bibnamefont {Ma}},\ }\bibfield  {title} {\bibinfo {title} {{Topological
  susceptibility near T$_c$ in SU(3) gauge theory}},\ }\href
  {https://doi.org/10.1016/j.physletb.2015.10.085} {\bibfield  {journal}
  {\bibinfo  {journal} {Phys. Lett. B}\ }\textbf {\bibinfo {volume} {752}},\
  \bibinfo {pages} {34} (\bibinfo {year} {2016})},\ \Eprint
  {https://arxiv.org/abs/1508.07704} {arXiv:1508.07704 [hep-lat]} \BibitemShut
  {NoStop}%
\bibitem [{\citenamefont {Jahn}\ \emph {et~al.}(2018)\citenamefont {Jahn},
  \citenamefont {Moore},\ and\ \citenamefont {Robaina}}]{Jahn:2018dke}%
  \BibitemOpen
  \bibfield  {author} {\bibinfo {author} {\bibfnamefont {P.~T.}\ \bibnamefont
  {Jahn}}, \bibinfo {author} {\bibfnamefont {G.~D.}\ \bibnamefont {Moore}},\
  and\ \bibinfo {author} {\bibfnamefont {D.}~\bibnamefont {Robaina}},\
  }\bibfield  {title} {\bibinfo {title} {{$\chi_{\textrm{top}}(T \gg
  T_{\textrm{c}})$ in pure-glue QCD through reweighting}},\ }\href
  {https://doi.org/10.1103/PhysRevD.98.054512} {\bibfield  {journal} {\bibinfo
  {journal} {Phys. Rev. D}\ }\textbf {\bibinfo {volume} {98}},\ \bibinfo
  {pages} {054512} (\bibinfo {year} {2018})},\ \Eprint
  {https://arxiv.org/abs/1806.01162} {arXiv:1806.01162 [hep-lat]} \BibitemShut
  {NoStop}%
\bibitem [{\citenamefont {Giusti}\ and\ \citenamefont
  {L\"uscher}(2019)}]{Giusti:2018cmp}%
  \BibitemOpen
  \bibfield  {author} {\bibinfo {author} {\bibfnamefont {L.}~\bibnamefont
  {Giusti}}\ and\ \bibinfo {author} {\bibfnamefont {M.}~\bibnamefont
  {L\"uscher}},\ }\bibfield  {title} {\bibinfo {title} {{Topological
  susceptibility at $T>T_{\rm c}$ from master-field simulations of the SU(3)
  gauge theory}},\ }\href {https://doi.org/10.1140/epjc/s10052-019-6706-7}
  {\bibfield  {journal} {\bibinfo  {journal} {Eur. Phys. J. C}\ }\textbf
  {\bibinfo {volume} {79}},\ \bibinfo {pages} {207} (\bibinfo {year} {2019})},\
  \Eprint {https://arxiv.org/abs/1812.02062} {arXiv:1812.02062 [hep-lat]}
  \BibitemShut {NoStop}%
\bibitem [{\citenamefont {Lucini}\ and\ \citenamefont
  {Teper}(2001)}]{Lucini:2001ej}%
  \BibitemOpen
  \bibfield  {author} {\bibinfo {author} {\bibfnamefont {B.}~\bibnamefont
  {Lucini}}\ and\ \bibinfo {author} {\bibfnamefont {M.}~\bibnamefont {Teper}},\
  }\bibfield  {title} {\bibinfo {title} {{SU(N) gauge theories in
  four-dimensions: Exploring the approach to N = infinity}},\ }\href
  {https://doi.org/10.1088/1126-6708/2001/06/050} {\bibfield  {journal}
  {\bibinfo  {journal} {JHEP}\ }\textbf {\bibinfo {volume} {06}},\ \bibinfo
  {pages} {050}},\ \Eprint {https://arxiv.org/abs/hep-lat/0103027}
  {arXiv:hep-lat/0103027} \BibitemShut {NoStop}%
\bibitem [{\citenamefont {Del~Debbio}\ \emph {et~al.}(2006)\citenamefont
  {Del~Debbio}, \citenamefont {Manca}, \citenamefont {Panagopoulos},
  \citenamefont {Skouroupathis},\ and\ \citenamefont
  {Vicari}}]{DelDebbio:2006yuf}%
  \BibitemOpen
  \bibfield  {author} {\bibinfo {author} {\bibfnamefont {L.}~\bibnamefont
  {Del~Debbio}}, \bibinfo {author} {\bibfnamefont {G.~M.}\ \bibnamefont
  {Manca}}, \bibinfo {author} {\bibfnamefont {H.}~\bibnamefont {Panagopoulos}},
  \bibinfo {author} {\bibfnamefont {A.}~\bibnamefont {Skouroupathis}},\ and\
  \bibinfo {author} {\bibfnamefont {E.}~\bibnamefont {Vicari}},\ }\bibfield
  {title} {\bibinfo {title} {{Theta-dependence of the spectrum of SU(N) gauge
  theories}},\ }\href {https://doi.org/10.1088/1126-6708/2006/06/005}
  {\bibfield  {journal} {\bibinfo  {journal} {JHEP}\ }\textbf {\bibinfo
  {volume} {06}},\ \bibinfo {pages} {005}},\ \Eprint
  {https://arxiv.org/abs/hep-th/0603041} {arXiv:hep-th/0603041} \BibitemShut
  {NoStop}%
\bibitem [{\citenamefont {Vicari}\ and\ \citenamefont
  {Panagopoulos}(2009)}]{Vicari:2008jw}%
  \BibitemOpen
  \bibfield  {author} {\bibinfo {author} {\bibfnamefont {E.}~\bibnamefont
  {Vicari}}\ and\ \bibinfo {author} {\bibfnamefont {H.}~\bibnamefont
  {Panagopoulos}},\ }\bibfield  {title} {\bibinfo {title} {{Theta dependence of
  SU(N) gauge theories in the presence of a topological term}},\ }\href
  {https://doi.org/10.1016/j.physrep.2008.10.001} {\bibfield  {journal}
  {\bibinfo  {journal} {Phys. Rept.}\ }\textbf {\bibinfo {volume} {470}},\
  \bibinfo {pages} {93} (\bibinfo {year} {2009})},\ \Eprint
  {https://arxiv.org/abs/0803.1593} {arXiv:0803.1593 [hep-th]} \BibitemShut
  {NoStop}%
\bibitem [{\citenamefont {Garcia~Perez}\ \emph {et~al.}(2009)\citenamefont
  {Garcia~Perez}, \citenamefont {Gonzalez-Arroyo},\ and\ \citenamefont
  {Sastre}}]{GarciaPerez:2009mg}%
  \BibitemOpen
  \bibfield  {author} {\bibinfo {author} {\bibfnamefont {M.}~\bibnamefont
  {Garcia~Perez}}, \bibinfo {author} {\bibfnamefont {A.}~\bibnamefont
  {Gonzalez-Arroyo}},\ and\ \bibinfo {author} {\bibfnamefont {A.}~\bibnamefont
  {Sastre}},\ }\bibfield  {title} {\bibinfo {title} {{Adjoint fermion
  zero-modes for SU(N) calorons}},\ }\href
  {https://doi.org/10.1088/1126-6708/2009/06/065} {\bibfield  {journal}
  {\bibinfo  {journal} {JHEP}\ }\textbf {\bibinfo {volume} {06}},\ \bibinfo
  {pages} {065}},\ \Eprint {https://arxiv.org/abs/0905.0645} {arXiv:0905.0645
  [hep-th]} \BibitemShut {NoStop}%
\bibitem [{\citenamefont {Fodor}\ \emph
  {et~al.}(2009{\natexlab{a}})\citenamefont {Fodor}, \citenamefont {Holland},
  \citenamefont {Kuti}, \citenamefont {Nogradi},\ and\ \citenamefont
  {Schroeder}}]{Fodor:2009nh}%
  \BibitemOpen
  \bibfield  {author} {\bibinfo {author} {\bibfnamefont {Z.}~\bibnamefont
  {Fodor}}, \bibinfo {author} {\bibfnamefont {K.}~\bibnamefont {Holland}},
  \bibinfo {author} {\bibfnamefont {J.}~\bibnamefont {Kuti}}, \bibinfo {author}
  {\bibfnamefont {D.}~\bibnamefont {Nogradi}},\ and\ \bibinfo {author}
  {\bibfnamefont {C.}~\bibnamefont {Schroeder}},\ }\bibfield  {title} {\bibinfo
  {title} {{Topology and higher dimensional representations}},\ }\href
  {https://doi.org/10.1088/1126-6708/2009/08/084} {\bibfield  {journal}
  {\bibinfo  {journal} {JHEP}\ }\textbf {\bibinfo {volume} {08}},\ \bibinfo
  {pages} {084}},\ \Eprint {https://arxiv.org/abs/0905.3586} {arXiv:0905.3586
  [hep-lat]} \BibitemShut {NoStop}%
\bibitem [{\citenamefont {Fodor}\ \emph
  {et~al.}(2009{\natexlab{b}})\citenamefont {Fodor}, \citenamefont {Holland},
  \citenamefont {Kuti}, \citenamefont {Nogradi},\ and\ \citenamefont
  {Schroeder}}]{Fodor:2009ar}%
  \BibitemOpen
  \bibfield  {author} {\bibinfo {author} {\bibfnamefont {Z.}~\bibnamefont
  {Fodor}}, \bibinfo {author} {\bibfnamefont {K.}~\bibnamefont {Holland}},
  \bibinfo {author} {\bibfnamefont {J.}~\bibnamefont {Kuti}}, \bibinfo {author}
  {\bibfnamefont {D.}~\bibnamefont {Nogradi}},\ and\ \bibinfo {author}
  {\bibfnamefont {C.}~\bibnamefont {Schroeder}},\ }\bibfield  {title} {\bibinfo
  {title} {{Chiral properties of SU(3) sextet fermions}},\ }\href
  {https://doi.org/10.1088/1126-6708/2009/11/103} {\bibfield  {journal}
  {\bibinfo  {journal} {JHEP}\ }\textbf {\bibinfo {volume} {11}},\ \bibinfo
  {pages} {103}},\ \Eprint {https://arxiv.org/abs/0908.2466} {arXiv:0908.2466
  [hep-lat]} \BibitemShut {NoStop}%
\bibitem [{\citenamefont {Panagopoulos}\ and\ \citenamefont
  {Vicari}(2011)}]{Panagopoulos:2011rb}%
  \BibitemOpen
  \bibfield  {author} {\bibinfo {author} {\bibfnamefont {H.}~\bibnamefont
  {Panagopoulos}}\ and\ \bibinfo {author} {\bibfnamefont {E.}~\bibnamefont
  {Vicari}},\ }\bibfield  {title} {\bibinfo {title} {{The 4D SU(3) gauge theory
  with an imaginary $\theta$ term}},\ }\href
  {https://doi.org/10.1007/JHEP11(2011)119} {\bibfield  {journal} {\bibinfo
  {journal} {JHEP}\ }\textbf {\bibinfo {volume} {11}},\ \bibinfo {pages}
  {119}},\ \Eprint {https://arxiv.org/abs/1109.6815} {arXiv:1109.6815
  [hep-lat]} \BibitemShut {NoStop}%
\bibitem [{\citenamefont {Lucini}\ and\ \citenamefont
  {Panero}(2013)}]{Lucini:2012gg}%
  \BibitemOpen
  \bibfield  {author} {\bibinfo {author} {\bibfnamefont {B.}~\bibnamefont
  {Lucini}}\ and\ \bibinfo {author} {\bibfnamefont {M.}~\bibnamefont
  {Panero}},\ }\bibfield  {title} {\bibinfo {title} {{SU(N) gauge theories at
  large N}},\ }\href {https://doi.org/10.1016/j.physrep.2013.01.001} {\bibfield
   {journal} {\bibinfo  {journal} {Phys. Rept.}\ }\textbf {\bibinfo {volume}
  {526}},\ \bibinfo {pages} {93} (\bibinfo {year} {2013})},\ \Eprint
  {https://arxiv.org/abs/1210.4997} {arXiv:1210.4997 [hep-th]} \BibitemShut
  {NoStop}%
\bibitem [{\citenamefont {Bonati}\ \emph {et~al.}(2013)\citenamefont {Bonati},
  \citenamefont {D'Elia}, \citenamefont {Panagopoulos},\ and\ \citenamefont
  {Vicari}}]{Bonati:2013tt}%
  \BibitemOpen
  \bibfield  {author} {\bibinfo {author} {\bibfnamefont {C.}~\bibnamefont
  {Bonati}}, \bibinfo {author} {\bibfnamefont {M.}~\bibnamefont {D'Elia}},
  \bibinfo {author} {\bibfnamefont {H.}~\bibnamefont {Panagopoulos}},\ and\
  \bibinfo {author} {\bibfnamefont {E.}~\bibnamefont {Vicari}},\ }\bibfield
  {title} {\bibinfo {title} {{Change of \ensuremath{\theta} Dependence in 4D
  SU(N) Gauge Theories Across the Deconfinement Transition}},\ }\href
  {https://doi.org/10.1103/PhysRevLett.110.252003} {\bibfield  {journal}
  {\bibinfo  {journal} {Phys. Rev. Lett.}\ }\textbf {\bibinfo {volume} {110}},\
  \bibinfo {pages} {252003} (\bibinfo {year} {2013})},\ \Eprint
  {https://arxiv.org/abs/1301.7640} {arXiv:1301.7640 [hep-lat]} \BibitemShut
  {NoStop}%
\bibitem [{\citenamefont {Bonati}\ \emph
  {et~al.}(2016{\natexlab{b}})\citenamefont {Bonati}, \citenamefont {D'Elia},\
  and\ \citenamefont {Scapellato}}]{Bonati:2015sqt}%
  \BibitemOpen
  \bibfield  {author} {\bibinfo {author} {\bibfnamefont {C.}~\bibnamefont
  {Bonati}}, \bibinfo {author} {\bibfnamefont {M.}~\bibnamefont {D'Elia}},\
  and\ \bibinfo {author} {\bibfnamefont {A.}~\bibnamefont {Scapellato}},\
  }\bibfield  {title} {\bibinfo {title} {{$\theta$ dependence in $SU(3)$
  Yang-Mills theory from analytic continuation}},\ }\href
  {https://doi.org/10.1103/PhysRevD.93.025028} {\bibfield  {journal} {\bibinfo
  {journal} {Phys. Rev. D}\ }\textbf {\bibinfo {volume} {93}},\ \bibinfo
  {pages} {025028} (\bibinfo {year} {2016}{\natexlab{b}})},\ \Eprint
  {https://arxiv.org/abs/1512.01544} {arXiv:1512.01544 [hep-lat]} \BibitemShut
  {NoStop}%
\bibitem [{\citenamefont {C\`e}\ \emph {et~al.}(2016)\citenamefont {C\`e},
  \citenamefont {Garc\'\i{}a~Vera}, \citenamefont {Giusti},\ and\ \citenamefont
  {Schaefer}}]{Ce:2016awn}%
  \BibitemOpen
  \bibfield  {author} {\bibinfo {author} {\bibfnamefont {M.}~\bibnamefont
  {C\`e}}, \bibinfo {author} {\bibfnamefont {M.}~\bibnamefont
  {Garc\'\i{}a~Vera}}, \bibinfo {author} {\bibfnamefont {L.}~\bibnamefont
  {Giusti}},\ and\ \bibinfo {author} {\bibfnamefont {S.}~\bibnamefont
  {Schaefer}},\ }\bibfield  {title} {\bibinfo {title} {{The topological
  susceptibility in the large-$N$ limit of SU($N$) Yang\textendash{}Mills
  theory}},\ }\href {https://doi.org/10.1016/j.physletb.2016.09.029} {\bibfield
   {journal} {\bibinfo  {journal} {Phys. Lett. B}\ }\textbf {\bibinfo {volume}
  {762}},\ \bibinfo {pages} {232} (\bibinfo {year} {2016})},\ \Eprint
  {https://arxiv.org/abs/1607.05939} {arXiv:1607.05939 [hep-lat]} \BibitemShut
  {NoStop}%
\bibitem [{\citenamefont {Bonati}\ \emph
  {et~al.}(2016{\natexlab{c}})\citenamefont {Bonati}, \citenamefont {D'Elia},
  \citenamefont {Rossi},\ and\ \citenamefont {Vicari}}]{Bonati:2016tvi}%
  \BibitemOpen
  \bibfield  {author} {\bibinfo {author} {\bibfnamefont {C.}~\bibnamefont
  {Bonati}}, \bibinfo {author} {\bibfnamefont {M.}~\bibnamefont {D'Elia}},
  \bibinfo {author} {\bibfnamefont {P.}~\bibnamefont {Rossi}},\ and\ \bibinfo
  {author} {\bibfnamefont {E.}~\bibnamefont {Vicari}},\ }\bibfield  {title}
  {\bibinfo {title} {{$\theta$ dependence of 4D $SU(N)$ gauge theories in the
  large-$N$ limit}},\ }\href {https://doi.org/10.1103/PhysRevD.94.085017}
  {\bibfield  {journal} {\bibinfo  {journal} {Phys. Rev. D}\ }\textbf {\bibinfo
  {volume} {94}},\ \bibinfo {pages} {085017} (\bibinfo {year}
  {2016}{\natexlab{c}})},\ \Eprint {https://arxiv.org/abs/1607.06360}
  {arXiv:1607.06360 [hep-lat]} \BibitemShut {NoStop}%
\bibitem [{\citenamefont {Kitano}\ \emph {et~al.}(2017)\citenamefont {Kitano},
  \citenamefont {Suyama},\ and\ \citenamefont {Yamada}}]{Kitano:2017jng}%
  \BibitemOpen
  \bibfield  {author} {\bibinfo {author} {\bibfnamefont {R.}~\bibnamefont
  {Kitano}}, \bibinfo {author} {\bibfnamefont {T.}~\bibnamefont {Suyama}},\
  and\ \bibinfo {author} {\bibfnamefont {N.}~\bibnamefont {Yamada}},\
  }\bibfield  {title} {\bibinfo {title} {{$\theta=\pi$ in $SU(N)/\mathbb{Z}_N$
  gauge theories}},\ }\href {https://doi.org/10.1007/JHEP09(2017)137}
  {\bibfield  {journal} {\bibinfo  {journal} {JHEP}\ }\textbf {\bibinfo
  {volume} {09}},\ \bibinfo {pages} {137}},\ \Eprint
  {https://arxiv.org/abs/1709.04225} {arXiv:1709.04225 [hep-th]} \BibitemShut
  {NoStop}%
\bibitem [{\citenamefont {Itou}(2019)}]{Itou:2018wkm}%
  \BibitemOpen
  \bibfield  {author} {\bibinfo {author} {\bibfnamefont {E.}~\bibnamefont
  {Itou}},\ }\bibfield  {title} {\bibinfo {title} {{Fractional instanton of the
  SU($3$) gauge theory in weak coupling regime}},\ }\href
  {https://doi.org/10.1007/JHEP05(2019)093} {\bibfield  {journal} {\bibinfo
  {journal} {JHEP}\ }\textbf {\bibinfo {volume} {05}},\ \bibinfo {pages}
  {093}},\ \Eprint {https://arxiv.org/abs/1811.05708} {arXiv:1811.05708
  [hep-th]} \BibitemShut {NoStop}%
\bibitem [{\citenamefont {Bonanno}\ \emph {et~al.}(2019)\citenamefont
  {Bonanno}, \citenamefont {Bonati},\ and\ \citenamefont
  {D'Elia}}]{Bonanno:2018xtd}%
  \BibitemOpen
  \bibfield  {author} {\bibinfo {author} {\bibfnamefont {C.}~\bibnamefont
  {Bonanno}}, \bibinfo {author} {\bibfnamefont {C.}~\bibnamefont {Bonati}},\
  and\ \bibinfo {author} {\bibfnamefont {M.}~\bibnamefont {D'Elia}},\
  }\bibfield  {title} {\bibinfo {title} {{Topological properties of $CP^{N-1}$
  models in the large-$N$ limit}},\ }\href
  {https://doi.org/10.1007/JHEP01(2019)003} {\bibfield  {journal} {\bibinfo
  {journal} {JHEP}\ }\textbf {\bibinfo {volume} {01}},\ \bibinfo {pages}
  {003}},\ \Eprint {https://arxiv.org/abs/1807.11357} {arXiv:1807.11357
  [hep-lat]} \BibitemShut {NoStop}%
\bibitem [{\citenamefont {Bonati}\ \emph {et~al.}(2018)\citenamefont {Bonati},
  \citenamefont {Cardinali},\ and\ \citenamefont {D'Elia}}]{Bonati:2018rfg}%
  \BibitemOpen
  \bibfield  {author} {\bibinfo {author} {\bibfnamefont {C.}~\bibnamefont
  {Bonati}}, \bibinfo {author} {\bibfnamefont {M.}~\bibnamefont {Cardinali}},\
  and\ \bibinfo {author} {\bibfnamefont {M.}~\bibnamefont {D'Elia}},\
  }\bibfield  {title} {\bibinfo {title} {{$\theta$ dependence in trace deformed
  $SU(3)$ Yang-Mills theory: a lattice study}},\ }\href
  {https://doi.org/10.1103/PhysRevD.98.054508} {\bibfield  {journal} {\bibinfo
  {journal} {Phys. Rev. D}\ }\textbf {\bibinfo {volume} {98}},\ \bibinfo
  {pages} {054508} (\bibinfo {year} {2018})},\ \Eprint
  {https://arxiv.org/abs/1807.06558} {arXiv:1807.06558 [hep-lat]} \BibitemShut
  {NoStop}%
\bibitem [{\citenamefont {Bonati}\ \emph {et~al.}(2020)\citenamefont {Bonati},
  \citenamefont {Cardinali}, \citenamefont {D'Elia},\ and\ \citenamefont
  {Mazziotti}}]{Bonati:2019kmf}%
  \BibitemOpen
  \bibfield  {author} {\bibinfo {author} {\bibfnamefont {C.}~\bibnamefont
  {Bonati}}, \bibinfo {author} {\bibfnamefont {M.}~\bibnamefont {Cardinali}},
  \bibinfo {author} {\bibfnamefont {M.}~\bibnamefont {D'Elia}},\ and\ \bibinfo
  {author} {\bibfnamefont {F.}~\bibnamefont {Mazziotti}},\ }\bibfield  {title}
  {\bibinfo {title} {{$\theta$-dependence and center symmetry in Yang-Mills
  theories}},\ }\href {https://doi.org/10.1103/PhysRevD.101.034508} {\bibfield
  {journal} {\bibinfo  {journal} {Phys. Rev. D}\ }\textbf {\bibinfo {volume}
  {101}},\ \bibinfo {pages} {034508} (\bibinfo {year} {2020})},\ \Eprint
  {https://arxiv.org/abs/1912.02662} {arXiv:1912.02662 [hep-lat]} \BibitemShut
  {NoStop}%
\bibitem [{\citenamefont {Bonanno}\ \emph {et~al.}(2021)\citenamefont
  {Bonanno}, \citenamefont {Bonati},\ and\ \citenamefont
  {D'Elia}}]{Bonanno:2020hht}%
  \BibitemOpen
  \bibfield  {author} {\bibinfo {author} {\bibfnamefont {C.}~\bibnamefont
  {Bonanno}}, \bibinfo {author} {\bibfnamefont {C.}~\bibnamefont {Bonati}},\
  and\ \bibinfo {author} {\bibfnamefont {M.}~\bibnamefont {D'Elia}},\
  }\bibfield  {title} {\bibinfo {title} {{Large-$N$ $SU(N)$ Yang-Mills theories
  with milder topological freezing}},\ }\href
  {https://doi.org/10.1007/JHEP03(2021)111} {\bibfield  {journal} {\bibinfo
  {journal} {JHEP}\ }\textbf {\bibinfo {volume} {03}},\ \bibinfo {pages}
  {111}},\ \Eprint {https://arxiv.org/abs/2012.14000} {arXiv:2012.14000
  [hep-lat]} \BibitemShut {NoStop}%
\bibitem [{\citenamefont {Kitano}\ \emph {et~al.}(2021)\citenamefont {Kitano},
  \citenamefont {Matsudo}, \citenamefont {Yamada},\ and\ \citenamefont
  {Yamazaki}}]{Kitano:2021jho}%
  \BibitemOpen
  \bibfield  {author} {\bibinfo {author} {\bibfnamefont {R.}~\bibnamefont
  {Kitano}}, \bibinfo {author} {\bibfnamefont {R.}~\bibnamefont {Matsudo}},
  \bibinfo {author} {\bibfnamefont {N.}~\bibnamefont {Yamada}},\ and\ \bibinfo
  {author} {\bibfnamefont {M.}~\bibnamefont {Yamazaki}},\ }\bibfield  {title}
  {\bibinfo {title} {{Peeking into the \ensuremath{\theta} vacuum}},\ }\href
  {https://doi.org/10.1016/j.physletb.2021.136657} {\bibfield  {journal}
  {\bibinfo  {journal} {Phys. Lett. B}\ }\textbf {\bibinfo {volume} {822}},\
  \bibinfo {pages} {136657} (\bibinfo {year} {2021})},\ \Eprint
  {https://arxiv.org/abs/2102.08784} {arXiv:2102.08784 [hep-lat]} \BibitemShut
  {NoStop}%
\bibitem [{\citenamefont {Bennett}\ \emph {et~al.}(2022)\citenamefont
  {Bennett}, \citenamefont {Hong}, \citenamefont {Lee}, \citenamefont {Lin},
  \citenamefont {Lucini}, \citenamefont {Piai},\ and\ \citenamefont
  {Vadacchino}}]{Bennett:2022gdz}%
  \BibitemOpen
  \bibfield  {author} {\bibinfo {author} {\bibfnamefont {E.}~\bibnamefont
  {Bennett}}, \bibinfo {author} {\bibfnamefont {D.~K.}\ \bibnamefont {Hong}},
  \bibinfo {author} {\bibfnamefont {J.-W.}\ \bibnamefont {Lee}}, \bibinfo
  {author} {\bibfnamefont {C.~J.~D.}\ \bibnamefont {Lin}}, \bibinfo {author}
  {\bibfnamefont {B.}~\bibnamefont {Lucini}}, \bibinfo {author} {\bibfnamefont
  {M.}~\bibnamefont {Piai}},\ and\ \bibinfo {author} {\bibfnamefont
  {D.}~\bibnamefont {Vadacchino}},\ }\bibfield  {title} {\bibinfo {title}
  {{Color dependence of the topological susceptibility in Yang-Mills
  theories}},\ }\href@noop {} {\  (\bibinfo {year} {2022})},\ \Eprint
  {https://arxiv.org/abs/2205.09254} {arXiv:2205.09254 [hep-lat]} \BibitemShut
  {NoStop}%
\bibitem [{\citenamefont {Boyd}\ \emph {et~al.}(1996)\citenamefont {Boyd},
  \citenamefont {Engels}, \citenamefont {Karsch}, \citenamefont {Laermann},
  \citenamefont {Legeland}, \citenamefont {Lutgemeier},\ and\ \citenamefont
  {Petersson}}]{Boyd:1996bx}%
  \BibitemOpen
  \bibfield  {author} {\bibinfo {author} {\bibfnamefont {G.}~\bibnamefont
  {Boyd}}, \bibinfo {author} {\bibfnamefont {J.}~\bibnamefont {Engels}},
  \bibinfo {author} {\bibfnamefont {F.}~\bibnamefont {Karsch}}, \bibinfo
  {author} {\bibfnamefont {E.}~\bibnamefont {Laermann}}, \bibinfo {author}
  {\bibfnamefont {C.}~\bibnamefont {Legeland}}, \bibinfo {author}
  {\bibfnamefont {M.}~\bibnamefont {Lutgemeier}},\ and\ \bibinfo {author}
  {\bibfnamefont {B.}~\bibnamefont {Petersson}},\ }\bibfield  {title} {\bibinfo
  {title} {{Thermodynamics of SU(3) lattice gauge theory}},\ }\href
  {https://doi.org/10.1016/0550-3213(96)00170-8} {\bibfield  {journal}
  {\bibinfo  {journal} {Nucl. Phys. B}\ }\textbf {\bibinfo {volume} {469}},\
  \bibinfo {pages} {419} (\bibinfo {year} {1996})},\ \Eprint
  {https://arxiv.org/abs/hep-lat/9602007} {arXiv:hep-lat/9602007} \BibitemShut
  {NoStop}%
\bibitem [{\citenamefont {Borsanyi}\ \emph {et~al.}(2012)\citenamefont
  {Borsanyi}, \citenamefont {Endrodi}, \citenamefont {Fodor}, \citenamefont
  {Katz},\ and\ \citenamefont {Szabo}}]{Borsanyi:2012ve}%
  \BibitemOpen
  \bibfield  {author} {\bibinfo {author} {\bibfnamefont {S.}~\bibnamefont
  {Borsanyi}}, \bibinfo {author} {\bibfnamefont {G.}~\bibnamefont {Endrodi}},
  \bibinfo {author} {\bibfnamefont {Z.}~\bibnamefont {Fodor}}, \bibinfo
  {author} {\bibfnamefont {S.~D.}\ \bibnamefont {Katz}},\ and\ \bibinfo
  {author} {\bibfnamefont {K.~K.}\ \bibnamefont {Szabo}},\ }\bibfield  {title}
  {\bibinfo {title} {{Precision SU(3) lattice thermodynamics for a large
  temperature range}},\ }\href {https://doi.org/10.1007/JHEP07(2012)056}
  {\bibfield  {journal} {\bibinfo  {journal} {JHEP}\ }\textbf {\bibinfo
  {volume} {07}},\ \bibinfo {pages} {056}},\ \Eprint
  {https://arxiv.org/abs/1204.6184} {arXiv:1204.6184 [hep-lat]} \BibitemShut
  {NoStop}%
\bibitem [{\citenamefont {Shirogane}\ \emph {et~al.}(2016)\citenamefont
  {Shirogane}, \citenamefont {Ejiri}, \citenamefont {Iwami}, \citenamefont
  {Kanaya},\ and\ \citenamefont {Kitazawa}}]{Shirogane:2016zbf}%
  \BibitemOpen
  \bibfield  {author} {\bibinfo {author} {\bibfnamefont {M.}~\bibnamefont
  {Shirogane}}, \bibinfo {author} {\bibfnamefont {S.}~\bibnamefont {Ejiri}},
  \bibinfo {author} {\bibfnamefont {R.}~\bibnamefont {Iwami}}, \bibinfo
  {author} {\bibfnamefont {K.}~\bibnamefont {Kanaya}},\ and\ \bibinfo {author}
  {\bibfnamefont {M.}~\bibnamefont {Kitazawa}},\ }\bibfield  {title} {\bibinfo
  {title} {{Latent heat at the first order phase transition point of SU(3)
  gauge theory}},\ }\href {https://doi.org/10.1103/PhysRevD.94.014506}
  {\bibfield  {journal} {\bibinfo  {journal} {Phys. Rev. D}\ }\textbf {\bibinfo
  {volume} {94}},\ \bibinfo {pages} {014506} (\bibinfo {year} {2016})},\
  \Eprint {https://arxiv.org/abs/1605.02997} {arXiv:1605.02997 [hep-lat]}
  \BibitemShut {NoStop}%
\bibitem [{\citenamefont {Caselle}\ \emph {et~al.}(2018)\citenamefont
  {Caselle}, \citenamefont {Nada},\ and\ \citenamefont
  {Panero}}]{Caselle:2018kap}%
  \BibitemOpen
  \bibfield  {author} {\bibinfo {author} {\bibfnamefont {M.}~\bibnamefont
  {Caselle}}, \bibinfo {author} {\bibfnamefont {A.}~\bibnamefont {Nada}},\ and\
  \bibinfo {author} {\bibfnamefont {M.}~\bibnamefont {Panero}},\ }\bibfield
  {title} {\bibinfo {title} {{QCD thermodynamics from lattice calculations with
  nonequilibrium methods: The SU(3) equation of state}},\ }\href
  {https://doi.org/10.1103/PhysRevD.98.054513} {\bibfield  {journal} {\bibinfo
  {journal} {Phys. Rev. D}\ }\textbf {\bibinfo {volume} {98}},\ \bibinfo
  {pages} {054513} (\bibinfo {year} {2018})},\ \Eprint
  {https://arxiv.org/abs/1801.03110} {arXiv:1801.03110 [hep-lat]} \BibitemShut
  {NoStop}%
\bibitem [{\citenamefont {Borsanyi}\ \emph {et~al.}(2022)\citenamefont
  {Borsanyi}, \citenamefont {Fodor}, \citenamefont {Godzieba}, \citenamefont
  {Kara}, \citenamefont {Parotto},\ and\ \citenamefont
  {Sexty}}]{Borsanyi:2022xml}%
  \BibitemOpen
  \bibfield  {author} {\bibinfo {author} {\bibfnamefont {S.}~\bibnamefont
  {Borsanyi}}, \bibinfo {author} {\bibfnamefont {Z.}~\bibnamefont {Fodor}},
  \bibinfo {author} {\bibfnamefont {D.~A.}\ \bibnamefont {Godzieba}}, \bibinfo
  {author} {\bibfnamefont {R.}~\bibnamefont {Kara}}, \bibinfo {author}
  {\bibfnamefont {P.}~\bibnamefont {Parotto}},\ and\ \bibinfo {author}
  {\bibfnamefont {D.}~\bibnamefont {Sexty}},\ }\bibfield  {title} {\bibinfo
  {title} {{Precision study of the continuum SU(3) Yang-Mills theory: how to
  use parallel tempering to improve on supercritical slowing down for first
  order phase transitions}},\ }\href@noop {} {\  (\bibinfo {year} {2022})},\
  \Eprint {https://arxiv.org/abs/2202.05234} {arXiv:2202.05234 [hep-lat]}
  \BibitemShut {NoStop}%
\bibitem [{Note4()}]{Note4}%
  \BibitemOpen
  \bibinfo {note} {There could be a dense liquid or crystal of instantons \cite
  {Carter:2001ih,Liu:2018znq}. However, as noted by Witten \cite
  {Witten:1978bc}, for the topological susceptibility not to be exponentially
  suppressed in $N$, it is necessary to show that the action of such a liquid
  or crystal vanishes at $\sim N$, and so instead is $\sim 1$. Consider how
  this could occur semi-classically. The effective action for a classical
  instanton of size $\rho $ is $\sim N$, and is a power series in $\lambda =
  g^2 N$: $${\protect \cal S}_{\protect \rm eff}(\rho \Lambda ) = N \protect
  \tmspace +\thickmuskip {.2777em} \protect \frac {8 \pi ^2}{\lambda } \left (
  1 + \lambda \protect \qopname \relax o{log}(\rho \Lambda ) + \protect \ldots
  \right ),$$ up to terms $\sim N^0$, where $\Lambda $ is the renormalization
  mass scale. At large $N$, one minimizes the effective action with respect to
  $\rho $, with the dominant configuration an instanton of a single scale size,
  $\rho _\infty $. This is not sufficient, though: it is necessary that the
  effective action at this scale size {\protect \it vanishes}, with ${\protect
  \cal S}_{\protect \rm eff}(\rho _\infty \Lambda ) = 0$, up to corrections
  $\sim 1$. It is not obvious how this might come about.}\BibitemShut {Stop}%
\bibitem [{\citenamefont {Witten}(1998)}]{Witten:1998uka}%
  \BibitemOpen
  \bibfield  {author} {\bibinfo {author} {\bibfnamefont {E.}~\bibnamefont
  {Witten}},\ }\bibfield  {title} {\bibinfo {title} {{Theta dependence in the
  large N limit of four-dimensional gauge theories}},\ }\href
  {https://doi.org/10.1103/PhysRevLett.81.2862} {\bibfield  {journal} {\bibinfo
   {journal} {Phys. Rev. Lett.}\ }\textbf {\bibinfo {volume} {81}},\ \bibinfo
  {pages} {2862} (\bibinfo {year} {1998})},\ \Eprint
  {https://arxiv.org/abs/hep-th/9807109} {arXiv:hep-th/9807109} \BibitemShut
  {NoStop}%
\bibitem [{\citenamefont {'t~Hooft}(1980)}]{tHooft:1980kjq}%
  \BibitemOpen
  \bibfield  {author} {\bibinfo {author} {\bibfnamefont {G.}~\bibnamefont
  {'t~Hooft}},\ }\bibfield  {title} {\bibinfo {title} {{Confinement and
  Topology in Nonabelian Gauge Theories}},\ }\href@noop {} {\bibfield
  {journal} {\bibinfo  {journal} {Acta Phys. Austriaca Suppl.}\ }\textbf
  {\bibinfo {volume} {22}},\ \bibinfo {pages} {531} (\bibinfo {year}
  {1980})}\BibitemShut {NoStop}%
\bibitem [{\citenamefont {'t~Hooft}(1981)}]{tHooft:1981nnx}%
  \BibitemOpen
  \bibfield  {author} {\bibinfo {author} {\bibfnamefont {G.}~\bibnamefont
  {'t~Hooft}},\ }\bibfield  {title} {\bibinfo {title} {{Some Twisted Selfdual
  Solutions for the Yang-Mills Equations on a Hypertorus}},\ }\href
  {https://doi.org/10.1007/BF01208900} {\bibfield  {journal} {\bibinfo
  {journal} {Commun. Math. Phys.}\ }\textbf {\bibinfo {volume} {81}},\ \bibinfo
  {pages} {267} (\bibinfo {year} {1981})}\BibitemShut {NoStop}%
\bibitem [{\citenamefont {van Baal}(1982)}]{vanBaal:1982ag}%
  \BibitemOpen
  \bibfield  {author} {\bibinfo {author} {\bibfnamefont {P.}~\bibnamefont {van
  Baal}},\ }\bibfield  {title} {\bibinfo {title} {{Some Results for SU(N) Gauge
  Fields on the Hypertorus}},\ }\href {https://doi.org/10.1007/BF01403503}
  {\bibfield  {journal} {\bibinfo  {journal} {Commun. Math. Phys.}\ }\textbf
  {\bibinfo {volume} {85}},\ \bibinfo {pages} {529} (\bibinfo {year}
  {1982})}\BibitemShut {NoStop}%
\bibitem [{\citenamefont {Sedlacek}(1982)}]{Sedlacek:1982cd}%
  \BibitemOpen
  \bibfield  {author} {\bibinfo {author} {\bibfnamefont {S.}~\bibnamefont
  {Sedlacek}},\ }\bibfield  {title} {\bibinfo {title} {{A direct method for
  minimizing the Yang-Mills functional over four manifolds}},\ }\href
  {https://doi.org/10.1007/BF01214887} {\bibfield  {journal} {\bibinfo
  {journal} {Commun. Math. Phys.}\ }\textbf {\bibinfo {volume} {86}},\ \bibinfo
  {pages} {515} (\bibinfo {year} {1982})}\BibitemShut {NoStop}%
\bibitem [{\citenamefont {Nash}(1983)}]{Nash:1982kp}%
  \BibitemOpen
  \bibfield  {author} {\bibinfo {author} {\bibfnamefont {C.}~\bibnamefont
  {Nash}},\ }\bibfield  {title} {\bibinfo {title} {{Gauge Potentials and
  Bundles Over the Four Torus}},\ }\href {https://doi.org/10.1007/BF01213211}
  {\bibfield  {journal} {\bibinfo  {journal} {Commun. Math. Phys.}\ }\textbf
  {\bibinfo {volume} {88}},\ \bibinfo {pages} {319} (\bibinfo {year}
  {1983})}\BibitemShut {NoStop}%
\bibitem [{\citenamefont {Killingback}(1984)}]{Killingback:1984en}%
  \BibitemOpen
  \bibfield  {author} {\bibinfo {author} {\bibfnamefont {T.~P.}\ \bibnamefont
  {Killingback}},\ }\bibfield  {title} {\bibinfo {title} {{The Gribov Ambiguity
  in Gauge Theories on the 4 Torus}},\ }\href
  {https://doi.org/10.1016/0370-2693(84)91878-1} {\bibfield  {journal}
  {\bibinfo  {journal} {Phys. Lett. B}\ }\textbf {\bibinfo {volume} {138}},\
  \bibinfo {pages} {87} (\bibinfo {year} {1984})}\BibitemShut {NoStop}%
\bibitem [{\citenamefont {Gonzalez-Arroyo}\ and\ \citenamefont
  {Martinez}(1996)}]{Gonzalez-Arroyo:1995ynx}%
  \BibitemOpen
  \bibfield  {author} {\bibinfo {author} {\bibfnamefont {A.}~\bibnamefont
  {Gonzalez-Arroyo}}\ and\ \bibinfo {author} {\bibfnamefont {P.}~\bibnamefont
  {Martinez}},\ }\bibfield  {title} {\bibinfo {title} {{Investigating
  Yang-Mills theory and confinement as a function of the spatial volume}},\
  }\href {https://doi.org/10.1016/0550-3213(95)00601-X} {\bibfield  {journal}
  {\bibinfo  {journal} {Nucl. Phys. B}\ }\textbf {\bibinfo {volume} {459}},\
  \bibinfo {pages} {337} (\bibinfo {year} {1996})},\ \Eprint
  {https://arxiv.org/abs/hep-lat/9507001} {arXiv:hep-lat/9507001} \BibitemShut
  {NoStop}%
\bibitem [{\citenamefont {Gonzalez-Arroyo}\ \emph {et~al.}(1995)\citenamefont
  {Gonzalez-Arroyo}, \citenamefont {Martinez},\ and\ \citenamefont
  {Montero}}]{Gonzalez-Arroyo:1995isl}%
  \BibitemOpen
  \bibfield  {author} {\bibinfo {author} {\bibfnamefont {A.}~\bibnamefont
  {Gonzalez-Arroyo}}, \bibinfo {author} {\bibfnamefont {P.}~\bibnamefont
  {Martinez}},\ and\ \bibinfo {author} {\bibfnamefont {A.}~\bibnamefont
  {Montero}},\ }\bibfield  {title} {\bibinfo {title} {{Gauge invariant
  structures and confinement}},\ }\href
  {https://doi.org/10.1016/0370-2693(95)01056-V} {\bibfield  {journal}
  {\bibinfo  {journal} {Phys. Lett. B}\ }\textbf {\bibinfo {volume} {359}},\
  \bibinfo {pages} {159} (\bibinfo {year} {1995})},\ \Eprint
  {https://arxiv.org/abs/hep-lat/9507006} {arXiv:hep-lat/9507006} \BibitemShut
  {NoStop}%
\bibitem [{\citenamefont {Gonzalez-Arroyo}\ and\ \citenamefont
  {Montero}(1996)}]{Gonzalez-Arroyo:1996eos}%
  \BibitemOpen
  \bibfield  {author} {\bibinfo {author} {\bibfnamefont {A.}~\bibnamefont
  {Gonzalez-Arroyo}}\ and\ \bibinfo {author} {\bibfnamefont {A.}~\bibnamefont
  {Montero}},\ }\bibfield  {title} {\bibinfo {title} {{Do classical
  configurations produce confinement?}},\ }\href
  {https://doi.org/10.1016/0370-2693(96)01107-0} {\bibfield  {journal}
  {\bibinfo  {journal} {Phys. Lett. B}\ }\textbf {\bibinfo {volume} {387}},\
  \bibinfo {pages} {823} (\bibinfo {year} {1996})},\ \Eprint
  {https://arxiv.org/abs/hep-th/9604017} {arXiv:hep-th/9604017} \BibitemShut
  {NoStop}%
\bibitem [{\citenamefont {Garcia~Perez}\ \emph {et~al.}(2000)\citenamefont
  {Garcia~Perez}, \citenamefont {Gonzalez-Arroyo},\ and\ \citenamefont
  {Pena}}]{GarciaPerez:2000aiw}%
  \BibitemOpen
  \bibfield  {author} {\bibinfo {author} {\bibfnamefont {M.}~\bibnamefont
  {Garcia~Perez}}, \bibinfo {author} {\bibfnamefont {A.}~\bibnamefont
  {Gonzalez-Arroyo}},\ and\ \bibinfo {author} {\bibfnamefont {C.}~\bibnamefont
  {Pena}},\ }\bibfield  {title} {\bibinfo {title} {{Perturbative construction
  of selfdual configurations on the torus}},\ }\href
  {https://doi.org/10.1088/1126-6708/2000/09/033} {\bibfield  {journal}
  {\bibinfo  {journal} {JHEP}\ }\textbf {\bibinfo {volume} {09}},\ \bibinfo
  {pages} {033}},\ \Eprint {https://arxiv.org/abs/hep-th/0007113}
  {arXiv:hep-th/0007113} \BibitemShut {NoStop}%
\bibitem [{\citenamefont {Gonz\'alez-Arroyo}(2020)}]{Gonzalez-Arroyo:2019wpu}%
  \BibitemOpen
  \bibfield  {author} {\bibinfo {author} {\bibfnamefont {A.}~\bibnamefont
  {Gonz\'alez-Arroyo}},\ }\bibfield  {title} {\bibinfo {title} {{Constructing
  SU(N) fractional instantons}},\ }\href
  {https://doi.org/10.1007/JHEP02(2020)137} {\bibfield  {journal} {\bibinfo
  {journal} {JHEP}\ }\textbf {\bibinfo {volume} {02}},\ \bibinfo {pages}
  {137}},\ \Eprint {https://arxiv.org/abs/1910.12565} {arXiv:1910.12565
  [hep-th]} \BibitemShut {NoStop}%
\bibitem [{\citenamefont {Dasilva~Gol\'an}\ and\ \citenamefont
  {Garc\'\i{}a~P\'erez}(2022)}]{DasilvaGolan:2022jlm}%
  \BibitemOpen
  \bibfield  {author} {\bibinfo {author} {\bibfnamefont {J.}~\bibnamefont
  {Dasilva~Gol\'an}}\ and\ \bibinfo {author} {\bibfnamefont {M.}~\bibnamefont
  {Garc\'\i{}a~P\'erez}},\ }\bibfield  {title} {\bibinfo {title} {{SU(N)
  fractional instantons and the Fibonacci sequence}},\ }\href
  {https://doi.org/10.1007/JHEP12(2022)109} {\bibfield  {journal} {\bibinfo
  {journal} {JHEP}\ }\textbf {\bibinfo {volume} {12}},\ \bibinfo {pages}
  {109}},\ \Eprint {https://arxiv.org/abs/2208.07133} {arXiv:2208.07133
  [hep-th]} \BibitemShut {NoStop}%
\bibitem [{\citenamefont {Gonzalez-Arroyo}(2023)}]{Gonzalez-Arroyo:2023kqv}%
  \BibitemOpen
  \bibfield  {author} {\bibinfo {author} {\bibfnamefont {A.}~\bibnamefont
  {Gonzalez-Arroyo}},\ }\bibfield  {title} {\bibinfo {title} {{On the
  fractional instanton liquid picture of the Yang-Mills vacuum and
  Confinement}},\ }\href@noop {} {\  (\bibinfo {year} {2023})},\ \Eprint
  {https://arxiv.org/abs/2302.12356} {arXiv:2302.12356 [hep-th]} \BibitemShut
  {NoStop}%
\bibitem [{\citenamefont {Lee}\ and\ \citenamefont {Yi}(1997)}]{Lee:1997vp}%
  \BibitemOpen
  \bibfield  {author} {\bibinfo {author} {\bibfnamefont {K.-M.}\ \bibnamefont
  {Lee}}\ and\ \bibinfo {author} {\bibfnamefont {P.}~\bibnamefont {Yi}},\
  }\bibfield  {title} {\bibinfo {title} {{Monopoles and instantons on partially
  compactified D-branes}},\ }\href {https://doi.org/10.1103/PhysRevD.56.3711}
  {\bibfield  {journal} {\bibinfo  {journal} {Phys. Rev. D}\ }\textbf {\bibinfo
  {volume} {56}},\ \bibinfo {pages} {3711} (\bibinfo {year} {1997})},\ \Eprint
  {https://arxiv.org/abs/hep-th/9702107} {arXiv:hep-th/9702107} \BibitemShut
  {NoStop}%
\bibitem [{\citenamefont {Lee}(1998)}]{Lee:1998vu}%
  \BibitemOpen
  \bibfield  {author} {\bibinfo {author} {\bibfnamefont {K.-M.}\ \bibnamefont
  {Lee}},\ }\bibfield  {title} {\bibinfo {title} {{Instantons and magnetic
  monopoles on R**3 x S**1 with arbitrary simple gauge groups}},\ }\href
  {https://doi.org/10.1016/S0370-2693(98)00283-4} {\bibfield  {journal}
  {\bibinfo  {journal} {Phys. Lett. B}\ }\textbf {\bibinfo {volume} {426}},\
  \bibinfo {pages} {323} (\bibinfo {year} {1998})},\ \Eprint
  {https://arxiv.org/abs/hep-th/9802012} {arXiv:hep-th/9802012} \BibitemShut
  {NoStop}%
\bibitem [{\citenamefont {Lee}\ and\ \citenamefont {Lu}(1998)}]{Lee:1998bb}%
  \BibitemOpen
  \bibfield  {author} {\bibinfo {author} {\bibfnamefont {K.-M.}\ \bibnamefont
  {Lee}}\ and\ \bibinfo {author} {\bibfnamefont {C.-h.}\ \bibnamefont {Lu}},\
  }\bibfield  {title} {\bibinfo {title} {{SU(2) calorons and magnetic
  monopoles}},\ }\href {https://doi.org/10.1103/PhysRevD.58.025011} {\bibfield
  {journal} {\bibinfo  {journal} {Phys. Rev. D}\ }\textbf {\bibinfo {volume}
  {58}},\ \bibinfo {pages} {025011} (\bibinfo {year} {1998})},\ \Eprint
  {https://arxiv.org/abs/hep-th/9802108} {arXiv:hep-th/9802108} \BibitemShut
  {NoStop}%
\bibitem [{\citenamefont {Kraan}\ and\ \citenamefont {van
  Baal}(1998{\natexlab{a}})}]{Kraan:1998pm}%
  \BibitemOpen
  \bibfield  {author} {\bibinfo {author} {\bibfnamefont {T.~C.}\ \bibnamefont
  {Kraan}}\ and\ \bibinfo {author} {\bibfnamefont {P.}~\bibnamefont {van
  Baal}},\ }\bibfield  {title} {\bibinfo {title} {{Periodic instantons with
  nontrivial holonomy}},\ }\href
  {https://doi.org/10.1016/S0550-3213(98)00590-2} {\bibfield  {journal}
  {\bibinfo  {journal} {Nucl. Phys. B}\ }\textbf {\bibinfo {volume} {533}},\
  \bibinfo {pages} {627} (\bibinfo {year} {1998}{\natexlab{a}})},\ \Eprint
  {https://arxiv.org/abs/hep-th/9805168} {arXiv:hep-th/9805168} \BibitemShut
  {NoStop}%
\bibitem [{\citenamefont {Kraan}\ and\ \citenamefont {van
  Baal}(1998{\natexlab{b}})}]{Kraan:1998sn}%
  \BibitemOpen
  \bibfield  {author} {\bibinfo {author} {\bibfnamefont {T.~C.}\ \bibnamefont
  {Kraan}}\ and\ \bibinfo {author} {\bibfnamefont {P.}~\bibnamefont {van
  Baal}},\ }\bibfield  {title} {\bibinfo {title} {{Monopole constituents inside
  SU(n) calorons}},\ }\href {https://doi.org/10.1016/S0370-2693(98)00799-0}
  {\bibfield  {journal} {\bibinfo  {journal} {Phys. Lett. B}\ }\textbf
  {\bibinfo {volume} {435}},\ \bibinfo {pages} {389} (\bibinfo {year}
  {1998}{\natexlab{b}})},\ \Eprint {https://arxiv.org/abs/hep-th/9806034}
  {arXiv:hep-th/9806034} \BibitemShut {NoStop}%
\bibitem [{\citenamefont {Garcia~Perez}\ \emph {et~al.}(1999)\citenamefont
  {Garcia~Perez}, \citenamefont {Gonzalez-Arroyo}, \citenamefont {Montero},\
  and\ \citenamefont {van Baal}}]{GarciaPerez:1999hs}%
  \BibitemOpen
  \bibfield  {author} {\bibinfo {author} {\bibfnamefont {M.}~\bibnamefont
  {Garcia~Perez}}, \bibinfo {author} {\bibfnamefont {A.}~\bibnamefont
  {Gonzalez-Arroyo}}, \bibinfo {author} {\bibfnamefont {A.}~\bibnamefont
  {Montero}},\ and\ \bibinfo {author} {\bibfnamefont {P.}~\bibnamefont {van
  Baal}},\ }\bibfield  {title} {\bibinfo {title} {{Calorons on the lattice: A
  New perspective}},\ }\href {https://doi.org/10.1088/1126-6708/1999/06/001}
  {\bibfield  {journal} {\bibinfo  {journal} {JHEP}\ }\textbf {\bibinfo
  {volume} {06}},\ \bibinfo {pages} {001}},\ \Eprint
  {https://arxiv.org/abs/hep-lat/9903022} {arXiv:hep-lat/9903022} \BibitemShut
  {NoStop}%
\bibitem [{\citenamefont {Diakonov}(2003)}]{Diakonov:2002fq}%
  \BibitemOpen
  \bibfield  {author} {\bibinfo {author} {\bibfnamefont {D.}~\bibnamefont
  {Diakonov}},\ }\bibfield  {title} {\bibinfo {title} {{Instantons at work}},\
  }\href {https://doi.org/10.1016/S0146-6410(03)90014-7} {\bibfield  {journal}
  {\bibinfo  {journal} {Prog. Part. Nucl. Phys.}\ }\textbf {\bibinfo {volume}
  {51}},\ \bibinfo {pages} {173} (\bibinfo {year} {2003})},\ \Eprint
  {https://arxiv.org/abs/hep-ph/0212026} {arXiv:hep-ph/0212026} \BibitemShut
  {NoStop}%
\bibitem [{\citenamefont {Eto}\ \emph {et~al.}(2005)\citenamefont {Eto},
  \citenamefont {Isozumi}, \citenamefont {Nitta}, \citenamefont {Ohashi},\ and\
  \citenamefont {Sakai}}]{Eto:2004rz}%
  \BibitemOpen
  \bibfield  {author} {\bibinfo {author} {\bibfnamefont {M.}~\bibnamefont
  {Eto}}, \bibinfo {author} {\bibfnamefont {Y.}~\bibnamefont {Isozumi}},
  \bibinfo {author} {\bibfnamefont {M.}~\bibnamefont {Nitta}}, \bibinfo
  {author} {\bibfnamefont {K.}~\bibnamefont {Ohashi}},\ and\ \bibinfo {author}
  {\bibfnamefont {N.}~\bibnamefont {Sakai}},\ }\bibfield  {title} {\bibinfo
  {title} {{Instantons in the Higgs phase}},\ }\href
  {https://doi.org/10.1103/PhysRevD.72.025011} {\bibfield  {journal} {\bibinfo
  {journal} {Phys. Rev. D}\ }\textbf {\bibinfo {volume} {72}},\ \bibinfo
  {pages} {025011} (\bibinfo {year} {2005})},\ \Eprint
  {https://arxiv.org/abs/hep-th/0412048} {arXiv:hep-th/0412048} \BibitemShut
  {NoStop}%
\bibitem [{\citenamefont {Eto}\ \emph {et~al.}(2006{\natexlab{a}})\citenamefont
  {Eto}, \citenamefont {Isozumi}, \citenamefont {Nitta}, \citenamefont
  {Ohashi},\ and\ \citenamefont {Sakai}}]{Eto:2006pg}%
  \BibitemOpen
  \bibfield  {author} {\bibinfo {author} {\bibfnamefont {M.}~\bibnamefont
  {Eto}}, \bibinfo {author} {\bibfnamefont {Y.}~\bibnamefont {Isozumi}},
  \bibinfo {author} {\bibfnamefont {M.}~\bibnamefont {Nitta}}, \bibinfo
  {author} {\bibfnamefont {K.}~\bibnamefont {Ohashi}},\ and\ \bibinfo {author}
  {\bibfnamefont {N.}~\bibnamefont {Sakai}},\ }\bibfield  {title} {\bibinfo
  {title} {{Solitons in the Higgs phase: The Moduli matrix approach}},\ }\href
  {https://doi.org/10.1088/0305-4470/39/26/R01} {\bibfield  {journal} {\bibinfo
   {journal} {J. Phys. A}\ }\textbf {\bibinfo {volume} {39}},\ \bibinfo {pages}
  {R315} (\bibinfo {year} {2006}{\natexlab{a}})},\ \Eprint
  {https://arxiv.org/abs/hep-th/0602170} {arXiv:hep-th/0602170} \BibitemShut
  {NoStop}%
\bibitem [{\citenamefont {Eto}\ \emph {et~al.}(2006{\natexlab{b}})\citenamefont
  {Eto}, \citenamefont {Fujimori}, \citenamefont {Isozumi}, \citenamefont
  {Nitta}, \citenamefont {Ohashi}, \citenamefont {Ohta},\ and\ \citenamefont
  {Sakai}}]{Eto:2006mz}%
  \BibitemOpen
  \bibfield  {author} {\bibinfo {author} {\bibfnamefont {M.}~\bibnamefont
  {Eto}}, \bibinfo {author} {\bibfnamefont {T.}~\bibnamefont {Fujimori}},
  \bibinfo {author} {\bibfnamefont {Y.}~\bibnamefont {Isozumi}}, \bibinfo
  {author} {\bibfnamefont {M.}~\bibnamefont {Nitta}}, \bibinfo {author}
  {\bibfnamefont {K.}~\bibnamefont {Ohashi}}, \bibinfo {author} {\bibfnamefont
  {K.}~\bibnamefont {Ohta}},\ and\ \bibinfo {author} {\bibfnamefont
  {N.}~\bibnamefont {Sakai}},\ }\bibfield  {title} {\bibinfo {title}
  {{Non-Abelian vortices on cylinder: Duality between vortices and walls}},\
  }\href {https://doi.org/10.1103/PhysRevD.73.085008} {\bibfield  {journal}
  {\bibinfo  {journal} {Phys. Rev. D}\ }\textbf {\bibinfo {volume} {73}},\
  \bibinfo {pages} {085008} (\bibinfo {year} {2006}{\natexlab{b}})},\ \Eprint
  {https://arxiv.org/abs/hep-th/0601181} {arXiv:hep-th/0601181} \BibitemShut
  {NoStop}%
\bibitem [{\citenamefont {Bruckmann}\ \emph {et~al.}(2009)\citenamefont
  {Bruckmann}, \citenamefont {Dinter}, \citenamefont {Ilgenfritz},
  \citenamefont {Muller-Preussker},\ and\ \citenamefont
  {Wagner}}]{Bruckmann:2009nw}%
  \BibitemOpen
  \bibfield  {author} {\bibinfo {author} {\bibfnamefont {F.}~\bibnamefont
  {Bruckmann}}, \bibinfo {author} {\bibfnamefont {S.}~\bibnamefont {Dinter}},
  \bibinfo {author} {\bibfnamefont {E.-M.}\ \bibnamefont {Ilgenfritz}},
  \bibinfo {author} {\bibfnamefont {M.}~\bibnamefont {Muller-Preussker}},\ and\
  \bibinfo {author} {\bibfnamefont {M.}~\bibnamefont {Wagner}},\ }\bibfield
  {title} {\bibinfo {title} {{Cautionary remarks on the moduli space metric for
  multi-dyon simulations}},\ }\href
  {https://doi.org/10.1103/PhysRevD.79.116007} {\bibfield  {journal} {\bibinfo
  {journal} {Phys. Rev. D}\ }\textbf {\bibinfo {volume} {79}},\ \bibinfo
  {pages} {116007} (\bibinfo {year} {2009})},\ \Eprint
  {https://arxiv.org/abs/0903.3075} {arXiv:0903.3075 [hep-ph]} \BibitemShut
  {NoStop}%
\bibitem [{\citenamefont {Diakonov}(2009)}]{Diakonov:2009jq}%
  \BibitemOpen
  \bibfield  {author} {\bibinfo {author} {\bibfnamefont {D.}~\bibnamefont
  {Diakonov}},\ }\bibfield  {title} {\bibinfo {title} {{Topology and
  confinement}},\ }\href {https://doi.org/10.1016/j.nuclphysbps.2009.10.010}
  {\bibfield  {journal} {\bibinfo  {journal} {Nucl. Phys. B Proc. Suppl.}\
  }\textbf {\bibinfo {volume} {195}},\ \bibinfo {pages} {5} (\bibinfo {year}
  {2009})},\ \Eprint {https://arxiv.org/abs/0906.2456} {arXiv:0906.2456
  [hep-ph]} \BibitemShut {NoStop}%
\bibitem [{\citenamefont {Poppitz}\ and\ \citenamefont
  {Unsal}(2009)}]{Poppitz:2008hr}%
  \BibitemOpen
  \bibfield  {author} {\bibinfo {author} {\bibfnamefont {E.}~\bibnamefont
  {Poppitz}}\ and\ \bibinfo {author} {\bibfnamefont {M.}~\bibnamefont
  {Unsal}},\ }\bibfield  {title} {\bibinfo {title} {{Index theorem for
  topological excitations on R**3 x S**1 and Chern-Simons theory}},\ }\href
  {https://doi.org/10.1088/1126-6708/2009/03/027} {\bibfield  {journal}
  {\bibinfo  {journal} {JHEP}\ }\textbf {\bibinfo {volume} {03}},\ \bibinfo
  {pages} {027}},\ \Eprint {https://arxiv.org/abs/0812.2085} {arXiv:0812.2085
  [hep-th]} \BibitemShut {NoStop}%
\bibitem [{\citenamefont {Anber}\ and\ \citenamefont
  {Poppitz}(2021)}]{Anber:2021upc}%
  \BibitemOpen
  \bibfield  {author} {\bibinfo {author} {\bibfnamefont {M.~M.}\ \bibnamefont
  {Anber}}\ and\ \bibinfo {author} {\bibfnamefont {E.}~\bibnamefont
  {Poppitz}},\ }\bibfield  {title} {\bibinfo {title} {{Nonperturbative effects
  in the Standard Model with gauged 1-form symmetry}},\ }\href
  {https://doi.org/10.1007/JHEP12(2021)055} {\bibfield  {journal} {\bibinfo
  {journal} {JHEP}\ }\textbf {\bibinfo {volume} {12}},\ \bibinfo {pages}
  {055}},\ \Eprint {https://arxiv.org/abs/2110.02981} {arXiv:2110.02981
  [hep-th]} \BibitemShut {NoStop}%
\bibitem [{\citenamefont {Polyakov}(1977)}]{Polyakov:1976fu}%
  \BibitemOpen
  \bibfield  {author} {\bibinfo {author} {\bibfnamefont {A.~M.}\ \bibnamefont
  {Polyakov}},\ }\bibfield  {title} {\bibinfo {title} {{Quark Confinement and
  Topology of Gauge Groups}},\ }\href
  {https://doi.org/10.1016/0550-3213(77)90086-4} {\bibfield  {journal}
  {\bibinfo  {journal} {Nucl. Phys. B}\ }\textbf {\bibinfo {volume} {120}},\
  \bibinfo {pages} {429} (\bibinfo {year} {1977})}\BibitemShut {NoStop}%
\bibitem [{\citenamefont {Dunne}\ \emph {et~al.}(2001)\citenamefont {Dunne},
  \citenamefont {Kogan}, \citenamefont {Kovner},\ and\ \citenamefont
  {Tekin}}]{Dunne:2000vp}%
  \BibitemOpen
  \bibfield  {author} {\bibinfo {author} {\bibfnamefont {G.~V.}\ \bibnamefont
  {Dunne}}, \bibinfo {author} {\bibfnamefont {I.~I.}\ \bibnamefont {Kogan}},
  \bibinfo {author} {\bibfnamefont {A.}~\bibnamefont {Kovner}},\ and\ \bibinfo
  {author} {\bibfnamefont {B.}~\bibnamefont {Tekin}},\ }\bibfield  {title}
  {\bibinfo {title} {{Deconfining phase transition in (2+1)-dimensions: The
  Georgi-Glashow model}},\ }\href
  {https://doi.org/10.1088/1126-6708/2001/01/032} {\bibfield  {journal}
  {\bibinfo  {journal} {JHEP}\ }\textbf {\bibinfo {volume} {01}},\ \bibinfo
  {pages} {032}},\ \Eprint {https://arxiv.org/abs/hep-th/0010201}
  {arXiv:hep-th/0010201} \BibitemShut {NoStop}%
\bibitem [{\citenamefont {Kogan}\ and\ \citenamefont
  {Kovner}(2002)}]{Kogan:2002au}%
  \BibitemOpen
  \bibfield  {author} {\bibinfo {author} {\bibfnamefont {I.~I.}\ \bibnamefont
  {Kogan}}\ and\ \bibinfo {author} {\bibfnamefont {A.}~\bibnamefont {Kovner}},\
  }\bibfield  {title} {\bibinfo {title} {{Monopoles, vortices and strings:
  Confinement and deconfinement in (2+1)-dimensions at weak coupling}}\ }\href
  {https://doi.org/10.1142/97898127772700003} {10.1142/97898127772700003}
  (\bibinfo {year} {2002}),\ \Eprint {https://arxiv.org/abs/hep-th/0205026}
  {arXiv:hep-th/0205026} \BibitemShut {NoStop}%
\bibitem [{\citenamefont {Kovchegov}\ and\ \citenamefont
  {Son}(2003)}]{Kovchegov:2002vi}%
  \BibitemOpen
  \bibfield  {author} {\bibinfo {author} {\bibfnamefont {Y.~V.}\ \bibnamefont
  {Kovchegov}}\ and\ \bibinfo {author} {\bibfnamefont {D.~T.}\ \bibnamefont
  {Son}},\ }\bibfield  {title} {\bibinfo {title} {{Critical temperature of the
  deconfining phase transition in (2+1)-d Georgi-Glashow model}},\ }\href
  {https://doi.org/10.1088/1126-6708/2003/01/050} {\bibfield  {journal}
  {\bibinfo  {journal} {JHEP}\ }\textbf {\bibinfo {volume} {01}},\ \bibinfo
  {pages} {050}},\ \Eprint {https://arxiv.org/abs/hep-th/0212230}
  {arXiv:hep-th/0212230} \BibitemShut {NoStop}%
\bibitem [{\citenamefont {Unsal}\ and\ \citenamefont
  {Yaffe}(2006)}]{Unsal:2006pj}%
  \BibitemOpen
  \bibfield  {author} {\bibinfo {author} {\bibfnamefont {M.}~\bibnamefont
  {Unsal}}\ and\ \bibinfo {author} {\bibfnamefont {L.~G.}\ \bibnamefont
  {Yaffe}},\ }\bibfield  {title} {\bibinfo {title} {{(In)validity of large N
  orientifold equivalence}},\ }\href
  {https://doi.org/10.1103/PhysRevD.74.105019} {\bibfield  {journal} {\bibinfo
  {journal} {Phys. Rev. D}\ }\textbf {\bibinfo {volume} {74}},\ \bibinfo
  {pages} {105019} (\bibinfo {year} {2006})},\ \Eprint
  {https://arxiv.org/abs/hep-th/0608180} {arXiv:hep-th/0608180} \BibitemShut
  {NoStop}%
\bibitem [{\citenamefont {Unsal}(2008)}]{Unsal:2007vu}%
  \BibitemOpen
  \bibfield  {author} {\bibinfo {author} {\bibfnamefont {M.}~\bibnamefont
  {Unsal}},\ }\bibfield  {title} {\bibinfo {title} {{Abelian duality,
  confinement, and chiral symmetry breaking in QCD(adj)}},\ }\href
  {https://doi.org/10.1103/PhysRevLett.100.032005} {\bibfield  {journal}
  {\bibinfo  {journal} {Phys. Rev. Lett.}\ }\textbf {\bibinfo {volume} {100}},\
  \bibinfo {pages} {032005} (\bibinfo {year} {2008})},\ \Eprint
  {https://arxiv.org/abs/0708.1772} {arXiv:0708.1772 [hep-th]} \BibitemShut
  {NoStop}%
\bibitem [{\citenamefont {Unsal}(2009)}]{Unsal:2007jx}%
  \BibitemOpen
  \bibfield  {author} {\bibinfo {author} {\bibfnamefont {M.}~\bibnamefont
  {Unsal}},\ }\bibfield  {title} {\bibinfo {title} {{Magnetic bion
  condensation: A New mechanism of confinement and mass gap in four
  dimensions}},\ }\href {https://doi.org/10.1103/PhysRevD.80.065001} {\bibfield
   {journal} {\bibinfo  {journal} {Phys. Rev. D}\ }\textbf {\bibinfo {volume}
  {80}},\ \bibinfo {pages} {065001} (\bibinfo {year} {2009})},\ \Eprint
  {https://arxiv.org/abs/0709.3269} {arXiv:0709.3269 [hep-th]} \BibitemShut
  {NoStop}%
\bibitem [{\citenamefont {Unsal}\ and\ \citenamefont
  {Yaffe}(2008)}]{Unsal:2008ch}%
  \BibitemOpen
  \bibfield  {author} {\bibinfo {author} {\bibfnamefont {M.}~\bibnamefont
  {Unsal}}\ and\ \bibinfo {author} {\bibfnamefont {L.~G.}\ \bibnamefont
  {Yaffe}},\ }\bibfield  {title} {\bibinfo {title} {{Center-stabilized
  Yang-Mills theory: Confinement and large N volume independence}},\ }\href
  {https://doi.org/10.1103/PhysRevD.78.065035} {\bibfield  {journal} {\bibinfo
  {journal} {Phys. Rev. D}\ }\textbf {\bibinfo {volume} {78}},\ \bibinfo
  {pages} {065035} (\bibinfo {year} {2008})},\ \Eprint
  {https://arxiv.org/abs/0803.0344} {arXiv:0803.0344 [hep-th]} \BibitemShut
  {NoStop}%
\bibitem [{\citenamefont {Shifman}\ and\ \citenamefont
  {Unsal}(2008)}]{Shifman:2008ja}%
  \BibitemOpen
  \bibfield  {author} {\bibinfo {author} {\bibfnamefont {M.}~\bibnamefont
  {Shifman}}\ and\ \bibinfo {author} {\bibfnamefont {M.}~\bibnamefont
  {Unsal}},\ }\bibfield  {title} {\bibinfo {title} {{QCD-like Theories on R(3)
  x S(1): A Smooth Journey from Small to Large r(S(1)) with Double-Trace
  Deformations}},\ }\href {https://doi.org/10.1103/PhysRevD.78.065004}
  {\bibfield  {journal} {\bibinfo  {journal} {Phys. Rev. D}\ }\textbf {\bibinfo
  {volume} {78}},\ \bibinfo {pages} {065004} (\bibinfo {year} {2008})},\
  \Eprint {https://arxiv.org/abs/0802.1232} {arXiv:0802.1232 [hep-th]}
  \BibitemShut {NoStop}%
\bibitem [{\citenamefont {Simic}\ and\ \citenamefont
  {Unsal}(2012)}]{Simic:2010sv}%
  \BibitemOpen
  \bibfield  {author} {\bibinfo {author} {\bibfnamefont {D.}~\bibnamefont
  {Simic}}\ and\ \bibinfo {author} {\bibfnamefont {M.}~\bibnamefont {Unsal}},\
  }\bibfield  {title} {\bibinfo {title} {{Deconfinement in Yang-Mills theory
  through toroidal compactification with deformation}},\ }\href
  {https://doi.org/10.1103/PhysRevD.85.105027} {\bibfield  {journal} {\bibinfo
  {journal} {Phys. Rev. D}\ }\textbf {\bibinfo {volume} {85}},\ \bibinfo
  {pages} {105027} (\bibinfo {year} {2012})},\ \Eprint
  {https://arxiv.org/abs/1010.5515} {arXiv:1010.5515 [hep-th]} \BibitemShut
  {NoStop}%
\bibitem [{\citenamefont {Anber}\ \emph {et~al.}(2012)\citenamefont {Anber},
  \citenamefont {Poppitz},\ and\ \citenamefont {Unsal}}]{Anber:2011gn}%
  \BibitemOpen
  \bibfield  {author} {\bibinfo {author} {\bibfnamefont {M.~M.}\ \bibnamefont
  {Anber}}, \bibinfo {author} {\bibfnamefont {E.}~\bibnamefont {Poppitz}},\
  and\ \bibinfo {author} {\bibfnamefont {M.}~\bibnamefont {Unsal}},\ }\bibfield
   {title} {\bibinfo {title} {{2d affine XY-spin model/4d gauge theory duality
  and deconfinement}},\ }\href {https://doi.org/10.1007/JHEP04(2012)040}
  {\bibfield  {journal} {\bibinfo  {journal} {JHEP}\ }\textbf {\bibinfo
  {volume} {04}},\ \bibinfo {pages} {040}},\ \Eprint
  {https://arxiv.org/abs/1112.6389} {arXiv:1112.6389 [hep-th]} \BibitemShut
  {NoStop}%
\bibitem [{\citenamefont {Poppitz}\ \emph {et~al.}(2012)\citenamefont
  {Poppitz}, \citenamefont {Sch\"afer},\ and\ \citenamefont
  {Unsal}}]{Poppitz:2012sw}%
  \BibitemOpen
  \bibfield  {author} {\bibinfo {author} {\bibfnamefont {E.}~\bibnamefont
  {Poppitz}}, \bibinfo {author} {\bibfnamefont {T.}~\bibnamefont {Sch\"afer}},\
  and\ \bibinfo {author} {\bibfnamefont {M.}~\bibnamefont {Unsal}},\ }\bibfield
   {title} {\bibinfo {title} {{Continuity, Deconfinement, and (Super)
  Yang-Mills Theory}},\ }\href {https://doi.org/10.1007/JHEP10(2012)115}
  {\bibfield  {journal} {\bibinfo  {journal} {JHEP}\ }\textbf {\bibinfo
  {volume} {10}},\ \bibinfo {pages} {115}},\ \Eprint
  {https://arxiv.org/abs/1205.0290} {arXiv:1205.0290 [hep-th]} \BibitemShut
  {NoStop}%
\bibitem [{\citenamefont {Anber}\ \emph {et~al.}(2013)\citenamefont {Anber},
  \citenamefont {Collier}, \citenamefont {Poppitz}, \citenamefont
  {Strimas-Mackey},\ and\ \citenamefont {Teeple}}]{Anber:2013doa}%
  \BibitemOpen
  \bibfield  {author} {\bibinfo {author} {\bibfnamefont {M.~M.}\ \bibnamefont
  {Anber}}, \bibinfo {author} {\bibfnamefont {S.}~\bibnamefont {Collier}},
  \bibinfo {author} {\bibfnamefont {E.}~\bibnamefont {Poppitz}}, \bibinfo
  {author} {\bibfnamefont {S.}~\bibnamefont {Strimas-Mackey}},\ and\ \bibinfo
  {author} {\bibfnamefont {B.}~\bibnamefont {Teeple}},\ }\bibfield  {title}
  {\bibinfo {title} {{Deconfinement in $\mathcal{N}=1$ super Yang-Mills theory
  on $\mathbb{R}^3 \times \mathbb{S}^1$ via dual-Coulomb gas and ''affine''
  XY-model}},\ }\href {https://doi.org/10.1007/JHEP11(2013)142} {\bibfield
  {journal} {\bibinfo  {journal} {JHEP}\ }\textbf {\bibinfo {volume} {11}},\
  \bibinfo {pages} {142}},\ \Eprint {https://arxiv.org/abs/1310.3522}
  {arXiv:1310.3522 [hep-th]} \BibitemShut {NoStop}%
\bibitem [{\citenamefont {Aitken}\ \emph {et~al.}(2017)\citenamefont {Aitken},
  \citenamefont {Cherman}, \citenamefont {Poppitz},\ and\ \citenamefont
  {Yaffe}}]{Aitken:2017ayq}%
  \BibitemOpen
  \bibfield  {author} {\bibinfo {author} {\bibfnamefont {K.}~\bibnamefont
  {Aitken}}, \bibinfo {author} {\bibfnamefont {A.}~\bibnamefont {Cherman}},
  \bibinfo {author} {\bibfnamefont {E.}~\bibnamefont {Poppitz}},\ and\ \bibinfo
  {author} {\bibfnamefont {L.~G.}\ \bibnamefont {Yaffe}},\ }\bibfield  {title}
  {\bibinfo {title} {{QCD on a small circle}},\ }\href
  {https://doi.org/10.1103/PhysRevD.96.096022} {\bibfield  {journal} {\bibinfo
  {journal} {Phys. Rev. D}\ }\textbf {\bibinfo {volume} {96}},\ \bibinfo
  {pages} {096022} (\bibinfo {year} {2017})},\ \Eprint
  {https://arxiv.org/abs/1707.08971} {arXiv:1707.08971 [hep-th]} \BibitemShut
  {NoStop}%
\bibitem [{\citenamefont {\"Unsal}(2021)}]{Unsal:2020yeh}%
  \BibitemOpen
  \bibfield  {author} {\bibinfo {author} {\bibfnamefont {M.}~\bibnamefont
  {\"Unsal}},\ }\bibfield  {title} {\bibinfo {title} {{Strongly coupled QFT
  dynamics via TQFT coupling}},\ }\href
  {https://doi.org/10.1007/JHEP11(2021)134} {\bibfield  {journal} {\bibinfo
  {journal} {JHEP}\ }\textbf {\bibinfo {volume} {11}},\ \bibinfo {pages}
  {134}},\ \Eprint {https://arxiv.org/abs/2007.03880} {arXiv:2007.03880
  [hep-th]} \BibitemShut {NoStop}%
\bibitem [{\citenamefont {Poppitz}(2022)}]{sym14010180}%
  \BibitemOpen
  \bibfield  {author} {\bibinfo {author} {\bibfnamefont {E.}~\bibnamefont
  {Poppitz}},\ }\bibfield  {title} {\bibinfo {title} {{Notes on Confinement on
  R3 x S1: From Yang-Mills, Super-Yang-Mills, and QCD(adj) to QCD(F)}},\
  }\bibfield  {journal} {\bibinfo  {journal} {Symmetry}\ }\textbf {\bibinfo
  {volume} {14}},\ \href {https://doi.org/10.3390/sym14010180}
  {10.3390/sym14010180} (\bibinfo {year} {2022}),\ \Eprint
  {https://arxiv.org/abs/2111.10423} {arXiv:2111.10423 [hep-th]} \BibitemShut
  {NoStop}%
\bibitem [{\citenamefont {Tanizaki}\ and\ \citenamefont
  {\"Unsal}(2022)}]{Tanizaki:2022ngt}%
  \BibitemOpen
  \bibfield  {author} {\bibinfo {author} {\bibfnamefont {Y.}~\bibnamefont
  {Tanizaki}}\ and\ \bibinfo {author} {\bibfnamefont {M.}~\bibnamefont
  {\"Unsal}},\ }\bibfield  {title} {\bibinfo {title} {{Center vortex and
  confinement in Yang-Mills theory and QCD with anomaly-preserving
  compactifications}},\ }\href {https://doi.org/10.1093/ptep/ptac042}
  {\bibfield  {journal} {\bibinfo  {journal} {PTEP}\ }\textbf {\bibinfo
  {volume} {2022}},\ \bibinfo {pages} {04} (\bibinfo {year} {2022})},\ \Eprint
  {https://arxiv.org/abs/2201.06166} {arXiv:2201.06166 [hep-th]} \BibitemShut
  {NoStop}%
\bibitem [{\citenamefont {Gross}(1978)}]{Gross:1977wu}%
  \BibitemOpen
  \bibfield  {author} {\bibinfo {author} {\bibfnamefont {D.~J.}\ \bibnamefont
  {Gross}},\ }\bibfield  {title} {\bibinfo {title} {{Meron Configurations in
  the Two-Dimensional O(3) Sigma Model}},\ }\href
  {https://doi.org/10.1016/0550-3213(78)90470-4} {\bibfield  {journal}
  {\bibinfo  {journal} {Nucl. Phys. B}\ }\textbf {\bibinfo {volume} {132}},\
  \bibinfo {pages} {439} (\bibinfo {year} {1978})}\BibitemShut {NoStop}%
\bibitem [{\citenamefont {Witten}(1979{\natexlab{b}})}]{Witten:1978bc}%
  \BibitemOpen
  \bibfield  {author} {\bibinfo {author} {\bibfnamefont {E.}~\bibnamefont
  {Witten}},\ }\bibfield  {title} {\bibinfo {title} {{Instantons, the Quark
  Model, and the 1/n Expansion}},\ }\href
  {https://doi.org/10.1016/0550-3213(79)90243-8} {\bibfield  {journal}
  {\bibinfo  {journal} {Nucl. Phys. B}\ }\textbf {\bibinfo {volume} {149}},\
  \bibinfo {pages} {285} (\bibinfo {year} {1979}{\natexlab{b}})}\BibitemShut
  {NoStop}%
\bibitem [{\citenamefont {D'Adda}\ \emph {et~al.}(1978)\citenamefont {D'Adda},
  \citenamefont {Luscher},\ and\ \citenamefont {Di~Vecchia}}]{DAdda:1978vbw}%
  \BibitemOpen
  \bibfield  {author} {\bibinfo {author} {\bibfnamefont {A.}~\bibnamefont
  {D'Adda}}, \bibinfo {author} {\bibfnamefont {M.}~\bibnamefont {Luscher}},\
  and\ \bibinfo {author} {\bibfnamefont {P.}~\bibnamefont {Di~Vecchia}},\
  }\bibfield  {title} {\bibinfo {title} {{A 1/n Expandable Series of Nonlinear
  Sigma Models with Instantons}},\ }\href
  {https://doi.org/10.1016/0550-3213(78)90432-7} {\bibfield  {journal}
  {\bibinfo  {journal} {Nucl. Phys. B}\ }\textbf {\bibinfo {volume} {146}},\
  \bibinfo {pages} {63} (\bibinfo {year} {1978})}\BibitemShut {NoStop}%
\bibitem [{\citenamefont {Di~Vecchia}(1979)}]{DiVecchia:1979pzw}%
  \BibitemOpen
  \bibfield  {author} {\bibinfo {author} {\bibfnamefont {P.}~\bibnamefont
  {Di~Vecchia}},\ }\bibfield  {title} {\bibinfo {title} {{An Effective
  Lagrangian With No U(1) Problem in {CP}$^{(n-1)}$ Models and {QCD}}},\ }\href
  {https://doi.org/10.1016/0370-2693(79)91271-1} {\bibfield  {journal}
  {\bibinfo  {journal} {Phys. Lett. B}\ }\textbf {\bibinfo {volume} {85}},\
  \bibinfo {pages} {357} (\bibinfo {year} {1979})}\BibitemShut {NoStop}%
\bibitem [{\citenamefont {Berg}\ and\ \citenamefont
  {Luscher}(1979)}]{Berg:1979uq}%
  \BibitemOpen
  \bibfield  {author} {\bibinfo {author} {\bibfnamefont {B.}~\bibnamefont
  {Berg}}\ and\ \bibinfo {author} {\bibfnamefont {M.}~\bibnamefont {Luscher}},\
  }\bibfield  {title} {\bibinfo {title} {{Computation of Quantum Fluctuations
  Around Multi-Instanton Fields from Exact Green's Functions: The CP**n-1
  Case}},\ }\href {https://doi.org/10.1007/BF01941324} {\bibfield  {journal}
  {\bibinfo  {journal} {Commun. Math. Phys.}\ }\textbf {\bibinfo {volume}
  {69}},\ \bibinfo {pages} {57} (\bibinfo {year} {1979})}\BibitemShut {NoStop}%
\bibitem [{\citenamefont {Fateev}\ \emph {et~al.}(1979)\citenamefont {Fateev},
  \citenamefont {Frolov},\ and\ \citenamefont {Schwarz}}]{Fateev:1979dc}%
  \BibitemOpen
  \bibfield  {author} {\bibinfo {author} {\bibfnamefont {V.~A.}\ \bibnamefont
  {Fateev}}, \bibinfo {author} {\bibfnamefont {I.~V.}\ \bibnamefont {Frolov}},\
  and\ \bibinfo {author} {\bibfnamefont {A.~S.}\ \bibnamefont {Schwarz}},\
  }\bibfield  {title} {\bibinfo {title} {{Quantum Fluctuations of Instantons in
  the Nonlinear Sigma Model}},\ }\href
  {https://doi.org/10.1016/0550-3213(79)90367-5} {\bibfield  {journal}
  {\bibinfo  {journal} {Nucl. Phys. B}\ }\textbf {\bibinfo {volume} {154}},\
  \bibinfo {pages} {1} (\bibinfo {year} {1979})}\BibitemShut {NoStop}%
\bibitem [{\citenamefont {Rothe}\ and\ \citenamefont
  {Swieca}(1980)}]{Rothe:1980rp}%
  \BibitemOpen
  \bibfield  {author} {\bibinfo {author} {\bibfnamefont {K.~D.}\ \bibnamefont
  {Rothe}}\ and\ \bibinfo {author} {\bibfnamefont {J.~A.}\ \bibnamefont
  {Swieca}},\ }\bibfield  {title} {\bibinfo {title} {{Fractional Winding
  Numbers and the $u$(1) Problem}},\ }\href
  {https://doi.org/10.1016/0550-3213(80)90137-6} {\bibfield  {journal}
  {\bibinfo  {journal} {Nucl. Phys. B}\ }\textbf {\bibinfo {volume} {168}},\
  \bibinfo {pages} {454} (\bibinfo {year} {1980})}\BibitemShut {NoStop}%
\bibitem [{\citenamefont {Zhitnitsky}(1988)}]{Zhitnitsky:1988zd}%
  \BibitemOpen
  \bibfield  {author} {\bibinfo {author} {\bibfnamefont {A.~R.}\ \bibnamefont
  {Zhitnitsky}},\ }\bibfield  {title} {\bibinfo {title} {{Fractional
  Topological Charge, Torons and Breaking of Discrete Chiral Symmetry in the
  Supersymmetric O(3) $\sigma$ Model}},\ }\href@noop {} {\bibfield  {journal}
  {\bibinfo  {journal} {Sov. Phys. JETP}\ }\textbf {\bibinfo {volume} {67}},\
  \bibinfo {pages} {1095} (\bibinfo {year} {1988})}\BibitemShut {NoStop}%
\bibitem [{\citenamefont {Ahmad}\ \emph {et~al.}(2005)\citenamefont {Ahmad},
  \citenamefont {Lenaghan},\ and\ \citenamefont {Thacker}}]{Ahmad:2005dr}%
  \BibitemOpen
  \bibfield  {author} {\bibinfo {author} {\bibfnamefont {S.}~\bibnamefont
  {Ahmad}}, \bibinfo {author} {\bibfnamefont {J.~T.}\ \bibnamefont
  {Lenaghan}},\ and\ \bibinfo {author} {\bibfnamefont {H.~B.}\ \bibnamefont
  {Thacker}},\ }\bibfield  {title} {\bibinfo {title} {{Coherent topological
  charge structure in CP(N-1) models and QCD}},\ }\href
  {https://doi.org/10.1103/PhysRevD.72.114511} {\bibfield  {journal} {\bibinfo
  {journal} {Phys. Rev. D}\ }\textbf {\bibinfo {volume} {72}},\ \bibinfo
  {pages} {114511} (\bibinfo {year} {2005})},\ \Eprint
  {https://arxiv.org/abs/hep-lat/0509066} {arXiv:hep-lat/0509066} \BibitemShut
  {NoStop}%
\bibitem [{\citenamefont {Lian}\ and\ \citenamefont
  {Thacker}(2007)}]{Lian:2006ky}%
  \BibitemOpen
  \bibfield  {author} {\bibinfo {author} {\bibfnamefont {Y.}~\bibnamefont
  {Lian}}\ and\ \bibinfo {author} {\bibfnamefont {H.~B.}\ \bibnamefont
  {Thacker}},\ }\bibfield  {title} {\bibinfo {title} {{Small Instantons in
  CP**1 and CP**2 Sigma Models}},\ }\href
  {https://doi.org/10.1103/PhysRevD.75.065031} {\bibfield  {journal} {\bibinfo
  {journal} {Phys. Rev. D}\ }\textbf {\bibinfo {volume} {75}},\ \bibinfo
  {pages} {065031} (\bibinfo {year} {2007})},\ \Eprint
  {https://arxiv.org/abs/hep-lat/0607026} {arXiv:hep-lat/0607026} \BibitemShut
  {NoStop}%
\bibitem [{\citenamefont {Bruckmann}(2008)}]{Bruckmann:2007zh}%
  \BibitemOpen
  \bibfield  {author} {\bibinfo {author} {\bibfnamefont {F.}~\bibnamefont
  {Bruckmann}},\ }\bibfield  {title} {\bibinfo {title} {{Instanton constituents
  in the O(3) model at finite temperature}},\ }\href
  {https://doi.org/10.1103/PhysRevLett.100.051602} {\bibfield  {journal}
  {\bibinfo  {journal} {Phys. Rev. Lett.}\ }\textbf {\bibinfo {volume} {100}},\
  \bibinfo {pages} {051602} (\bibinfo {year} {2008})},\ \Eprint
  {https://arxiv.org/abs/0707.0775} {arXiv:0707.0775 [hep-th]} \BibitemShut
  {NoStop}%
\bibitem [{\citenamefont {Brendel}\ \emph {et~al.}(2009)\citenamefont
  {Brendel}, \citenamefont {Bruckmann}, \citenamefont {Janssen}, \citenamefont
  {Wipf},\ and\ \citenamefont {Wozar}}]{Brendel:2009mp}%
  \BibitemOpen
  \bibfield  {author} {\bibinfo {author} {\bibfnamefont {W.}~\bibnamefont
  {Brendel}}, \bibinfo {author} {\bibfnamefont {F.}~\bibnamefont {Bruckmann}},
  \bibinfo {author} {\bibfnamefont {L.}~\bibnamefont {Janssen}}, \bibinfo
  {author} {\bibfnamefont {A.}~\bibnamefont {Wipf}},\ and\ \bibinfo {author}
  {\bibfnamefont {C.}~\bibnamefont {Wozar}},\ }\bibfield  {title} {\bibinfo
  {title} {{Instanton constituents and fermionic zero modes in twisted CP**n
  models}},\ }\href {https://doi.org/10.1016/j.physletb.2009.04.055} {\bibfield
   {journal} {\bibinfo  {journal} {Phys. Lett. B}\ }\textbf {\bibinfo {volume}
  {676}},\ \bibinfo {pages} {116} (\bibinfo {year} {2009})},\ \Eprint
  {https://arxiv.org/abs/0902.2328} {arXiv:0902.2328 [hep-th]} \BibitemShut
  {NoStop}%
\bibitem [{\citenamefont {Atiyah}\ \emph {et~al.}(1978)\citenamefont {Atiyah},
  \citenamefont {Hitchin}, \citenamefont {Drinfeld},\ and\ \citenamefont
  {Manin}}]{Atiyah:1978ri}%
  \BibitemOpen
  \bibfield  {author} {\bibinfo {author} {\bibfnamefont {M.~F.}\ \bibnamefont
  {Atiyah}}, \bibinfo {author} {\bibfnamefont {N.~J.}\ \bibnamefont {Hitchin}},
  \bibinfo {author} {\bibfnamefont {V.~G.}\ \bibnamefont {Drinfeld}},\ and\
  \bibinfo {author} {\bibfnamefont {Y.~I.}\ \bibnamefont {Manin}},\ }\bibfield
  {title} {\bibinfo {title} {{Construction of Instantons}},\ }\href
  {https://doi.org/10.1016/0375-9601(78)90141-X} {\bibfield  {journal}
  {\bibinfo  {journal} {Phys. Lett. A}\ }\textbf {\bibinfo {volume} {65}},\
  \bibinfo {pages} {185} (\bibinfo {year} {1978})}\BibitemShut {NoStop}%
\bibitem [{\citenamefont {Edwards}\ \emph {et~al.}(1998)\citenamefont
  {Edwards}, \citenamefont {Heller},\ and\ \citenamefont
  {Narayanan}}]{Edwards:1998dj}%
  \BibitemOpen
  \bibfield  {author} {\bibinfo {author} {\bibfnamefont {R.~G.}\ \bibnamefont
  {Edwards}}, \bibinfo {author} {\bibfnamefont {U.~M.}\ \bibnamefont
  {Heller}},\ and\ \bibinfo {author} {\bibfnamefont {R.}~\bibnamefont
  {Narayanan}},\ }\bibfield  {title} {\bibinfo {title} {{Evidence for
  fractional topological charge in SU(2) pure Yang-Mills theory}},\ }\href
  {https://doi.org/10.1016/S0370-2693(98)00951-4} {\bibfield  {journal}
  {\bibinfo  {journal} {Phys. Lett. B}\ }\textbf {\bibinfo {volume} {438}},\
  \bibinfo {pages} {96} (\bibinfo {year} {1998})},\ \Eprint
  {https://arxiv.org/abs/hep-lat/9806011} {arXiv:hep-lat/9806011} \BibitemShut
  {NoStop}%
\bibitem [{\citenamefont {Biddle}\ \emph
  {et~al.}(2022{\natexlab{a}})\citenamefont {Biddle}, \citenamefont {Kamleh},\
  and\ \citenamefont {Leinweber}}]{Biddle:2022acd}%
  \BibitemOpen
  \bibfield  {author} {\bibinfo {author} {\bibfnamefont {J.~C.}\ \bibnamefont
  {Biddle}}, \bibinfo {author} {\bibfnamefont {W.}~\bibnamefont {Kamleh}},\
  and\ \bibinfo {author} {\bibfnamefont {D.~B.}\ \bibnamefont {Leinweber}},\
  }\bibfield  {title} {\bibinfo {title} {{Impact of dynamical fermions on the
  center vortex gluon propagator}},\ }\href
  {https://doi.org/10.1103/PhysRevD.106.014506} {\bibfield  {journal} {\bibinfo
   {journal} {Phys. Rev. D}\ }\textbf {\bibinfo {volume} {106}},\ \bibinfo
  {pages} {014506} (\bibinfo {year} {2022}{\natexlab{a}})},\ \Eprint
  {https://arxiv.org/abs/2206.02320} {arXiv:2206.02320 [hep-lat]} \BibitemShut
  {NoStop}%
\bibitem [{\citenamefont {Biddle}\ \emph
  {et~al.}(2022{\natexlab{b}})\citenamefont {Biddle}, \citenamefont {Kamleh},\
  and\ \citenamefont {Leinweber}}]{Biddle:2022zgw}%
  \BibitemOpen
  \bibfield  {author} {\bibinfo {author} {\bibfnamefont {J.~C.}\ \bibnamefont
  {Biddle}}, \bibinfo {author} {\bibfnamefont {W.}~\bibnamefont {Kamleh}},\
  and\ \bibinfo {author} {\bibfnamefont {D.~B.}\ \bibnamefont {Leinweber}},\
  }\bibfield  {title} {\bibinfo {title} {{Static quark potential from center
  vortices in the presence of dynamical fermions}},\ }\href
  {https://doi.org/10.1103/PhysRevD.106.054505} {\bibfield  {journal} {\bibinfo
   {journal} {Phys. Rev. D}\ }\textbf {\bibinfo {volume} {106}},\ \bibinfo
  {pages} {054505} (\bibinfo {year} {2022}{\natexlab{b}})},\ \Eprint
  {https://arxiv.org/abs/2206.00844} {arXiv:2206.00844 [hep-lat]} \BibitemShut
  {NoStop}%
\bibitem [{\citenamefont {Leinweber}\ \emph {et~al.}(2022)\citenamefont
  {Leinweber}, \citenamefont {Biddle},\ and\ \citenamefont
  {Kamleh}}]{Leinweber:2022dpp}%
  \BibitemOpen
  \bibfield  {author} {\bibinfo {author} {\bibfnamefont {D.}~\bibnamefont
  {Leinweber}}, \bibinfo {author} {\bibfnamefont {J.}~\bibnamefont {Biddle}},\
  and\ \bibinfo {author} {\bibfnamefont {W.}~\bibnamefont {Kamleh}},\
  }\bibfield  {title} {\bibinfo {title} {{Centre vortex structure of QCD-vacuum
  fields and confinement}},\ }\href
  {https://doi.org/10.21468/SciPostPhysProc.6.004} {\bibfield  {journal}
  {\bibinfo  {journal} {SciPost Phys. Proc.}\ }\textbf {\bibinfo {volume}
  {6}},\ \bibinfo {pages} {004} (\bibinfo {year} {2022})},\ \Eprint
  {https://arxiv.org/abs/2205.12518} {arXiv:2205.12518 [hep-lat]} \BibitemShut
  {NoStop}%
\bibitem [{\citenamefont {Biddle}\ \emph {et~al.}(2023)\citenamefont {Biddle},
  \citenamefont {Kamleh},\ and\ \citenamefont {Leinweber}}]{Biddle:2023lod}%
  \BibitemOpen
  \bibfield  {author} {\bibinfo {author} {\bibfnamefont {J.~C.}\ \bibnamefont
  {Biddle}}, \bibinfo {author} {\bibfnamefont {W.}~\bibnamefont {Kamleh}},\
  and\ \bibinfo {author} {\bibfnamefont {D.~B.}\ \bibnamefont {Leinweber}},\
  }\bibfield  {title} {\bibinfo {title} {{Centre vortex structure in the
  presence of dynamical fermions}},\ }\href@noop {} {\  (\bibinfo {year}
  {2023})},\ \Eprint {https://arxiv.org/abs/2302.05897} {arXiv:2302.05897
  [hep-lat]} \BibitemShut {NoStop}%
\bibitem [{Note5()}]{Note5}%
  \BibitemOpen
  \bibinfo {note} {Notice that the constraint ${\protect \bar {z}}\cdot z = 1$
  defines a sphere $S^{2 N -1}$, while the gauge symmetry removes a phase,
  i.e., $S^1$, which leaves $S^{2 N -1}/ S^1$. This is another way to define
  $\protect \mathbb {CP}^{N-1} $.}\BibitemShut {Stop}%
\bibitem [{\citenamefont {Jackiw}(1996)}]{Jackiw:1996ec}%
  \BibitemOpen
  \bibfield  {author} {\bibinfo {author} {\bibfnamefont {R.}~\bibnamefont
  {Jackiw}},\ }\bibfield  {title} {\bibinfo {title} {{Gauge theories in the
  momentum / curvature representation}}\ }(\bibinfo {year} {1996})\ \Eprint
  {https://arxiv.org/abs/hep-th/9604040} {arXiv:hep-th/9604040} \BibitemShut
  {NoStop}%
\bibitem [{\citenamefont {Goddard}\ \emph {et~al.}(1977)\citenamefont
  {Goddard}, \citenamefont {Nuyts},\ and\ \citenamefont
  {Olive}}]{Goddard:1976qe}%
  \BibitemOpen
  \bibfield  {author} {\bibinfo {author} {\bibfnamefont {P.}~\bibnamefont
  {Goddard}}, \bibinfo {author} {\bibfnamefont {J.}~\bibnamefont {Nuyts}},\
  and\ \bibinfo {author} {\bibfnamefont {D.~I.}\ \bibnamefont {Olive}},\
  }\bibfield  {title} {\bibinfo {title} {{Gauge Theories and Magnetic
  Charge}},\ }\href {https://doi.org/10.1016/0550-3213(77)90221-8} {\bibfield
  {journal} {\bibinfo  {journal} {Nucl. Phys. B}\ }\textbf {\bibinfo {volume}
  {125}},\ \bibinfo {pages} {1} (\bibinfo {year} {1977})}\BibitemShut {NoStop}%
\bibitem [{\citenamefont {Ginsparg}\ and\ \citenamefont
  {Wilson}(1982)}]{Ginsparg:1981bj}%
  \BibitemOpen
  \bibfield  {author} {\bibinfo {author} {\bibfnamefont {P.~H.}\ \bibnamefont
  {Ginsparg}}\ and\ \bibinfo {author} {\bibfnamefont {K.~G.}\ \bibnamefont
  {Wilson}},\ }\bibfield  {title} {\bibinfo {title} {{A Remnant of Chiral
  Symmetry on the Lattice}},\ }\href {https://doi.org/10.1103/PhysRevD.25.2649}
  {\bibfield  {journal} {\bibinfo  {journal} {Phys. Rev. D}\ }\textbf {\bibinfo
  {volume} {25}},\ \bibinfo {pages} {2649} (\bibinfo {year}
  {1982})}\BibitemShut {NoStop}%
\bibitem [{\citenamefont {Kaplan}(1992)}]{Kaplan:1992bt}%
  \BibitemOpen
  \bibfield  {author} {\bibinfo {author} {\bibfnamefont {D.~B.}\ \bibnamefont
  {Kaplan}},\ }\bibfield  {title} {\bibinfo {title} {{A Method for simulating
  chiral fermions on the lattice}},\ }\href
  {https://doi.org/10.1016/0370-2693(92)91112-M} {\bibfield  {journal}
  {\bibinfo  {journal} {Phys. Lett. B}\ }\textbf {\bibinfo {volume} {288}},\
  \bibinfo {pages} {342} (\bibinfo {year} {1992})},\ \Eprint
  {https://arxiv.org/abs/hep-lat/9206013} {arXiv:hep-lat/9206013} \BibitemShut
  {NoStop}%
\bibitem [{\citenamefont {Narayanan}\ and\ \citenamefont
  {Neuberger}(1993{\natexlab{a}})}]{Narayanan:1993zzh}%
  \BibitemOpen
  \bibfield  {author} {\bibinfo {author} {\bibfnamefont {R.}~\bibnamefont
  {Narayanan}}\ and\ \bibinfo {author} {\bibfnamefont {H.}~\bibnamefont
  {Neuberger}},\ }\bibfield  {title} {\bibinfo {title} {{Infinitely many
  regulator fields for chiral fermions}},\ }\href
  {https://doi.org/10.1016/0370-2693(93)90636-V} {\bibfield  {journal}
  {\bibinfo  {journal} {Phys. Lett. B}\ }\textbf {\bibinfo {volume} {302}},\
  \bibinfo {pages} {62} (\bibinfo {year} {1993}{\natexlab{a}})},\ \Eprint
  {https://arxiv.org/abs/hep-lat/9212019} {arXiv:hep-lat/9212019} \BibitemShut
  {NoStop}%
\bibitem [{\citenamefont {Narayanan}\ and\ \citenamefont
  {Neuberger}(1994)}]{Narayanan:1993sk}%
  \BibitemOpen
  \bibfield  {author} {\bibinfo {author} {\bibfnamefont {R.}~\bibnamefont
  {Narayanan}}\ and\ \bibinfo {author} {\bibfnamefont {H.}~\bibnamefont
  {Neuberger}},\ }\bibfield  {title} {\bibinfo {title} {{Chiral determinant as
  an overlap of two vacua}},\ }\href
  {https://doi.org/10.1016/0550-3213(94)90393-X} {\bibfield  {journal}
  {\bibinfo  {journal} {Nucl. Phys. B}\ }\textbf {\bibinfo {volume} {412}},\
  \bibinfo {pages} {574} (\bibinfo {year} {1994})},\ \Eprint
  {https://arxiv.org/abs/hep-lat/9307006} {arXiv:hep-lat/9307006} \BibitemShut
  {NoStop}%
\bibitem [{\citenamefont {Narayanan}\ and\ \citenamefont
  {Neuberger}(1993{\natexlab{b}})}]{Narayanan:1993ss}%
  \BibitemOpen
  \bibfield  {author} {\bibinfo {author} {\bibfnamefont {R.}~\bibnamefont
  {Narayanan}}\ and\ \bibinfo {author} {\bibfnamefont {H.}~\bibnamefont
  {Neuberger}},\ }\bibfield  {title} {\bibinfo {title} {{Chiral fermions on the
  lattice}},\ }\href {https://doi.org/10.1103/PhysRevLett.71.3251} {\bibfield
  {journal} {\bibinfo  {journal} {Phys. Rev. Lett.}\ }\textbf {\bibinfo
  {volume} {71}},\ \bibinfo {pages} {3251} (\bibinfo {year}
  {1993}{\natexlab{b}})},\ \Eprint {https://arxiv.org/abs/hep-lat/9308011}
  {arXiv:hep-lat/9308011} \BibitemShut {NoStop}%
\bibitem [{\citenamefont {Narayanan}\ and\ \citenamefont
  {Neuberger}(1995)}]{Narayanan:1994gw}%
  \BibitemOpen
  \bibfield  {author} {\bibinfo {author} {\bibfnamefont {R.}~\bibnamefont
  {Narayanan}}\ and\ \bibinfo {author} {\bibfnamefont {H.}~\bibnamefont
  {Neuberger}},\ }\bibfield  {title} {\bibinfo {title} {{A Construction of
  lattice chiral gauge theories}},\ }\href
  {https://doi.org/10.1016/0550-3213(95)00111-5} {\bibfield  {journal}
  {\bibinfo  {journal} {Nucl. Phys. B}\ }\textbf {\bibinfo {volume} {443}},\
  \bibinfo {pages} {305} (\bibinfo {year} {1995})},\ \Eprint
  {https://arxiv.org/abs/hep-th/9411108} {arXiv:hep-th/9411108} \BibitemShut
  {NoStop}%
\bibitem [{\citenamefont {Luscher}(1998)}]{Luscher:1998pqa}%
  \BibitemOpen
  \bibfield  {author} {\bibinfo {author} {\bibfnamefont {M.}~\bibnamefont
  {Luscher}},\ }\bibfield  {title} {\bibinfo {title} {{Exact chiral symmetry on
  the lattice and the Ginsparg-Wilson relation}},\ }\href
  {https://doi.org/10.1016/S0370-2693(98)00423-7} {\bibfield  {journal}
  {\bibinfo  {journal} {Phys. Lett. B}\ }\textbf {\bibinfo {volume} {428}},\
  \bibinfo {pages} {342} (\bibinfo {year} {1998})},\ \Eprint
  {https://arxiv.org/abs/hep-lat/9802011} {arXiv:hep-lat/9802011} \BibitemShut
  {NoStop}%
\bibitem [{Note6()}]{Note6}%
  \BibitemOpen
  \bibinfo {note} {Fodor {\protect \it et al.} performed simulations in a
  $SU(3)$ gauge theory using an external quark propagator in the sextet
  representation \cite {Fodor:2009nh}. As they discuss, the sextet
  representation is sensitive to the presence of objects with topological
  charge $1/5$, and for which they see no evidence. However, this does not
  exclude the appearance of objects with charge $1/3$.}\BibitemShut {Stop}%
\bibitem [{\citenamefont {Rothe}\ and\ \citenamefont
  {Swieca}(2003)}]{Horvath:2002yn}%
  \BibitemOpen
  \bibfield  {author} {\bibinfo {author} {\bibfnamefont {K.~D.}\ \bibnamefont
  {Rothe}}\ and\ \bibinfo {author} {\bibfnamefont {J.~A.}\ \bibnamefont
  {Swieca}},\ }\bibfield  {title} {\bibinfo {title} {{On the local structure of
  topological charge fluctuations in QCD}},\ }\href
  {https://doi.org/10.1103/PhysRevD.67.011501} {\bibfield  {journal} {\bibinfo
  {journal} {Phys. Rev. D}\ }\textbf {\bibinfo {volume} {67}},\ \bibinfo
  {pages} {011501} (\bibinfo {year} {2003})},\ \Eprint
  {https://arxiv.org/abs/hep-lat/0203027} {arXiv:hep-lat/0203027} \BibitemShut
  {NoStop}%
\bibitem [{\citenamefont {Horvath}\ \emph {et~al.}(2005)\citenamefont {Horvath}
  \emph {et~al.}}]{Horvath:2005rv}%
  \BibitemOpen
  \bibfield  {author} {\bibinfo {author} {\bibfnamefont {I.}~\bibnamefont
  {Horvath}} \emph {et~al.},\ }\bibfield  {title} {\bibinfo {title}
  {{Inherently global nature of topological charge fluctuations in QCD}},\
  }\href {https://doi.org/10.1016/j.physletb.2005.03.004} {\bibfield  {journal}
  {\bibinfo  {journal} {Phys. Lett. B}\ }\textbf {\bibinfo {volume} {612}},\
  \bibinfo {pages} {21} (\bibinfo {year} {2005})},\ \Eprint
  {https://arxiv.org/abs/hep-lat/0501025} {arXiv:hep-lat/0501025} \BibitemShut
  {NoStop}%
\bibitem [{\citenamefont {Thacker}(2010)}]{Thacker:2010zk}%
  \BibitemOpen
  \bibfield  {author} {\bibinfo {author} {\bibfnamefont {H.~B.}\ \bibnamefont
  {Thacker}},\ }\bibfield  {title} {\bibinfo {title} {{Tachyonic crystals and
  the laminar instability of the perturbative vacuum in asymptotically free
  gauge theories}},\ }\href {https://doi.org/10.1103/PhysRevD.81.125006}
  {\bibfield  {journal} {\bibinfo  {journal} {Phys. Rev. D}\ }\textbf {\bibinfo
  {volume} {81}},\ \bibinfo {pages} {125006} (\bibinfo {year} {2010})},\
  \Eprint {https://arxiv.org/abs/1001.4215} {arXiv:1001.4215 [hep-th]}
  \BibitemShut {NoStop}%
\bibitem [{\citenamefont {Thacker}\ \emph {et~al.}(2011)\citenamefont
  {Thacker}, \citenamefont {Xiong},\ and\ \citenamefont
  {Kamat}}]{Thacker:2011sz}%
  \BibitemOpen
  \bibfield  {author} {\bibinfo {author} {\bibfnamefont {H.~B.}\ \bibnamefont
  {Thacker}}, \bibinfo {author} {\bibfnamefont {C.}~\bibnamefont {Xiong}},\
  and\ \bibinfo {author} {\bibfnamefont {A.}~\bibnamefont {Kamat}},\ }\bibfield
   {title} {\bibinfo {title} {{Chiral quark dynamics and topological charge:
  The role of the Ramond-Ramond U(1) Gauge Field in Holographic QCD}},\ }\href
  {https://doi.org/10.1103/PhysRevD.84.105011} {\bibfield  {journal} {\bibinfo
  {journal} {Phys. Rev. D}\ }\textbf {\bibinfo {volume} {84}},\ \bibinfo
  {pages} {105011} (\bibinfo {year} {2011})},\ \Eprint
  {https://arxiv.org/abs/1104.3063} {arXiv:1104.3063 [hep-th]} \BibitemShut
  {NoStop}%
\bibitem [{\citenamefont {Alexandru}\ and\ \citenamefont
  {Horv\'ath}(2021)}]{Alexandru:2021pap}%
  \BibitemOpen
  \bibfield  {author} {\bibinfo {author} {\bibfnamefont {A.}~\bibnamefont
  {Alexandru}}\ and\ \bibinfo {author} {\bibfnamefont {I.}~\bibnamefont
  {Horv\'ath}},\ }\bibfield  {title} {\bibinfo {title} {{Unusual Features of
  QCD Low-Energy Modes in the Infrared Phase}},\ }\href
  {https://doi.org/10.1103/PhysRevLett.127.052303} {\bibfield  {journal}
  {\bibinfo  {journal} {Phys. Rev. Lett.}\ }\textbf {\bibinfo {volume} {127}},\
  \bibinfo {pages} {052303} (\bibinfo {year} {2021})},\ \Eprint
  {https://arxiv.org/abs/2103.05607} {arXiv:2103.05607 [hep-lat]} \BibitemShut
  {NoStop}%
\bibitem [{Note7()}]{Note7}%
  \BibitemOpen
  \bibinfo {note} {We also comment that by looking at the spectra of higher
  spin mesons, the region between $300 \leq T \leq 600$~MeV is still far from a
  perturbative quark-gluon plasma \cite {Glozman:2022lda}.}\BibitemShut {Stop}%
\bibitem [{\citenamefont {Bazavov}\ \emph {et~al.}(2019)\citenamefont {Bazavov}
  \emph {et~al.}}]{HotQCD:2018pds}%
  \BibitemOpen
  \bibfield  {author} {\bibinfo {author} {\bibfnamefont {A.}~\bibnamefont
  {Bazavov}} \emph {et~al.} (\bibinfo {collaboration} {HotQCD}),\ }\bibfield
  {title} {\bibinfo {title} {{Chiral crossover in QCD at zero and non-zero
  chemical potentials}},\ }\href
  {https://doi.org/10.1016/j.physletb.2019.05.013} {\bibfield  {journal}
  {\bibinfo  {journal} {Phys. Lett. B}\ }\textbf {\bibinfo {volume} {795}},\
  \bibinfo {pages} {15} (\bibinfo {year} {2019})},\ \Eprint
  {https://arxiv.org/abs/1812.08235} {arXiv:1812.08235 [hep-lat]} \BibitemShut
  {NoStop}%
\bibitem [{\citenamefont {Borsanyi}\ \emph {et~al.}(2020)\citenamefont
  {Borsanyi}, \citenamefont {Fodor}, \citenamefont {Guenther}, \citenamefont
  {Kara}, \citenamefont {Katz}, \citenamefont {Parotto}, \citenamefont
  {Pasztor}, \citenamefont {Ratti},\ and\ \citenamefont
  {Szabo}}]{Borsanyi:2020fev}%
  \BibitemOpen
  \bibfield  {author} {\bibinfo {author} {\bibfnamefont {S.}~\bibnamefont
  {Borsanyi}}, \bibinfo {author} {\bibfnamefont {Z.}~\bibnamefont {Fodor}},
  \bibinfo {author} {\bibfnamefont {J.~N.}\ \bibnamefont {Guenther}}, \bibinfo
  {author} {\bibfnamefont {R.}~\bibnamefont {Kara}}, \bibinfo {author}
  {\bibfnamefont {S.~D.}\ \bibnamefont {Katz}}, \bibinfo {author}
  {\bibfnamefont {P.}~\bibnamefont {Parotto}}, \bibinfo {author} {\bibfnamefont
  {A.}~\bibnamefont {Pasztor}}, \bibinfo {author} {\bibfnamefont
  {C.}~\bibnamefont {Ratti}},\ and\ \bibinfo {author} {\bibfnamefont {K.~K.}\
  \bibnamefont {Szabo}},\ }\bibfield  {title} {\bibinfo {title} {{QCD Crossover
  at Finite Chemical Potential from Lattice Simulations}},\ }\href
  {https://doi.org/10.1103/PhysRevLett.125.052001} {\bibfield  {journal}
  {\bibinfo  {journal} {Phys. Rev. Lett.}\ }\textbf {\bibinfo {volume} {125}},\
  \bibinfo {pages} {052001} (\bibinfo {year} {2020})},\ \Eprint
  {https://arxiv.org/abs/2002.02821} {arXiv:2002.02821 [hep-lat]} \BibitemShut
  {NoStop}%
\bibitem [{\citenamefont {Guenther}(2022)}]{Guenther:2022wcr}%
  \BibitemOpen
  \bibfield  {author} {\bibinfo {author} {\bibfnamefont {J.~N.}\ \bibnamefont
  {Guenther}},\ }\bibfield  {title} {\bibinfo {title} {{An overview of the QCD
  phase diagram at finite $T$ and $\mu$}},\ }in\ \href@noop {} {\emph {\bibinfo
  {booktitle} {{38th International Symposium on Lattice Field Theory}}}}\
  (\bibinfo {year} {2022})\ \Eprint {https://arxiv.org/abs/2201.02072}
  {arXiv:2201.02072 [hep-lat]} \BibitemShut {NoStop}%
\bibitem [{\citenamefont {McLerran}\ and\ \citenamefont
  {Pisarski}(2007)}]{McLerran:2007qj}%
  \BibitemOpen
  \bibfield  {author} {\bibinfo {author} {\bibfnamefont {L.}~\bibnamefont
  {McLerran}}\ and\ \bibinfo {author} {\bibfnamefont {R.~D.}\ \bibnamefont
  {Pisarski}},\ }\bibfield  {title} {\bibinfo {title} {{Phases of cold, dense
  quarks at large N(c)}},\ }\href
  {https://doi.org/10.1016/j.nuclphysa.2007.08.013} {\bibfield  {journal}
  {\bibinfo  {journal} {Nucl. Phys. A}\ }\textbf {\bibinfo {volume} {796}},\
  \bibinfo {pages} {83} (\bibinfo {year} {2007})},\ \Eprint
  {https://arxiv.org/abs/0706.2191} {arXiv:0706.2191 [hep-ph]} \BibitemShut
  {NoStop}%
\bibitem [{\citenamefont {Lajer}\ \emph {et~al.}(2022)\citenamefont {Lajer},
  \citenamefont {Konik}, \citenamefont {Pisarski},\ and\ \citenamefont
  {Tsvelik}}]{Lajer:2021kcz}%
  \BibitemOpen
  \bibfield  {author} {\bibinfo {author} {\bibfnamefont {M.}~\bibnamefont
  {Lajer}}, \bibinfo {author} {\bibfnamefont {R.~M.}\ \bibnamefont {Konik}},
  \bibinfo {author} {\bibfnamefont {R.~D.}\ \bibnamefont {Pisarski}},\ and\
  \bibinfo {author} {\bibfnamefont {A.~M.}\ \bibnamefont {Tsvelik}},\
  }\bibfield  {title} {\bibinfo {title} {{When cold, dense quarks in 1+1 and
  3+1 dimensions are not a Fermi liquid}},\ }\href
  {https://doi.org/10.1103/PhysRevD.105.054035} {\bibfield  {journal} {\bibinfo
   {journal} {Phys. Rev. D}\ }\textbf {\bibinfo {volume} {105}},\ \bibinfo
  {pages} {054035} (\bibinfo {year} {2022})},\ \Eprint
  {https://arxiv.org/abs/2112.10238} {arXiv:2112.10238 [hep-th]} \BibitemShut
  {NoStop}%
\bibitem [{\citenamefont {Gorda}\ \emph {et~al.}(2018)\citenamefont {Gorda},
  \citenamefont {Kurkela}, \citenamefont {Romatschke}, \citenamefont
  {S\"appi},\ and\ \citenamefont {Vuorinen}}]{Gorda:2018gpy}%
  \BibitemOpen
  \bibfield  {author} {\bibinfo {author} {\bibfnamefont {T.}~\bibnamefont
  {Gorda}}, \bibinfo {author} {\bibfnamefont {A.}~\bibnamefont {Kurkela}},
  \bibinfo {author} {\bibfnamefont {P.}~\bibnamefont {Romatschke}}, \bibinfo
  {author} {\bibfnamefont {M.}~\bibnamefont {S\"appi}},\ and\ \bibinfo {author}
  {\bibfnamefont {A.}~\bibnamefont {Vuorinen}},\ }\bibfield  {title} {\bibinfo
  {title} {{Next-to-Next-to-Next-to-Leading Order Pressure of Cold Quark
  Matter: Leading Logarithm}},\ }\href
  {https://doi.org/10.1103/PhysRevLett.121.202701} {\bibfield  {journal}
  {\bibinfo  {journal} {Phys. Rev. Lett.}\ }\textbf {\bibinfo {volume} {121}},\
  \bibinfo {pages} {202701} (\bibinfo {year} {2018})},\ \Eprint
  {https://arxiv.org/abs/1807.04120} {arXiv:1807.04120 [hep-ph]} \BibitemShut
  {NoStop}%
\bibitem [{\citenamefont {Gorda}\ \emph
  {et~al.}(2021{\natexlab{a}})\citenamefont {Gorda}, \citenamefont {Kurkela},
  \citenamefont {Paatelainen}, \citenamefont {S\"appi},\ and\ \citenamefont
  {Vuorinen}}]{Gorda:2021znl}%
  \BibitemOpen
  \bibfield  {author} {\bibinfo {author} {\bibfnamefont {T.}~\bibnamefont
  {Gorda}}, \bibinfo {author} {\bibfnamefont {A.}~\bibnamefont {Kurkela}},
  \bibinfo {author} {\bibfnamefont {R.}~\bibnamefont {Paatelainen}}, \bibinfo
  {author} {\bibfnamefont {S.}~\bibnamefont {S\"appi}},\ and\ \bibinfo {author}
  {\bibfnamefont {A.}~\bibnamefont {Vuorinen}},\ }\bibfield  {title} {\bibinfo
  {title} {{Soft Interactions in Cold Quark Matter}},\ }\href
  {https://doi.org/10.1103/PhysRevLett.127.162003} {\bibfield  {journal}
  {\bibinfo  {journal} {Phys. Rev. Lett.}\ }\textbf {\bibinfo {volume} {127}},\
  \bibinfo {pages} {162003} (\bibinfo {year} {2021}{\natexlab{a}})},\ \Eprint
  {https://arxiv.org/abs/2103.05658} {arXiv:2103.05658 [hep-ph]} \BibitemShut
  {NoStop}%
\bibitem [{\citenamefont {Gorda}\ \emph
  {et~al.}(2021{\natexlab{b}})\citenamefont {Gorda}, \citenamefont {Kurkela},
  \citenamefont {Paatelainen}, \citenamefont {S\"appi},\ and\ \citenamefont
  {Vuorinen}}]{Gorda:2021kme}%
  \BibitemOpen
  \bibfield  {author} {\bibinfo {author} {\bibfnamefont {T.}~\bibnamefont
  {Gorda}}, \bibinfo {author} {\bibfnamefont {A.}~\bibnamefont {Kurkela}},
  \bibinfo {author} {\bibfnamefont {R.}~\bibnamefont {Paatelainen}}, \bibinfo
  {author} {\bibfnamefont {S.}~\bibnamefont {S\"appi}},\ and\ \bibinfo {author}
  {\bibfnamefont {A.}~\bibnamefont {Vuorinen}},\ }\bibfield  {title} {\bibinfo
  {title} {{Cold quark matter at N3LO: Soft contributions}},\ }\href
  {https://doi.org/10.1103/PhysRevD.104.074015} {\bibfield  {journal} {\bibinfo
   {journal} {Phys. Rev. D}\ }\textbf {\bibinfo {volume} {104}},\ \bibinfo
  {pages} {074015} (\bibinfo {year} {2021}{\natexlab{b}})},\ \Eprint
  {https://arxiv.org/abs/2103.07427} {arXiv:2103.07427 [hep-ph]} \BibitemShut
  {NoStop}%
\bibitem [{\citenamefont {Pisarski}(2000)}]{Pisarski:1999gq}%
  \BibitemOpen
  \bibfield  {author} {\bibinfo {author} {\bibfnamefont {R.~D.}\ \bibnamefont
  {Pisarski}},\ }\bibfield  {title} {\bibinfo {title} {{Critical line for H
  superfluidity in strange quark matter?}},\ }\href
  {https://doi.org/10.1103/PhysRevC.62.035202} {\bibfield  {journal} {\bibinfo
  {journal} {Phys. Rev. C}\ }\textbf {\bibinfo {volume} {62}},\ \bibinfo
  {pages} {035202} (\bibinfo {year} {2000})},\ \Eprint
  {https://arxiv.org/abs/nucl-th/9912070} {arXiv:nucl-th/9912070} \BibitemShut
  {NoStop}%
\end{thebibliography}
%

\end{document}